\documentclass[a4paper,prl,aps,showpacs,twocolumn,superscriptaddress,footinbib]		{revtex4-1}
\usepackage	[pdftex, bookmarks, colorlinks, breaklinks]							{hyperref}

\usepackage				{xr}						% this allows cross-referencing between the two papers
\usepackage	[T1]			{fontenc}
\usepackage	[latin1]		{inputenc}
\usepackage	[english]		{babel}
\usepackage	[export]		{adjustbox}
\usepackage	[caption=false]	{subfig}
\usepackage				{graphicx, epstopdf, epsfig, rotating, floatpag}
\usepackage				{xcolor, color, verbatim, varioref, units, cancel}
\usepackage				{booktabs, tabu, longtable, dcolumn}
\usepackage				{amsmath, amssymb, amsfonts, nicefrac, xfrac, gensymb, braket, bm}%, bbm} 
\usepackage				{paralist, enumerate, lipsum}
\usepackage				{dsfont}					% for identity matrix using \mathds{1}
\usepackage                             {footmisc}					%for referring to the smae foot note twice

\hypersetup				{colorlinks=true, linkcolor=black, citecolor=blue, filecolor=dullmagenta, urlcolor=blue}	
\begin{document}

\title{Correlated Quantum Tunnelling of Monopoles in Spin Ice}

\author	{Bruno Tomasello}
\email	{brunotomasello83@gmail.com, tomasello@ill.fr}
\affiliation	{SEPnet and Hubbard Theory Consortium, University of Kent, Canterbury CT2 7NH, U.K.}
\affiliation	{ISIS facility, STFC Rutherford Appleton Laboratory, Harwell Campus, Didcot OX11 0QX,  U.K.}
\affiliation	{Institut Laue-Langevin, CS 20156, 71 avenue des Martyrs, 38042 Grenoble Cedex 9, France}

\author	{Claudio Castelnovo}
\affiliation	{TCM group, Cavendish Laboratory, University of Cambridge, Cambridge CB3 0HE, U.K.}

\author	{Roderich Moessner}
\affiliation	{Max-Planck-Institut f\"{u}r Physik komplexer Systeme, 01187 Dresden, Germany}

\author	{Jorge Quintanilla}
\email	{j.quintanilla@kent.ac.uk}
\affiliation	{SEPnet and Hubbard Theory Consortium, University of Kent, Canterbury CT2 7NH, U.K.}
\affiliation	{ISIS facility, STFC Rutherford Appleton Laboratory, Harwell Campus, Didcot OX11 0QX,  U.K.}

\begin{abstract} 
	The spin ice materials Ho$_{2}$Ti$_{2}$O$_{7}$ and Dy$_{2}$Ti$_{2}$O$_{7}$ are by now perhaps the best-studied classical frustrated magnets. A crucial step towards the understanding of their low temperature behaviour -- both regarding their unusual dynamical properties and the possibility of observing their quantum coherent time evolution -- is a quantitative understanding of the spin-flip processes which underpin the hopping of magnetic monopoles. We attack this problem in the framework of a quantum treatment of a single-ion subject to the crystal, exchange and dipolar fields from neighbouring ions. By studying the fundamental quantum mechanical mechanisms, we discover a bimodal distribution of hopping rates which depends on the local spin configuration, in broad agreement with rates extracted from experiment. Applying the same analysis to Pr$_{2}$Sn$_{2}$O$_{7}$ and Pr$_{2}$Zr$_{2}$O$_{7}$, we find an even more pronounced separation of timescales signalling the likelihood of coherent many-body dynamics. 
\end{abstract}

\maketitle

%%%%%%%%%%%%%%%%%%%%%%%%%%%%%%%%%%%%%%%%%
\emph{Introduction ---} 
	Some of the most exciting discoveries in strongly correlated systems in recent years are related to topological phases of matter. These are novel types of vacua hosting quasiparticle excitations charged under an emergent gauge field. In contrast to Fermi \cite{Landau:1956,Landau:1957} and Luttinger liquids~\cite{Luttinger:1963}, where quasiparticles provide a complete description of all low energy states, here one needs to keep track of the joint time evolution of quasiparticles and gauge fields. Doing this in full generality is a highly non-trivial task~\cite{Szabo:2019}, and remains largely unexplored, despite the huge interest in the context of, e.g., parton theories of correlated quantum matter~\cite{Beran:1996,Grusdt:2018}. In practice, one can instead resort to largely uncontrolled approximation schemes such as mean-field treatments in which the particle moves in an averaged background gauge field.

	The spin ice compounds Ho$_{2}$Ti$_{2}$O$_{7}$ (HTO) and Dy$_{2}$Ti$_{2}$O$_{7}$ (DTO)~\cite{Bramwell:2001} are generally believed to host a topological Coulomb spin liquid~\cite{Castelnovo:2008}. Here, the emergent gauge field is particularly simple to visualise, as its gauge flux is encoded in the spins themselves. The motion of a quasiparticle amounts to spin flip processes, which are subject to the energetics and quantum dynamics of the underlying many-body Hamiltonian. Despite the elegance of this physical scenario, a comprehensive understanding of the large variety of experimental timescales~\cite{Snyder:2003, Bramwell:2009, Giblin:2011, Jaubert:2011, Matsuhira:2011, Snyder:2001, Sala:2012, Paulsen:2014} is still lacking. Puzzling timescales are not uncommon in geometrically frustrated magnets~\cite{Nambu:2015}.

\begin{figure}[ht!]
	\centering
	\subfloat	[\label{fig:TwoTetrahedra(a)}]
			{\includegraphics[trim=	0mm		0mm		0mm		0mm, clip, width=.48\linewidth]{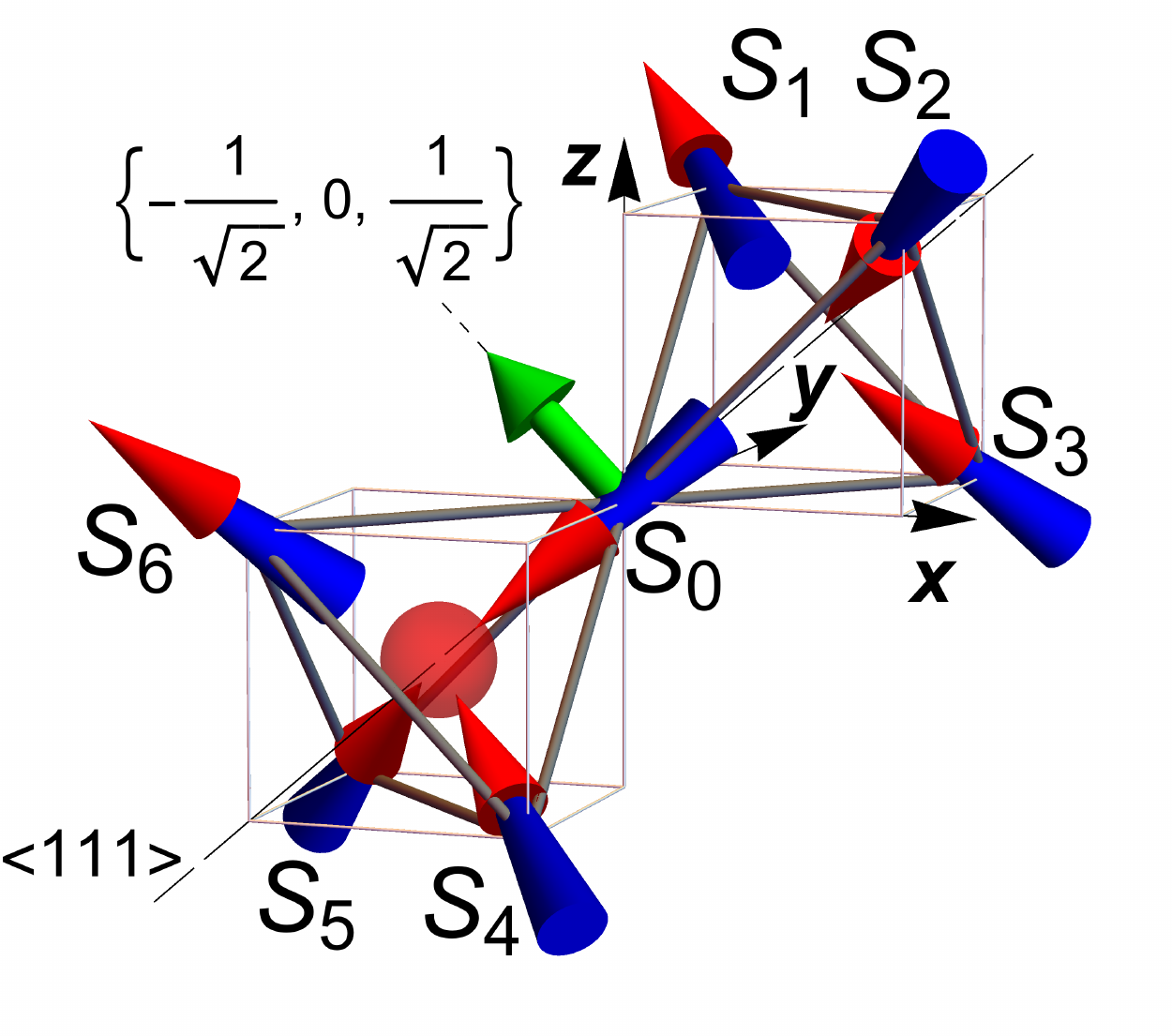}}
	\quad
	\subfloat	[\label{fig:TwoTetrahedra(b)}]
			{\includegraphics[trim=	0mm		0mm		0mm		0mm, clip, width=.48\linewidth]{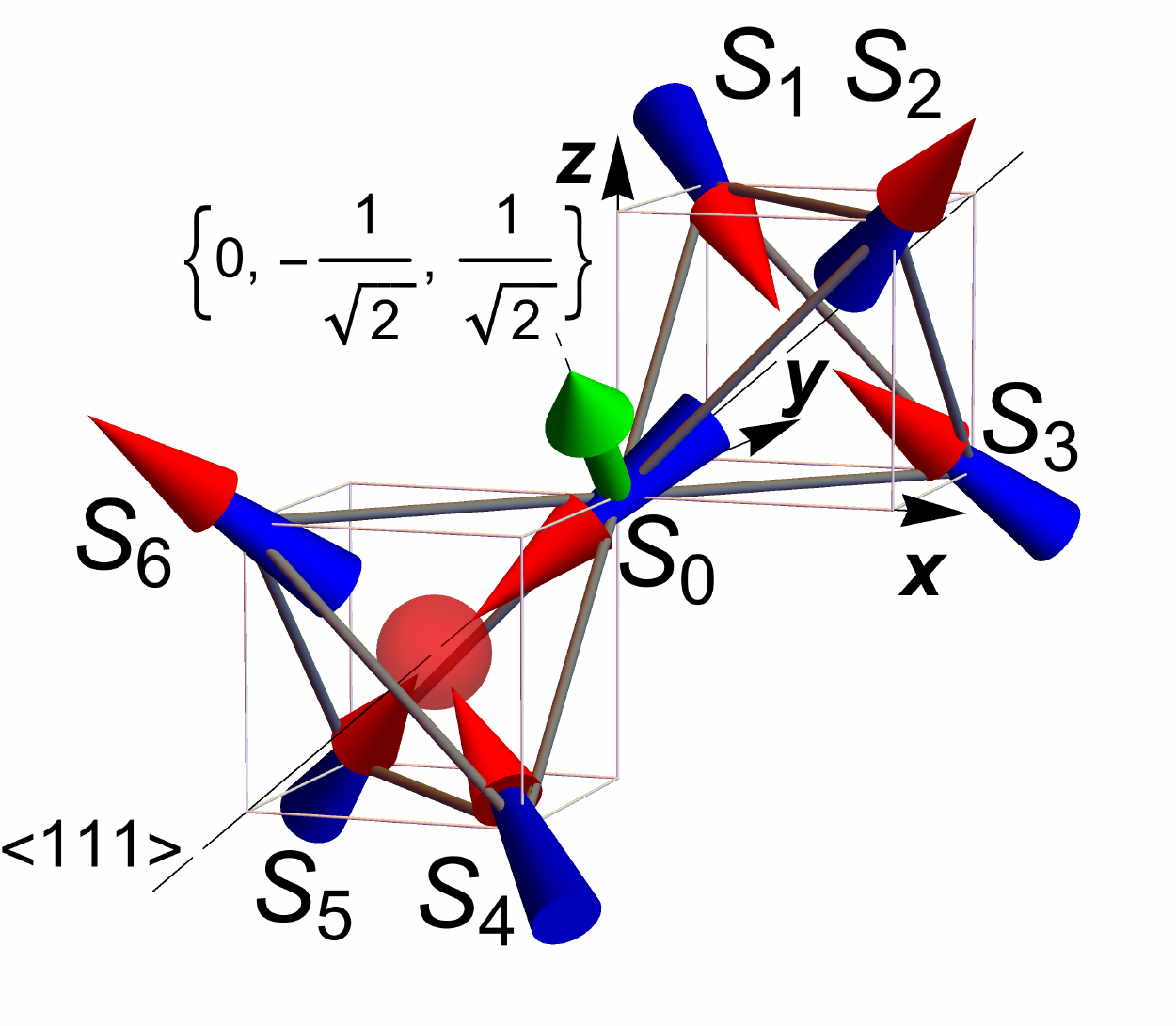}}

	\subfloat	[\label{fig:TwoTetrahedra(c)}]
			{\includegraphics[trim=	0mm		0mm		0mm		0mm, clip, width=.48\linewidth]{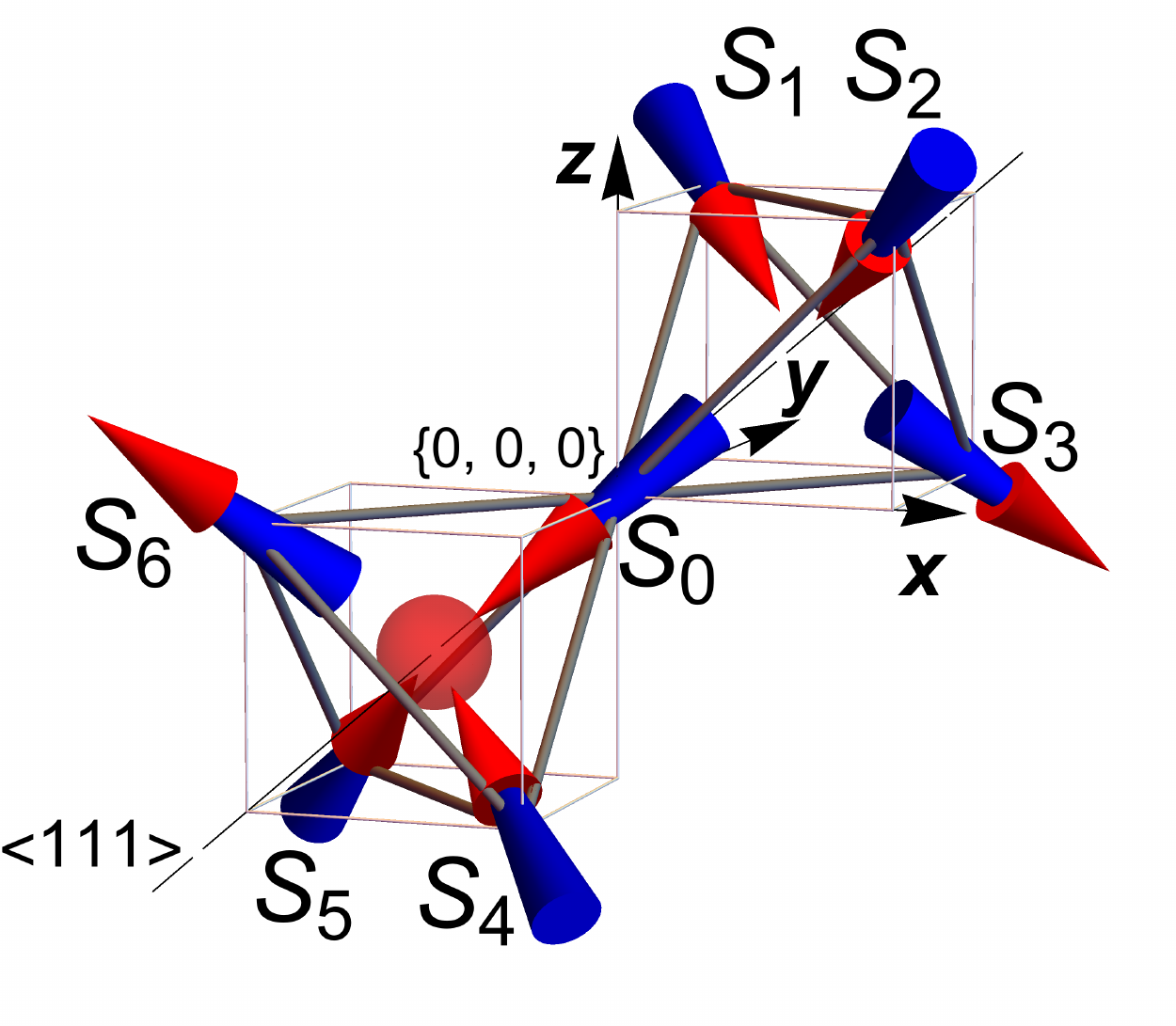}}
	\quad
	\subfloat	[\label{fig:PairedHistogram}]	
			{\includegraphics[trim=	0mm		0mm		0mm		0mm, clip, width=.48\linewidth]{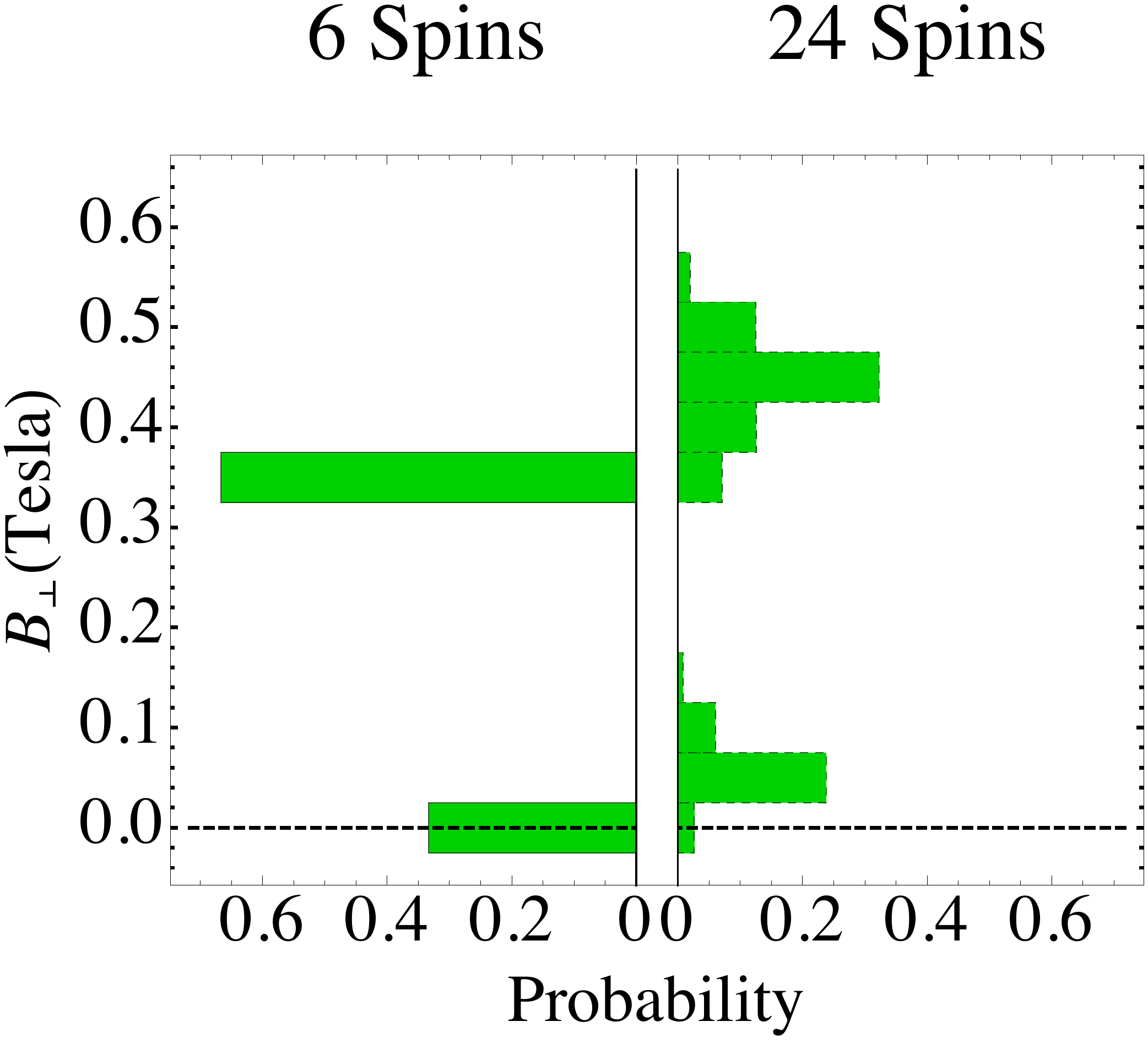}}
	\caption{
		(a-c) {Choice of all} 3 inequivalent low-energy configurations of a 2-tetrahedron system hosting only one monopole (red sphere). {Both dipolar and exchange f}ields on the central site $0$ due to its $j=1,2, \dots, 6$ n.n. (nearest neighbours) are purely transverse to $\mathbf{z}_{0}\propto \langle111\rangle$ in $2/3$ of the cases (a,b -- vectors in green), and are identically null in the remaining $1/3$ (c). (d) Histogram of the dipolar fields resulting on $0$ from its 6 n.n. spins (left panel), and from the inclusion of further 18 spins in the n.n.n tetrahedra (right panel). This verifies that the bimodal distribution (2:1) is {largely} unaffected when we consider an 8-tetrahedron system (24 surrounding spins).
	}	
	\label{fig:TwoTetrahedra}
\end{figure}

	In this paper, we report an analysis of the elementary building block of quasiparticle motion, with the detailed microscopic knowledge on spin ice available in the literature~\cite{Bramwell:2001,Castelnovo:2008} as foundation. We study the local dynamics of emergent monopole excitations, which has a quantum mechanical (tunnelling) origin, rooted in the transverse terms of the dipolar and exchange interactions between rare-earth (RE) ions~\cite{Bovo:2013,Tomasello:2015}. We focus on elastic processes (monopole hopping), since inelastic ones (monopole creation/annihilation) are suppressed at low temperatures. The key question is: how do the predominantly off-diagonal terms (`transverse fields') necessary to induce monopole hopping arise in a material whose statistics are excellently described by a classical Ising model? Our central result is that there is a fundamental feedback mechanism between spin dynamics, monopole quasiparticles, and the local spin environment. Whereas the vast majority of spins in the sample experience longitudinal fields, which justify a classical description, some of the spins adjacent to a monopole experience predominantly transverse fields. As illustrated in Fig.~\ref{fig:TwoTetrahedra}, a monopole has 3 available lattice bonds to hop across, and, statistically, we find a bimodal distribution of transverse fields, and thence of quasiparticle hopping rates, in ratio 2:1 (fast:slow). We posit that these $\tau^{\rm fast}$ and $\tau^{\rm slow}$ are the fundamental (tunnelling) timescales underlying a broad range of dynamic phenomena in spin ices~\cite{Snyder:2003, Bramwell:2009, Giblin:2011, Jaubert:2011, Matsuhira:2011, Snyder:2001, Sala:2012, Paulsen:2014} and find they are  consistent with experimental timescales (Table \ref{tab:Timescales}). We extend our calculations to `quantum spin ices' Pr$_{2}$Sn$_{2}$O$_{7}$ (PSO) and Pr$_{2}$Zr$_{2}$O$_{7}$ (PZO) and find, as expected, much faster timescales and also, more surprisingly, much greater separation between $\tau^{\rm fast}$ and $\tau^{\rm slow}$. Finally we argue that decoherence may play an essential role in the emergence of slow, classical spin flips out of fast, quantum tunnelling and we provide a simple model based on the Zeno effect.  Interestingly the large separation of timescales in PSO and PZO implies a coexistence in these systems of coherent and incoherent processes.

%%%%%%%%%%%%%%%%%%%%%%%%%%%%%%%%%%%%%%%%%
\emph{Model ---} 
	Spin ices are magnets where anisotropic Ising-like spins reside on a pyrochlore lattice of corner-sharing tetrahedra. The exchange and dipolar interactions between the spins are largely frustrated, and at low temperatures the ground state is described by an extensively degenerate manifold of configurations obeying the so called `ice rules' (in each tetrahedron, 2 spins point `in', towards its centre, and 2 point `out')~\cite{Bramwell:2001}. The lowest excitations above such ground state are effective magnetic monopoles with Coulomb interactions~\cite{Castelnovo:2008}. 

	The local Ising anisotropies originate from the strong crystalline-electric-fields (CEF) acting on the $J$-manifold of the RE$^{3+}$ ions (for Ho$^{3+}$, Dy$^{3+}$ and Pr$^{3+}$ ions, the total angular momentum quantum number is, respectively, $J=8,15/2$ and $4$)~\cite{Tomasello:2015, Ruminy:2016}. Low energy dynamics between the single-ion states of the ground-state doublet, $\ket{-}_{i}$ and $ \ket{+}_{i}$ (labelled by $S_{i}=-1,1$), necessarily involve transitions via the CEF excited states, with energies $\Delta E \gtrsim 10^2$~K~\cite{Rosenkranz:2000, Ehlers:2003}. In the temperature range where the monopole description is valid ($T \lesssim 1$~K), thermal activation of CEF excited states is negligible so that quantum tunnelling must underpin the spin dynamics~\cite{Snyder:2003}. This provides a mechanism for the flipping of the minority of spins that are not frozen by a local (longitudinal) combined dipolar and effective exchange field. These are of course the flippable spins next to a monopole. 

	We focus on a given spin, say {at} $i=0$, to study the single spin-flip dynamics which amounts to the hopping of a monopole. Our Hamiltonian, 
\begin{gather}
	\hat{\mathcal{H}}(0)
	=	
	\hat{\mathcal{H}}_{\rm CEF}
	+ 
	\hat{\mathcal{H}}_{\rm dip}(0) 
	+ 
	\hat{\mathcal{H}}_{\rm exc}(0) 
 	, 
\label{eq:HamiltonianMain}
\end{gather}
describes a RE-ion at site $0$ of an $N$-site pyrochlore system. $\hat{\mathcal{H}}$ acts on the Hilbert space of the RE$^{3+}$ of interest with total angular momentum quantum number $J$. (We work in the $2J+1$ dimensional $\ket{M} \equiv \ket{J,M}$ eigenbasis of $\hat{J}_{z}$ for the local quantisation axis $\mathbf{z}_{0} \propto \langle111\rangle$.) $\hat{\mathcal{H}}_{\rm CEF}$ is the crystal-field Hamiltonian~\cite{Tomasello:2015}, and $\hat{\mathcal{H}}_{\rm dip}(0)$ and $\hat{\mathcal{H}}_{\rm exc}(0)$ describe, respectively, dipolar and exchange interactions with other RE$^{3+}$~ions.

	Our main approximation for Eq.~\eqref{eq:HamiltonianMain} is that each of the other $N-1$ spins in the system is projected onto \emph{one} of its own CEF ground-state doublet states, $\ket{\pm}_{j}$ (i.e. $S_{j}=\pm 1$ for $j\neq 0$). We thus ignore joint dynamical correlations that may develop in the simultaneous motion of all flippable spins (e.g., spins $S_{0}, S_{4}, S_{5}$ in Fig.~\ref{fig:TwoTetrahedra} ought to be studied together, as potentially entangled spins~\footnote{The fourth spin {$S_{6}$} remains static because its reversal would produce higher energy defects (double monopoles)}). Notice however that the simultaneous re-orientation of two or three of the spins $S_{0}, S_{4}, S_{5}$ in Fig.~\ref{fig:TwoTetrahedra} produces higher excitated states; and that the re-orientation of one of them generates a longitudinal field pinning the other two in their initial state. This observation supports the validity of our approximation.

%%%%%%%%%%%%%%%%%%%%%%%%%%%%%%%%%%%%%%%%%
\emph{Dipolar interactions ---} 
	Let us consider the simplest pyrochlore system of interest for the hopping of a monopole: two adjacent tetrahedra, where only one tetrahedron hosts a monopole and flipping the central spin $S_{0}$ allows the monopole to hop to the other tetrahedron. There are only 3 symmetry-inequivalent such spin configurations, illustrated in Fig.~\ref{fig:TwoTetrahedra}. 

	Using the conventional dipolar interaction~\cite{Bramwell:2001}, it is straightforward to show that the 6 neighbouring spins exert a field on the central site given by 
\begin{equation}
	\mathbf{B}_{\rm dip}^{\{\rm 6\}}(0)		
	=
	\frac{\mu_0 |\mathbf{m}|}{4 \pi r_{\rm nn}^{3}}
	\sum_{j=1,2,3}
	( \hat{\mathbf{z}}_{j} + \sqrt{6} \hat{\mathbf{r}}_{j}  )
	\left( S_{j} + S_{j+3}  \right)
	. 
	\label{eq:Bdip}
\end{equation}
As represented in Figs.~\ref{fig:TwoTetrahedra(a)}-\ref{fig:TwoTetrahedra(c)}, spin pairs $(S_{j}, S_{j+3})$ with $j=1,2,3$, have the same anisotropies $\hat{\mathbf{z}}_{j} = \hat{\mathbf{z}}_{j+3}$ and opposite n.n. positions ${\mathbf{r}}_{j} \equiv {\mathbf{r}}_{0j}=r_{\rm nn} \hat{\mathbf{r}}_{j} = -{\mathbf{r}}_{0j+3}$, with $|\hat{\mathbf{z}}_{j}|=|\hat{\mathbf{r}}_{j}|=~1$. 

	Using Eq.~\eqref{eq:Bdip}, we find that all 2-tetrahedron configurations hosting one monopole next to a flippable spin (Fig.~\ref{fig:TwoTetrahedra}) have vanishing longitudinal component, $B_{\|}= \mathbf{B}_{\rm dip}^{\{\rm 6\}}(0) \cdot \hat{\mathbf{z}}_{0}=0$, as expected for flippable spins. Remarkably, such spin configurations do not all give the same (transverse) fields: in $2/3$ of the cases, Figs.~\ref{fig:TwoTetrahedra(a)}-\ref{fig:TwoTetrahedra(b)}, $\mathbf{B}_{\bot}$ is finite and points along {one of} the high-symmetry CEF angles $\phi_{n}$ of Ref.~\cite{Tomasello:2015}; whereas in the remaining $1/3$, Fig.~\ref{fig:TwoTetrahedra(c)}, {$\mathbf{B}_{\rm dip}^{\{\rm 6\}}(0)=0$} identically, since ${\mathbf{r}}_{0}$ is a centre of inversion~\cite{pie}.

	Because of corrections to the projective equivalence~\cite{Isakov:2005}, dipolar-fields from farther spins ($N-1>6$) can alter our conclusions from the 2-tetrahedron system. To check this, in Fig.~\ref{fig:PairedHistogram} we compare the field-distribution of $\mathbf{B}_{\rm dip}^{\{6\}}$ (left panel) and $\mathbf{B}_{\rm dip}^{\{24\}}$ (right panel) compatible with the monopole constraint in Figs.~\ref{fig:TwoTetrahedra(a)}-\ref{fig:TwoTetrahedra(c)} ($\mathbf{B}_{\rm dip}^{\{24\}}$ includes the farther 18 spins belonging to the next 6 tetrahedra adjacent to the 2 tetrahedra at $\mathbf{r}_{1}, \dots, \mathbf{r}_{6}$). The histograms in Fig.~\ref{fig:PairedHistogram} show that the bimodal field-distribution is {qualitatively unchanged}: the two values for $B_{\bot}^{\{6\}}$ evolve into two well separated distributions for $B_{\bot}^{\{24\}}\approx 0.45$~Tesla in $2/3$ of the cases, and $B_{\bot}^{\{24\}} \approx 0.03$~Tesla in $1/3$. {Note that the differences in values between $\mathbf{B}_{\rm dip}^{\{6\}}$ and $\mathbf{B}_{\rm dip}^{\{24\}}$, and in particular the non-zero spreads, are due to quadrupolar corrections, which are well-captured by the $24$ spin calculation. Indeed, the histograms in Fig.~\ref{fig:PairedHistogram} agree quantitatively with Monte Carlo simulations on larger systems~\cite{Sala:MonteCarlo}.}

	These results imply that the associated spin dynamics is remarkably correlated to the local environment. For a flippable spin next to a monopole, two very distinct flipping rates, $\tau^{\rm fast}$ and $\tau^{\rm slow}$, appear, with a 2:1 ratio. In the following we show that the same comes to pass for the full fledged form of the exchange interactions.

%%%%%%%%%%%%%%%%%%%%%%%%%%%%%%%%%%%%%%%%%
\emph{Exchange interactions ---} 
	To achieve a realistic model of exchange couplings in RE$^{3+}$~pyrochlores, we first write
\begingroup\belowdisplayskip=\belowdisplayshortskip
\begin{gather}	
	\begin{split}
		\hat{\mathcal{H}}_{f\!f} 
		&
		(\mathbf{r},\! \mathbf{r'})
				=
		\mathcal{E}_{\rm exc} 
		\!\!\!\!
		\sum_{\substack
						{
							m_{1},		m_{2}\\
							\sigma_{1},	\sigma_{2}
						}
		}		
		\!													
		\sum_{\substack
						{
							m'_{1},			m'_{2}\\
							\sigma'_{1},	\sigma'_{2}	
						}
		}															
		\!\!\!
		\hat{f}^{\dagger	}_{\mathbf{r},	m_{1},\sigma_{1}}								
		\hat{f}^{				}_{\mathbf{r},	m_{2},\sigma_{2}}		
		\hat{f}^{\dagger	}_{\mathbf{r'},	m'_{1},\sigma'_{1}}				
		\hat{f}^{				}_{\mathbf{r'},	m'_{2},\sigma'_{2}}					
		\\
		\times \,
		&
		x^{\vert m_{1} \vert+\vert m'_{1} 
		   \vert+\vert m_{2} \vert+\vert m'_{2} \vert}
		\\
		\times \,
		&
		\Bigg[
		a	\,	
			\delta_{  \substack{ m_{1}\!, m_{2}	 \\ \sigma_{1}\!, \sigma_{2} } }
			\delta_{  \substack{ m'_{1}\!, m'_{2}	 \\ \sigma'_{1}\!, \sigma'_{2} } }						
			\!\!+\!
			( \mathcal{R}^{\dagger}_{\mathbf{r}}	\mathcal{R}_{\mathbf{r'}} )_{ \substack{ m_{1},	m'_{2} \\ \sigma_{1}, \sigma'_{2} } }
			( \mathcal{R}^{\dagger}_{\mathbf{r'}}	\mathcal{R}_{\mathbf{r}} )_{ \substack{ m'_{1},	m_{2} \\ \sigma'_{1}, \sigma_{2}} }
		 \Bigg]	
		\, ,
	\end{split}
\label{eq:Gen17OnodaCompactMainText}
\end{gather}
the Hamiltonian for the oxygen-mediated super-exchange of $f$-electrons between two n.n. RE$^{3+}$~ions at $\mathbf{r}$ and $ \mathbf{r'}$. Eq.~\eqref{eq:Gen17OnodaCompactMainText} generalises Eq.~(18) in Ref.~\cite{Onoda:2011} -- originally written for Pr$^{3+}$ pyrochlores -- to any RE$^{3+}$~pyrochlore (details in Ref.~\cite{Tomasello:PhDThesis}). The operator $\hat{f}^{\dagger}_{\mathbf{r}, m, \sigma}$ ($\hat{f}^{}_{\mathbf{r}, m, \sigma}$) creates (annihilates) at $\mathbf{r}$ an $f$-electron with orbital and spin magnetic quantum numbers $m=0,\pm1$  and $m_{s} = {\sigma}/{2}$, respectively ($\sigma= \pm1$). $\mathcal{R}^{\dagger}_{\mathbf{r}} \mathcal{R}_{\mathbf{r'}}$ matches the local systems of coordinates at $\mathbf{r}$ and $\mathbf{r'}$. The parameters $\mathcal{E}_{\rm exc}= 2 \frac{V_{pf\sigma}^4} {(n U - \Delta)^2} \left( \frac{1}{n U - \Delta}+ \frac{1}{U} \right)$, $a={U}/{(\Delta - U(n+1))}$, and $x= V_{pf\pi}/V_{pf\sigma}$ contain the complex relationships between the $n$ electrons in the $f$-shell, their (repulsive) Coulomb energy $U$, the change in energy $\Delta$ for the removal of an electron from the oxygen, and, most importantly, the Slater-Koster hybridisation parameters $V_{m=\pm1}=V_{pf\pi}$ {and} $  V_{m=0}= V_{pf\sigma}$. 

	We then project each n.n. ion, as we did for the case of dipolar interactions, to obtain the exchange Hamiltonian
\begin{gather}
		\hat{\mathcal{H}}_{\rm exc}(0)=
		\sum_{j=1}^{6}
		\bra{\pm}_{j} \hat{\mathcal{H}}_{f\!f} (\mathbf{r}_{0},\mathbf{r}_{j}) \ket{\pm}_{j},
\label{eq:HamExcQuasiQuantMainText}
\end{gather}
which operates in the  $2 J+1$ dimensional Hilbert space of the central `single-ion'. This is a highly anisotropic Hamiltonian that cannot be easily interpreted as a  field distribution ($\hat{\mathcal{H}}_{\rm exc} \neq - g_{J} \mu_{\mathrm{B}}\hat{\mathbf{J}} \cdot \mathbf{B}_{\rm exc}^{\{6\}}$). We study it by considering $\braket{ \hat{\mathbf{J}}}_{}=\braket{\psi| \hat{\mathbf{J}} |\psi}_{}$, where $\ket{\psi}$ is the ground state of $\hat{\mathcal{H}}_{\rm CFx }(0)= \hat{\mathcal{H}}_{\rm CEF}+\hat{\mathcal{H}}_{\rm exc}(0)$. Once again, the spin configurations in Fig.~\ref{fig:TwoTetrahedra} exhibit a bimodal behaviour: in $2/3$ of the cases, $\braket{ \hat{\mathbf{J}}}_{}$ is purely transverse to $\mathbf{z}_{0}$ at $\phi_{n}$ angles (e.g., Figs.~\ref{fig:TwoTetrahedra(a)}-\ref{fig:TwoTetrahedra(b)}); in $1/3$ of the cases $\braket{ \hat{\mathbf{J}}}_{}=0$ (e.g., Fig.~\ref{fig:TwoTetrahedra(c)})~\cite{pie}. As a matter of fact, for this inversion-symmetric case, $\hat{\mathcal{H}}_{\rm exc}(0)$ is diagonal in the $\ket{M}$ basis, and symmetric under $M \leftrightarrow -M$.

	The above behaviour holds for any $\mathcal{E}_{\rm exc}$, $a$ and $x$, as long as $\hat{\mathcal{H}}_{\mathrm{exc}}$ is a small perturbation to $\hat{\mathcal{H}}_{\rm CEF}$. Notwithstanding, we summarise here how we set these parameters to obtain quantitative results (further details {can be found} in the Supplemental Material~\footnote{See Supplemental Material (SM) at [URL will be inserted by publisher] for more details.}). Firstly, a relationship between $a$ and $x$ is found by requiring the diagonal part of $\hat{\mathcal{H}}_{f\!f}(\mathbf{r},\mathbf{r'})$, projected onto the ground state CEF doublet manifold of each of the two spins involved, to be proportional to $\sigma_{\mathbf{r}}^{z} \otimes \sigma_{\mathbf{r'}}^{z}$, which is a central feature in both classical and quantum (n.n.) spin-ice Hamiltonians~\cite{Bramwell:2001,Onoda:2011}. Then the behaviour of $\hat{\mathcal{H}}_{\rm exc}(0)$ projected on the GS doublet of the central spin is compared with $\hat{\mathcal{H}}_{\mathrm{Ising}}= J_{\mathrm{nn}} \sum_{j=1}^{6} \sigma_{\mathbf{r_0}}^{z} \otimes \sigma_{\mathbf{r_j}}^{z}$ for given configurations of the 6 outer spins. {The} value of $J_{\mathrm{nn}}$, obtained experimentally, sets {therefore} $\mathcal{E}_{\rm exc}$ as a function of $x$. 

	We are finally left with only one parameter, $x$, which was argued in Ref.~\cite{Rau:2015} to vary in the range $(-1,0)$. We study its effect by looking at the behaviour of the central spin under the single-ion Hamiltonian $\hat{\mathcal{H}}_{\rm CEF}+\hat{\mathcal{H}}_{\rm exc}(0)$ derived from a 2-tetrahedron system where the 6 outer spins are projected on different CEF GS configurations. We observe no appreciable change in HTO and DTO: the results are consistent throughout the range with the known sign of the interaction and a nearly fully polarised GS dipole moment. The same is not true for PSO and PZO, and in particular the correct sign of the exchange interactions (opposite to HTO and DTO) occurs only for $x \in (-1,-0.3)$ in PSO and for $x \in (-1,-0.6)$ in PZO. Beyond these values we observe a reversal of the GS dipole moment, with corresponding closing of the gap, signalling a change between ferromagnetic and antiferromagnetic nature of the exchange interactions. Such phenomena are worthy of further investigation in their own right, but are beyond the scope of the present work. {Here, we find it sufficient} to set $x=-1$ for all systems.

%%%%%%%%%%%%%%%%%%%%%%%%%%%%%%%%%%%%%%%%%
\emph{Spin dynamics and timescales ---} 
	We study the unitary spin dynamics of the central spin under the influence of both dipolar and exchange interactions by means of Eq.~\eqref{eq:HamiltonianMain}. In Fig.~\ref{fig:TimeplotQuench} the magnetic moment $\braket{ \hat{\mathbf{J}}}_{} \approx \braket{\hat{J}_{z}}$, initialised in $\ket{-}$, completely reverses direction ($\ket{+}$) with a precession timescale $\tau = \pi \hbar / \Delta E_{\pm}$ ($\tau^{\rm fast}$ and $\tau^{\rm slow}$ correspond to {the cases illustrated in Fig.~\ref{fig:TwoTetrahedra(a)}-\ref{fig:TwoTetrahedra(b)} and Fig.~\ref{fig:TwoTetrahedra(c)}, respectively).} 
For HTO and DTO, Figs.~\ref{fig:TimeplotQuenchExcHTO}-\ref{fig:TimeplotQuenchExcDTO}, it is worth noting that, in spite of having weak transverse terms~\cite{Rau:2015}, exchange does make a quantitative difference (e.g., {in DTO} dipolar fields alone give $\tau^{\rm fast} \approx 33$~$\mu$s{\color{black}, in contrast to $1.5 \ \mu$s for the full Hamiltonian}). In PSO and PZO (Figs.~\ref{fig:TimeplotQuenchExcPSO}-\ref{fig:TimeplotQuenchExcPZO}), $\tau^{\rm fast}$ is several orders of magnitude shorter than for HTO and DTO, consistent with the expectation of much stronger quantum fluctuations. 

\begin{figure}
	\centering
	\subfloat	[HTO, Dip. + Exc. \label{fig:TimeplotQuenchExcHTO}]
					{\includegraphics[trim=	0mm		15		45		0, clip,  width=.48\linewidth]{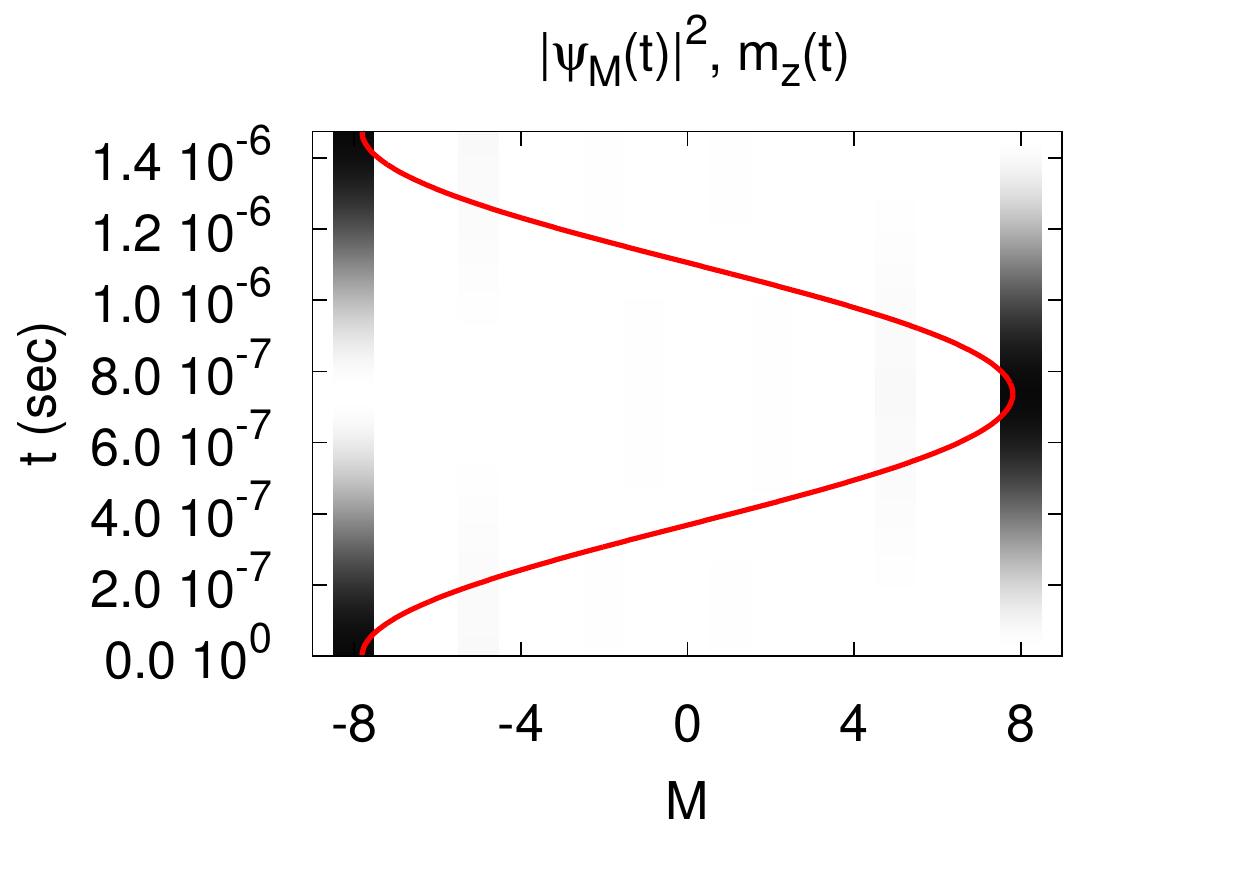}}
	\quad
	\subfloat	[DTO, Dip. + Exc. \label{fig:TimeplotQuenchExcDTO}]
					{\includegraphics[trim=	0mm		15		45		0, clip,  width=.48\linewidth]{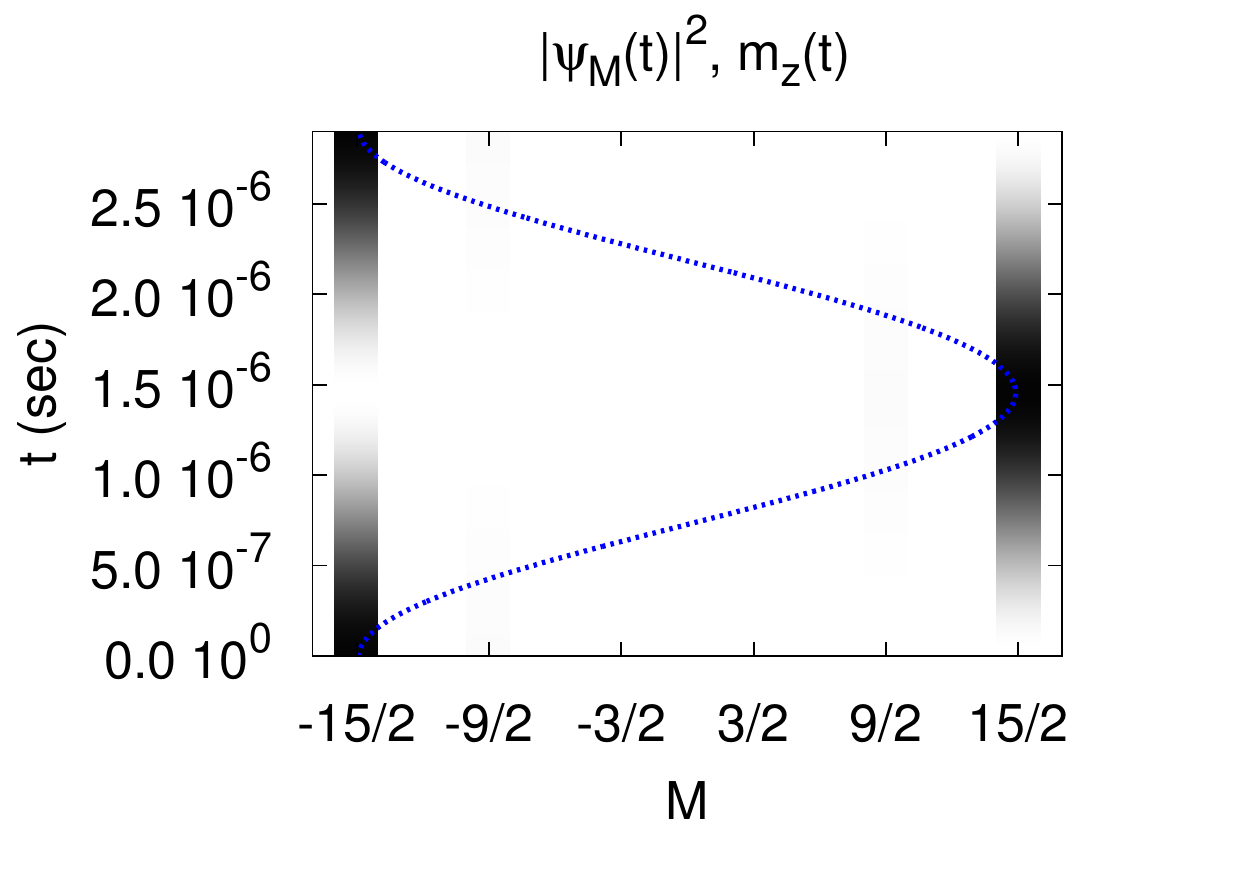}}
					
	\subfloat	[PSO, Dip. + Exc. \label{fig:TimeplotQuenchExcPSO}]
					{\includegraphics[trim=	0mm		15		45		0, clip,  width=.48\linewidth]{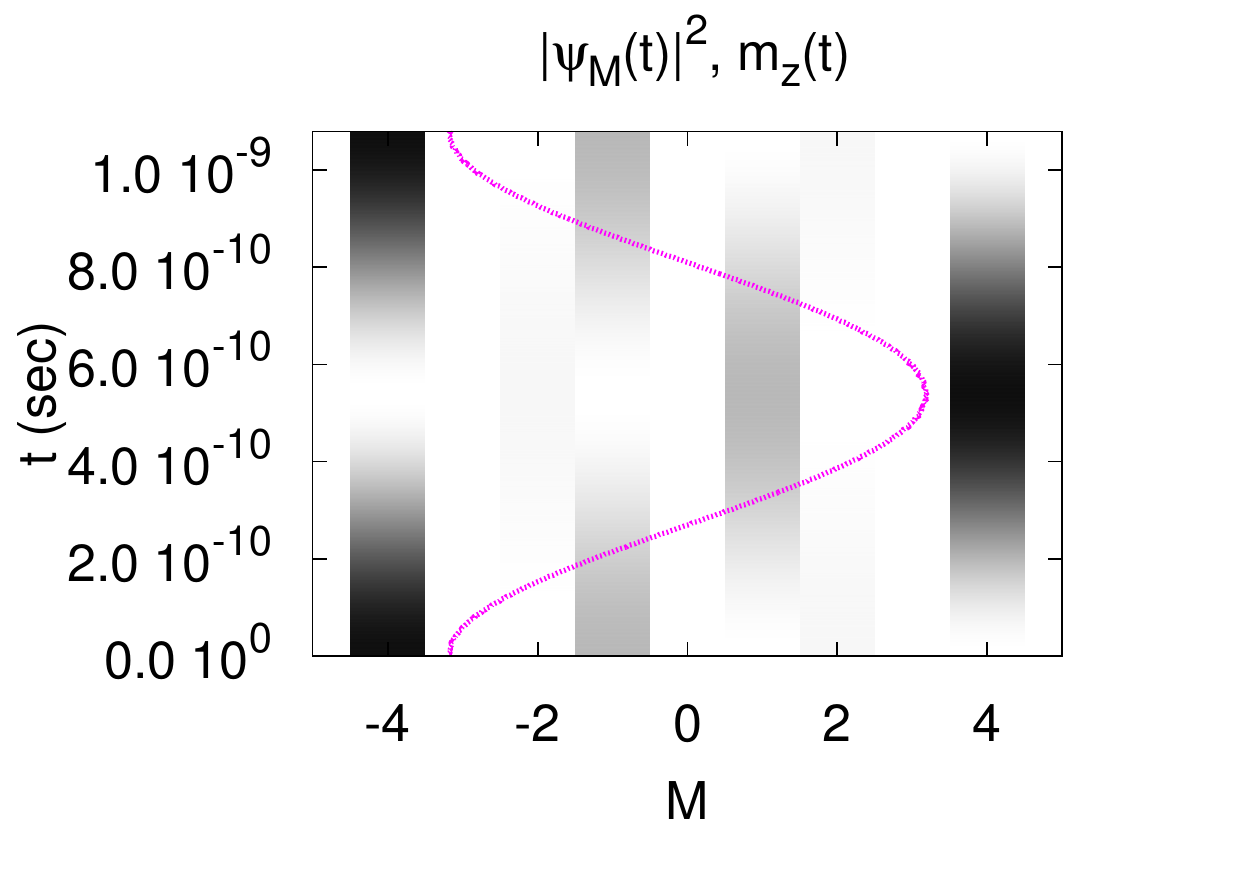}}
	\quad
	\subfloat	[PZO, Dip. + Exc. \label{fig:TimeplotQuenchExcPZO}]
					{\includegraphics[trim=	0mm		15		45		0, clip,  width=.48\linewidth]{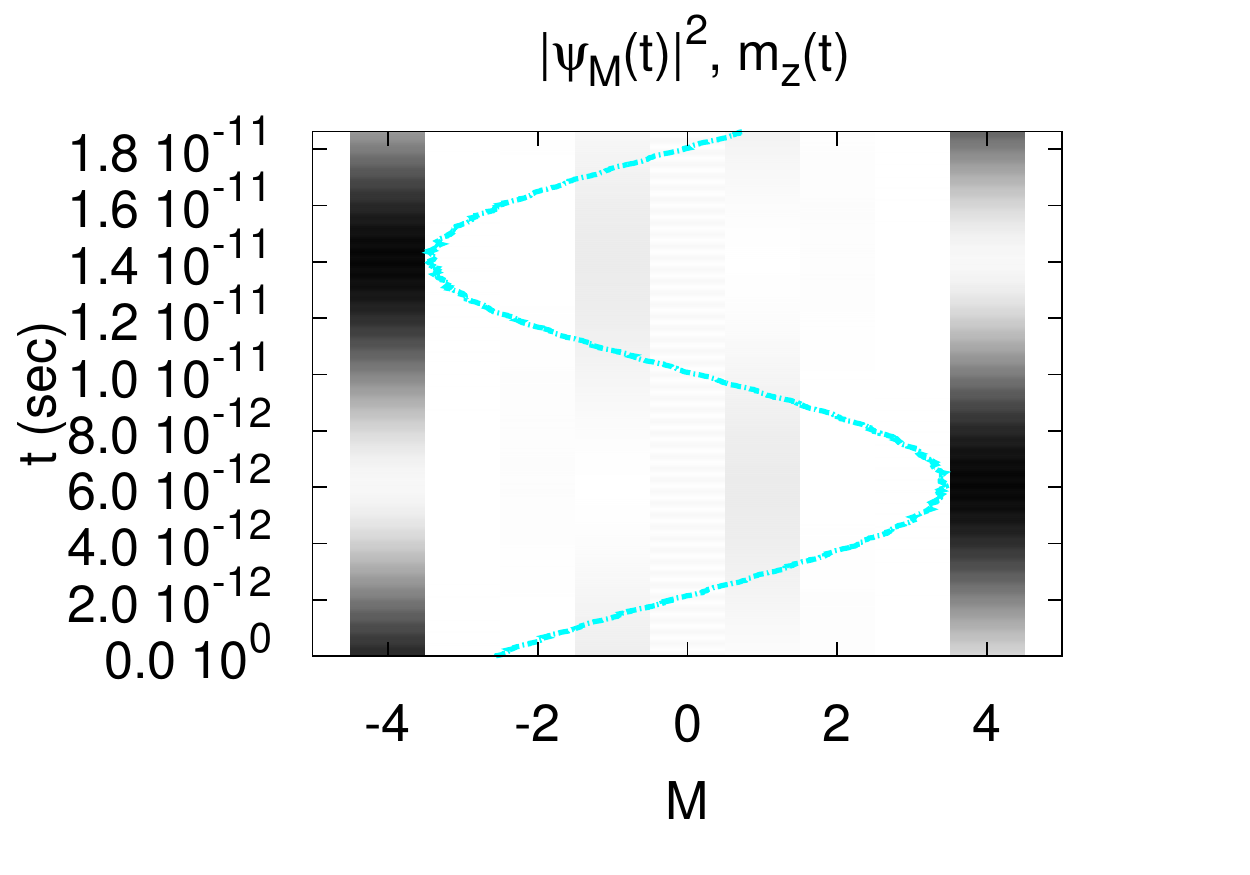}}					
	\caption{
		Time-evolution of the probability density $P_{M}(t)=\left| \braket{ M_{}| \psi(t)} \right|^{2}$ (black regions, $P_{M}(t) \approx 1$) as a function of $M=-J, \dots, J-1, J$ and $t$~(sec), as discussed in the main text. On each density plot is overlaid the curve $\braket{\hat{J}_{z}(t)}$. 
	}
	\label{fig:TimeplotQuench}
\end{figure}

\begin{table}[ht]	\centering
\setlength{\tabcolsep}{0.12cm}
\def\arraystretch{1.2}%
\begin{tabular}{ l | c | c | c | c | }
	$ \left[x=-1\right]$	& HTO 		& DTO		& PSO				& PZO 				\\
	\hline
	\hline
		$a$
					& -0.17		& -0.17		& -0.20  				& -0.19 				\\
	\hline
		$\mathcal{E}_{\rm exc}$
					& 0.15		& 0.22	 	& -32	  			& -87 				\\
	\hline
	\hline
		$\tau^{\rm fast}$
					& $0.74 \, 10^{-6}$	& $1.5 \, 10^{-6}$	& 	$5.4 \, 10^{-10} $		& $7.9 \, 10^{-12}$		\\	
	\hline
		$\tau^{\rm slow}$
					& $2.0  \, 10^{-4}$		& $4.4 \, 10^{-2}$	& $1.0 \, 10^{-4}$		& $1.0 \, 10^{-4}$		\\
	\hline
	\hline
		$\tau^{\rm exp} \approx $
					&  $ 10^{-5}$~\cite{Ehlers:2003}	&	$10^{-2}$~\cite{Snyder:2004}	& $10^{-11}$~\cite{Zhou:2008}	& $10^{-12}$~\cite{Kimura:2013}	\\	
	\hline
\end{tabular}
	\caption{
		Key-quantities in different compounds. Rows 2-3: estimations of $a$ (dimensionless) and $\mathcal{E}_{\rm exc}$ (meV) in Eq.~\eqref{eq:Gen17OnodaCompactMainText}. Rows 4-5: corresponding fast and slow \emph{tunnelling} timescales -- unitary evolution under Eq.~\eqref{eq:HamiltonianMain} with experimental parameters from Refs.~\cite{Bramwell:2001,Ruminy:2016,Princep:2013,Kimura:2013}. Row 6: experimental timescales from the literature.  All timescales $\tau$ are {expressed} in seconds.
	}
	\label{tab:Timescales}
\end{table}

%%%%%%%%%%%%%%%%%%%%%%%%%%%%%%%%%%%%%%%%%
\emph{Experimental timescales \emph{(HTO, DTO)} ---} 
	It is worth contrasting our timescales in Tab.~\ref{tab:Timescales} with the experimental ones in the literature. On the one hand, $\mu$SR experiments report $\mu$s timescales due to persistent spin dynamics~\cite{Lago:2007,Dunsiger:2011, Blundell:2012}. On the other, AC-susceptibility measurements report ms timescales for relaxation in DTO~\cite{Matsuhira:2011,Snyder:2001}. Suggestively, such two timescales have the same order of magnitude as $\tau^{\rm fast}$ and $\tau^{\rm slow}$, respectively, and it is tempting to look for a direct correspondence. However, we hitherto neglected any source of decoherence. Even the fast timescales for HTO and DTO in our work are relatively `slow' in that respect. Experiments on molecular spin-qubits with Ho$^{3+}$ ions in the same point symmetry $D_{3d}$ yield decoherence times up to $50~\mu$s at best~\cite{Shiddiq:2016}. In conventional spin ice experimental settings, we therefore expect decoherence timescales faster than spin ice dynamics, by orders of magnitude. The effective spin-flip time can then be much longer than the precession time, due to the environment projecting the time-evolved state back to its initial one -- a phenomenon called quantum Zeno effect~\cite{Misra:1977,Gherardini:201601:NJP}. The effect can be illustrated by considering a toy model -- details can be found in the Supplemental Materials~\cite{Note2} -- of a precessing pseudospin-$1/2$ degree of freedom (with precession time $\tau$ caused by a transverse field), coupled to an effective bath at exponentially-distributed random times (with an `observation' time constant $\tau_{\rm o}$). The bath takes the form of projections onto the states $\ket{\pm}$. {A} fast-decoherence regime $\tau_{\rm o} \ll \tau$ results in a large effective spin-flip timescale $\braket{\Delta t} \sim \tau^2 / \tau_{\rm o}$. In HTO and DTO, this could reconcile the AC experimental timescales ($\sim$ ms) with $\tau^{\rm fast}$ ($\sim \mu$s). These values require spin-decoherence timescales of order $0.1-1$~ns, which are plausible. Notice that, under this assumption, $\tau^{\rm slow}$ becomes of the order of $1$~s.

%%%%%%%%%%%%%%%%%%%%%%%%%%%%%%%%%%%%%%%%%
\emph{Conclusions ---} 
	Our work highlights an intriguing and hitherto poorly understood correlation in the dynamics of spin ice models and materials, whereby a monopole alters locally the spin background, and the latter (pre)determines whether and how fast the former can hop. Specifically, the arrival of a monopole in a tetrahedron quenches the longitudinal fields acting on its spins, inducing a bimodal distribution of temperature-independent spin-tunnelling timescales dictated solely by the CEF, dipolar and exchange transverse terms.

	We believe that this is crucial for understanding the mechanism of hopping of a monopole~\cite{Jaubert:2009} and, more in general, of the dynamics below the `quantum-classical' crossover at $T\approx13$~K~\cite{Snyder:2001,Ehlers:2003,Snyder:2004}. While at higher temperatures, monopole diffusion is well-understood in terms of thermal population of CEF levels \cite{Ehlers:2003}, the predictions of that theory differ considerably from the behaviour observed at lower temperatures. In contrast, the broad agreement of the experimental low-temperature timescales with our predictions, and their temperature independence, provide evidence in favour of our theory.

	In classical spin ice (HTO and DTO) our finding of $\tau^{\rm fast}$ and $\tau^{\rm slow}$ may have important implications on the response and equilibration properties. For instance, Monte Carlo simulations used to model AC susceptibility~\cite{Jaubert:2009, Yaraskavitch:2012, Revell:2013} ought to be modified to account for the two timescales discussed in our work, which can lead to measurable effects~\footnote{Thus far, experimental AC susceptibility is only in coarse agreement with Monte Carlo simulations.}. Ascertaining the extent to which the modified scenario improves agreement with experiments deserves dedicated studies, beyond the scope of this paper. Furthermore, that $\tau^{\rm fast}$ and $\tau^{\rm slow}$ turn out to be longer than the expected decoherence times in HTO and DTO is one of our principal {conclusions}. It implies that any theory of the long experimental timescales and other {puzzling} dynamical phenomena (e.g., the falling out of equilibrium at $600$~mK) must take decoherence into account. Our simple one, based on the Zeno effect, shows how this might work in terms of one adjustable parameter $\tau_{\rm o}$. A predictive theory of $\tau_{\rm o}$ 	will probably need to include degrees of freedom `extrinsic' to the spins (which thus become an open quantum system decohering {\it via} quantum dissipation induced by their environment~\cite{BreuerPetruccioneBook:2002}). This remains an outstanding challenge in the field.

	In quantum spin ice (PSO and PZO) the separation in timescales becomes remarkably large, with $\tau^{\rm fast}$ potentially shorter than the decoherence time scale, consistently with expectations. {This is in agreement, for example, with Ref.~\cite{Zhou:2008}, where timescales for PSO -- extracted from quasi-elastic neutron scattering experiments -- were contrasted to the ones for DTO and HTO.} The fast timescales in PSO and PZO suggest the need for a few-body dynamical description incorporating coherent longer-range hopping processes for the monopoles (e.g., flips of $S_0$ and $S_{3}$ in Fig.~\ref{fig:TwoTetrahedra(a)}) which cannot be described within our current approximations and is thus a further challenge for future work. Our results also imply that the dynamics corresponding to the slow time scale will not occur quantum-coherently. It will be interesting to determine which observable consequences this may have. In particular, it provides us with the intriguing possibility of an approximate modelling of these systems that, {\it \`{a} la} Born-Oppenheimer, focuses on a quantum mechanical description of the fast processes, and a classical stochastic description of the slow ones, thus substantially reducing the complexity of a three dimensional strongly correlated quantum system~\footnote{On average, there are two fast spins per tetrahedron, and the corresponding quantum many body system is a network of coordination 2 (1D {on average}) with occasional 3-way branching and end points.}. 
	
	Finally, we note that since a monopole breaks the inversion symmetry of the tetrahedron, it is expected to generate an \emph{electric} dipole moment~\cite{Khomskii:2012}. This may alter the crystal electric fields and split the g.s. doublet in non-Kramers systems, potentially affecting the hopping timescales in HTO, PSO and PZO. In contrast, DTO and other Kramers system should remain unperturbed.

%%%%%%%%%%%%%%%%%%%%%%%%%%%%%%%%%%%%%%%%%
\begin{acknowledgements}
\emph{Acknowledgments ---} 
	This work was supported in part by Engineering and Physical Sciences Research Council (EPSRC) Grants No.EP/G049394/1, No.EP/K028960/1 and No.EP/M007065/1 (C.C.), and in part by the DFG through SFB 1143 (project-id 247310070) and ct.qmat (EXC 2147, project-id 39085490), and by the EPSRC NetworkPlus on ``Emergence and Physics far from Equilibrium''. B.T. was supported by a PhD studentship from HEFCE and SEPnet. J.Q. was supported by a fellowship from STFC and SEPnet and by EPSRC Grant No.EP/P00749X/1. Statement of compliance with EPSRC policy framework on research data: this publication reports theoretical work that does not require supporting research data. The authors thank S.T.~Bramwell, J.~Chalker, T.~Fennel, J.~Gardner, S.~Giblin, S.~Gopalakrishnan, C.~Hooley, V.~Oganesyan, G.L.~Pascut, S.~Ramos, G.~Sala, P.~Strange and T.~Ziman for useful discussions. B.T. is grateful for the hospitality at the MPIPKS in Dresden and the TCM group at the Cavendish Laboratory in Cambridge.
\end{acknowledgements}

%\bibliography{BiblioPaper02v17}{}

\begin{thebibliography}{47}%
\makeatletter
\providecommand \@ifxundefined [1]{%
 \@ifx{#1\undefined}
}%
\providecommand \@ifnum [1]{%
 \ifnum #1\expandafter \@firstoftwo
 \else \expandafter \@secondoftwo
 \fi
}%
\providecommand \@ifx [1]{%
 \ifx #1\expandafter \@firstoftwo
 \else \expandafter \@secondoftwo
 \fi
}%
\providecommand \natexlab [1]{#1}%
\providecommand \enquote  [1]{``#1''}%
\providecommand \bibnamefont  [1]{#1}%
\providecommand \bibfnamefont [1]{#1}%
\providecommand \citenamefont [1]{#1}%
\providecommand \href@noop [0]{\@secondoftwo}%
\providecommand \href [0]{\begingroup \@sanitize@url \@href}%
\providecommand \@href[1]{\@@startlink{#1}\@@href}%
\providecommand \@@href[1]{\endgroup#1\@@endlink}%
\providecommand \@sanitize@url [0]{\catcode `\\12\catcode `\$12\catcode
  `\&12\catcode `\#12\catcode `\^12\catcode `\_12\catcode `\%12\relax}%
\providecommand \@@startlink[1]{}%
\providecommand \@@endlink[0]{}%
\providecommand \url  [0]{\begingroup\@sanitize@url \@url }%
\providecommand \@url [1]{\endgroup\@href {#1}{\urlprefix }}%
\providecommand \urlprefix  [0]{URL }%
\providecommand \Eprint [0]{\href }%
\providecommand \doibase [0]{http://dx.doi.org/}%
\providecommand \selectlanguage [0]{\@gobble}%
\providecommand \bibinfo  [0]{\@secondoftwo}%
\providecommand \bibfield  [0]{\@secondoftwo}%
\providecommand \translation [1]{[#1]}%
\providecommand \BibitemOpen [0]{}%
\providecommand \bibitemStop [0]{}%
\providecommand \bibitemNoStop [0]{.\EOS\space}%
\providecommand \EOS [0]{\spacefactor3000\relax}%
\providecommand \BibitemShut  [1]{\csname bibitem#1\endcsname}%
\let\auto@bib@innerbib\@empty
%</preamble>
\bibitem [{\citenamefont {Landau}(1956)}]{Landau:1956}%
  \BibitemOpen
  \bibfield  {author} {\bibinfo {author} {\bibfnamefont {L.~D.}\ \bibnamefont
  {Landau}},\ }\href {http://www.jetp.ac.ru/cgi-bin/e/index/e/3/6/p920?a=list}
  {\bibfield  {journal} {\bibinfo  {journal} {Sov. Phys. JETP}\ }\textbf
  {\bibinfo {volume} {3}},\ \bibinfo {pages} {920} (\bibinfo {year}
  {1956})}\BibitemShut {NoStop}%
\bibitem [{\citenamefont {Landau}(1957)}]{Landau:1957}%
  \BibitemOpen
  \bibfield  {author} {\bibinfo {author} {\bibfnamefont {L.~D.}\ \bibnamefont
  {Landau}},\ }\href {http://www.jetp.ac.ru/cgi-bin/e/index/e/5/1/p101?a=list}
  {\bibfield  {journal} {\bibinfo  {journal} {Sov. Phys. JETP}\ }\textbf
  {\bibinfo {volume} {32}},\ \bibinfo {pages} {59} (\bibinfo {year}
  {1957})}\BibitemShut {NoStop}%
\bibitem [{\citenamefont {Luttinger}(1963)}]{Luttinger:1963}%
  \BibitemOpen
  \bibfield  {author} {\bibinfo {author} {\bibfnamefont {J.~M.}\ \bibnamefont
  {Luttinger}},\ }\bibfield  {booktitle} {\emph {\bibinfo {booktitle} {Journal
  of Mathematical Physics}},\ }\href {\doibase 10.1063/1.1704046} {\bibfield
  {journal} {\bibinfo  {journal} {Journal of Mathematical Physics}\ }\textbf
  {\bibinfo {volume} {4}},\ \bibinfo {pages} {1154} (\bibinfo {year}
  {1963})}\BibitemShut {NoStop}%
\bibitem [{\citenamefont {Szab{\'o}}\ and\ \citenamefont
  {Castelnovo}(2019)}]{Szabo:2019}%
  \BibitemOpen
  \bibfield  {author} {\bibinfo {author} {\bibfnamefont {A.}~\bibnamefont
  {Szab{\'o}}}\ and\ \bibinfo {author} {\bibfnamefont {C.}~\bibnamefont
  {Castelnovo}},\ }\href {\doibase 10.1103/PhysRevB.100.014417} {\bibfield
  {journal} {\bibinfo  {journal} {Physical Review B}\ }\textbf {\bibinfo
  {volume} {100}},\ \bibinfo {pages} {014417} (\bibinfo {year}
  {2019})}\BibitemShut {NoStop}%
\bibitem [{\citenamefont {B{\ifmmode\acute{e}\else\'{e}\fi}ran}\ \emph
  {et~al.}(1996)\citenamefont {B{\ifmmode\acute{e}\else\'{e}\fi}ran},
  \citenamefont {Poilblanc},\ and\ \citenamefont {Laughlin}}]{Beran:1996}%
  \BibitemOpen
  \bibfield  {author} {\bibinfo {author} {\bibfnamefont {P.}~\bibnamefont
  {B{\ifmmode\acute{e}\else\'{e}\fi}ran}}, \bibinfo {author} {\bibfnamefont
  {D.}~\bibnamefont {Poilblanc}}, \ and\ \bibinfo {author} {\bibfnamefont
  {R.~B.}\ \bibnamefont {Laughlin}},\ }\href {\doibase
  10.1016/0550-3213(96)00196-4} {\bibfield  {journal} {\bibinfo  {journal}
  {Nucl. Phys. B}\ }\textbf {\bibinfo {volume} {473}},\ \bibinfo {pages} {707}
  (\bibinfo {year} {1996})}\BibitemShut {NoStop}%
\bibitem [{\citenamefont {Grusdt}\ \emph {et~al.}(2018)\citenamefont {Grusdt},
  \citenamefont {K{\ifmmode\acute{a}\else\'{a}\fi}nasz-Nagy}, \citenamefont
  {Bohrdt}, \citenamefont {Chiu}, \citenamefont {Ji}, \citenamefont {Greiner},
  \citenamefont {Greif},\ and\ \citenamefont {Demler}}]{Grusdt:2018}%
  \BibitemOpen
  \bibfield  {author} {\bibinfo {author} {\bibfnamefont {F.}~\bibnamefont
  {Grusdt}}, \bibinfo {author} {\bibfnamefont {M.}~\bibnamefont
  {K{\ifmmode\acute{a}\else\'{a}\fi}nasz-Nagy}}, \bibinfo {author}
  {\bibfnamefont {A.}~\bibnamefont {Bohrdt}}, \bibinfo {author} {\bibfnamefont
  {C.~S.}\ \bibnamefont {Chiu}}, \bibinfo {author} {\bibfnamefont
  {G.}~\bibnamefont {Ji}}, \bibinfo {author} {\bibfnamefont {M.}~\bibnamefont
  {Greiner}}, \bibinfo {author} {\bibfnamefont {D.}~\bibnamefont {Greif}}, \
  and\ \bibinfo {author} {\bibfnamefont {E.}~\bibnamefont {Demler}},\ }\href
  {\doibase 10.1103/PhysRevX.8.011046} {\bibfield  {journal} {\bibinfo
  {journal} {Phys. Rev. X}\ }\textbf {\bibinfo {volume} {8}},\ \bibinfo {pages}
  {011046} (\bibinfo {year} {2018})}\BibitemShut {NoStop}%
\bibitem [{\citenamefont {Bramwell}\ and\ \citenamefont
  {Gingras}(2001)}]{Bramwell:2001}%
  \BibitemOpen
  \bibfield  {author} {\bibinfo {author} {\bibfnamefont {S.~T.}\ \bibnamefont
  {Bramwell}}\ and\ \bibinfo {author} {\bibfnamefont {M.~J.~P.}\ \bibnamefont
  {Gingras}},\ }\href
  {http://www.sciencemag.org/content/294/5546/1495.abstract} {\bibfield
  {journal} {\bibinfo  {journal} {Science}\ }\textbf {\bibinfo {volume}
  {294}},\ \bibinfo {pages} {1495} (\bibinfo {year} {2001})}\BibitemShut
  {NoStop}%
\bibitem [{\citenamefont {Castelnovo}\ \emph {et~al.}(2008)\citenamefont
  {Castelnovo}, \citenamefont {Moessner},\ and\ \citenamefont
  {Sondhi}}]{Castelnovo:2008}%
  \BibitemOpen
  \bibfield  {author} {\bibinfo {author} {\bibfnamefont {C.}~\bibnamefont
  {Castelnovo}}, \bibinfo {author} {\bibfnamefont {R.}~\bibnamefont
  {Moessner}}, \ and\ \bibinfo {author} {\bibfnamefont {S.~L.}\ \bibnamefont
  {Sondhi}},\ }\href {http://dx.doi.org/10.1038/nature06433} {\bibfield
  {journal} {\bibinfo  {journal} {Nature}\ }\textbf {\bibinfo {volume} {451}},\
  \bibinfo {pages} {42} (\bibinfo {year} {2008})}\BibitemShut {NoStop}%
\bibitem [{\citenamefont {Snyder}\ \emph {et~al.}(2003)\citenamefont {Snyder},
  \citenamefont {Ueland}, \citenamefont {Slusky}, \citenamefont {Karunadasa},
  \citenamefont {Cava}, \citenamefont {Mizel},\ and\ \citenamefont
  {Schiffer}}]{Snyder:2003}%
  \BibitemOpen
  \bibfield  {author} {\bibinfo {author} {\bibfnamefont {J.}~\bibnamefont
  {Snyder}}, \bibinfo {author} {\bibfnamefont {B.~G.}\ \bibnamefont {Ueland}},
  \bibinfo {author} {\bibfnamefont {J.~S.}\ \bibnamefont {Slusky}}, \bibinfo
  {author} {\bibfnamefont {H.}~\bibnamefont {Karunadasa}}, \bibinfo {author}
  {\bibfnamefont {R.~J.}\ \bibnamefont {Cava}}, \bibinfo {author}
  {\bibfnamefont {A.}~\bibnamefont {Mizel}}, \ and\ \bibinfo {author}
  {\bibfnamefont {P.}~\bibnamefont {Schiffer}},\ }\href {\doibase
  10.1103/PhysRevLett.91.107201} {\bibfield  {journal} {\bibinfo  {journal}
  {Physical Review Letters}\ }\textbf {\bibinfo {volume} {91}},\ \bibinfo
  {pages} {107201} (\bibinfo {year} {2003})}\BibitemShut {NoStop}%
\bibitem [{\citenamefont {Bramwell}\ \emph {et~al.}(2009)\citenamefont
  {Bramwell}, \citenamefont {Giblin}, \citenamefont {Calder}, \citenamefont
  {Aldus}, \citenamefont {Prabhakaran},\ and\ \citenamefont
  {Fennell}}]{Bramwell:2009}%
  \BibitemOpen
  \bibfield  {author} {\bibinfo {author} {\bibfnamefont {S.~T.}\ \bibnamefont
  {Bramwell}}, \bibinfo {author} {\bibfnamefont {S.~R.}\ \bibnamefont
  {Giblin}}, \bibinfo {author} {\bibfnamefont {S.}~\bibnamefont {Calder}},
  \bibinfo {author} {\bibfnamefont {R.}~\bibnamefont {Aldus}}, \bibinfo
  {author} {\bibfnamefont {D.}~\bibnamefont {Prabhakaran}}, \ and\ \bibinfo
  {author} {\bibfnamefont {T.}~\bibnamefont {Fennell}},\ }\href
  {http://dx.doi.org/10.1038/nature08500} {\bibfield  {journal} {\bibinfo
  {journal} {Nature}\ }\textbf {\bibinfo {volume} {461}},\ \bibinfo {pages}
  {956} (\bibinfo {year} {2009})}\BibitemShut {NoStop}%
\bibitem [{\citenamefont {Giblin}\ \emph {et~al.}(2011)\citenamefont {Giblin},
  \citenamefont {Bramwell}, \citenamefont {Holdsworth}, \citenamefont
  {Prabhakaran},\ and\ \citenamefont {Terry}}]{Giblin:2011}%
  \BibitemOpen
  \bibfield  {author} {\bibinfo {author} {\bibfnamefont {S.~R.}\ \bibnamefont
  {Giblin}}, \bibinfo {author} {\bibfnamefont {S.~T.}\ \bibnamefont
  {Bramwell}}, \bibinfo {author} {\bibfnamefont {P.~C.~W.}\ \bibnamefont
  {Holdsworth}}, \bibinfo {author} {\bibfnamefont {D.}~\bibnamefont
  {Prabhakaran}}, \ and\ \bibinfo {author} {\bibfnamefont {I.}~\bibnamefont
  {Terry}},\ }\href {http://dx.doi.org/10.1038/nphys1896} {\bibfield  {journal}
  {\bibinfo  {journal} {Nat Phys}\ }\textbf {\bibinfo {volume} {7}},\ \bibinfo
  {pages} {252} (\bibinfo {year} {2011})}\BibitemShut {NoStop}%
\bibitem [{\citenamefont {Jaubert}\ and\ \citenamefont
  {Holdsworth}(2011)}]{Jaubert:2011}%
  \BibitemOpen
  \bibfield  {author} {\bibinfo {author} {\bibfnamefont {L.~D.~C.}\
  \bibnamefont {Jaubert}}\ and\ \bibinfo {author} {\bibfnamefont {P.~C.~W.}\
  \bibnamefont {Holdsworth}},\ }\href
  {http://stacks.iop.org/0953-8984/23/i=16/a=164222} {\bibfield  {journal}
  {\bibinfo  {journal} {Journal of Physics: Condensed Matter}\ }\textbf
  {\bibinfo {volume} {23}},\ \bibinfo {pages} {164222} (\bibinfo {year}
  {2011})}\BibitemShut {NoStop}%
\bibitem [{\citenamefont {Matsuhira}\ \emph {et~al.}(2011)\citenamefont
  {Matsuhira}, \citenamefont {Paulsen}, \citenamefont {Lhotel}, \citenamefont
  {Sekine}, \citenamefont {Hiroi},\ and\ \citenamefont
  {Takagi}}]{Matsuhira:2011}%
  \BibitemOpen
  \bibfield  {author} {\bibinfo {author} {\bibfnamefont {K.}~\bibnamefont
  {Matsuhira}}, \bibinfo {author} {\bibfnamefont {C.}~\bibnamefont {Paulsen}},
  \bibinfo {author} {\bibfnamefont {E.}~\bibnamefont {Lhotel}}, \bibinfo
  {author} {\bibfnamefont {C.}~\bibnamefont {Sekine}}, \bibinfo {author}
  {\bibfnamefont {Z.}~\bibnamefont {Hiroi}}, \ and\ \bibinfo {author}
  {\bibfnamefont {S.}~\bibnamefont {Takagi}},\ }\href
  {http://jpsj.ipap.jp/link?JPSJ/80/123711/} {\bibfield  {journal} {\bibinfo
  {journal} {Journal of the Physical Society of Japan}\ }\textbf {\bibinfo
  {volume} {80}},\ \bibinfo {pages} {123711} (\bibinfo {year}
  {2011})}\BibitemShut {NoStop}%
\bibitem [{\citenamefont {Snyder}\ \emph {et~al.}(2001)\citenamefont {Snyder},
  \citenamefont {Slusky}, \citenamefont {Cava},\ and\ \citenamefont
  {Schiffer}}]{Snyder:2001}%
  \BibitemOpen
  \bibfield  {author} {\bibinfo {author} {\bibfnamefont {J.}~\bibnamefont
  {Snyder}}, \bibinfo {author} {\bibfnamefont {J.~S.}\ \bibnamefont {Slusky}},
  \bibinfo {author} {\bibfnamefont {R.~J.}\ \bibnamefont {Cava}}, \ and\
  \bibinfo {author} {\bibfnamefont {P.}~\bibnamefont {Schiffer}},\ }\href
  {http://dx.doi.org/10.1038/35092516} {\bibfield  {journal} {\bibinfo
  {journal} {Nature}\ }\textbf {\bibinfo {volume} {413}},\ \bibinfo {pages}
  {48} (\bibinfo {year} {2001})}\BibitemShut {NoStop}%
\bibitem [{\citenamefont {Sala}\ \emph {et~al.}(2012)\citenamefont {Sala},
  \citenamefont {Castelnovo}, \citenamefont {Moessner}, \citenamefont {Sondhi},
  \citenamefont {Kitagawa}, \citenamefont {Takigawa}, \citenamefont
  {Higashinaka},\ and\ \citenamefont {Maeno}}]{Sala:2012}%
  \BibitemOpen
  \bibfield  {author} {\bibinfo {author} {\bibfnamefont {G.}~\bibnamefont
  {Sala}}, \bibinfo {author} {\bibfnamefont {C.}~\bibnamefont {Castelnovo}},
  \bibinfo {author} {\bibfnamefont {R.}~\bibnamefont {Moessner}}, \bibinfo
  {author} {\bibfnamefont {S.~L.}\ \bibnamefont {Sondhi}}, \bibinfo {author}
  {\bibfnamefont {K.}~\bibnamefont {Kitagawa}}, \bibinfo {author}
  {\bibfnamefont {M.}~\bibnamefont {Takigawa}}, \bibinfo {author}
  {\bibfnamefont {R.}~\bibnamefont {Higashinaka}}, \ and\ \bibinfo {author}
  {\bibfnamefont {Y.}~\bibnamefont {Maeno}},\ }\href
  {http://link.aps.org/doi/10.1103/PhysRevLett.108.217203} {\bibfield
  {journal} {\bibinfo  {journal} {Physical Review Letters}\ }\textbf {\bibinfo
  {volume} {108}},\ \bibinfo {pages} {217203} (\bibinfo {year}
  {2012})}\BibitemShut {NoStop}%
\bibitem [{\citenamefont {Paulsen}\ \emph {et~al.}(2014)\citenamefont
  {Paulsen}, \citenamefont {Jackson}, \citenamefont {Lhotel}, \citenamefont
  {Canals}, \citenamefont {Prabhakaran}, \citenamefont {Matsuhira},
  \citenamefont {Giblin},\ and\ \citenamefont {Bramwell}}]{Paulsen:2014}%
  \BibitemOpen
  \bibfield  {author} {\bibinfo {author} {\bibfnamefont {C.}~\bibnamefont
  {Paulsen}}, \bibinfo {author} {\bibfnamefont {M.~J.}\ \bibnamefont
  {Jackson}}, \bibinfo {author} {\bibfnamefont {E.}~\bibnamefont {Lhotel}},
  \bibinfo {author} {\bibfnamefont {B.}~\bibnamefont {Canals}}, \bibinfo
  {author} {\bibfnamefont {D.}~\bibnamefont {Prabhakaran}}, \bibinfo {author}
  {\bibfnamefont {K.}~\bibnamefont {Matsuhira}}, \bibinfo {author}
  {\bibfnamefont {S.~R.}\ \bibnamefont {Giblin}}, \ and\ \bibinfo {author}
  {\bibfnamefont {S.~T.}\ \bibnamefont {Bramwell}},\ }\href
  {http://dx.doi.org/10.1038/nphys2847} {\bibfield  {journal} {\bibinfo
  {journal} {Nat Phys}\ }\textbf {\bibinfo {volume} {10}},\ \bibinfo {pages}
  {135} (\bibinfo {year} {2014})}\BibitemShut {NoStop}%
\bibitem [{\citenamefont {Nambu}\ \emph {et~al.}(2015)\citenamefont {Nambu},
  \citenamefont {Gardner}, \citenamefont {MacLaughlin}, \citenamefont {Stock},
  \citenamefont {Endo}, \citenamefont {Jonas}, \citenamefont {Sato},
  \citenamefont {Nakatsuji},\ and\ \citenamefont {Broholm}}]{Nambu:2015}%
  \BibitemOpen
  \bibfield  {author} {\bibinfo {author} {\bibfnamefont {Y.}~\bibnamefont
  {Nambu}}, \bibinfo {author} {\bibfnamefont {J.~S.}\ \bibnamefont {Gardner}},
  \bibinfo {author} {\bibfnamefont {D.~E.}\ \bibnamefont {MacLaughlin}},
  \bibinfo {author} {\bibfnamefont {C.}~\bibnamefont {Stock}}, \bibinfo
  {author} {\bibfnamefont {H.}~\bibnamefont {Endo}}, \bibinfo {author}
  {\bibfnamefont {S.}~\bibnamefont {Jonas}}, \bibinfo {author} {\bibfnamefont
  {T.~J.}\ \bibnamefont {Sato}}, \bibinfo {author} {\bibfnamefont
  {S.}~\bibnamefont {Nakatsuji}}, \ and\ \bibinfo {author} {\bibfnamefont
  {C.}~\bibnamefont {Broholm}},\ }\href {\doibase
  10.1103/PhysRevLett.115.127202} {\bibfield  {journal} {\bibinfo  {journal}
  {Physical Review Letters}\ }\textbf {\bibinfo {volume} {115}},\ \bibinfo
  {pages} {127202} (\bibinfo {year} {2015})}\BibitemShut {NoStop}%
\bibitem [{\citenamefont {Bovo}\ \emph {et~al.}(2013)\citenamefont {Bovo},
  \citenamefont {Bloxsom}, \citenamefont {Prabhakaran}, \citenamefont
  {Aeppli},\ and\ \citenamefont {Bramwell}}]{Bovo:2013}%
  \BibitemOpen
  \bibfield  {author} {\bibinfo {author} {\bibfnamefont {L.}~\bibnamefont
  {Bovo}}, \bibinfo {author} {\bibfnamefont {J.~A.}\ \bibnamefont {Bloxsom}},
  \bibinfo {author} {\bibfnamefont {D.}~\bibnamefont {Prabhakaran}}, \bibinfo
  {author} {\bibfnamefont {G.}~\bibnamefont {Aeppli}}, \ and\ \bibinfo {author}
  {\bibfnamefont {S.~T.}\ \bibnamefont {Bramwell}},\ }\href
  {http://dx.doi.org/10.1038/ncomms2551} {\bibfield  {journal} {\bibinfo
  {journal} {Nat Commun}\ }\textbf {\bibinfo {volume} {4}},\ \bibinfo {pages}
  {1535} (\bibinfo {year} {2013})}\BibitemShut {NoStop}%
\bibitem [{\citenamefont {Tomasello}\ \emph {et~al.}(2015)\citenamefont
  {Tomasello}, \citenamefont {Castelnovo}, \citenamefont {Moessner},\ and\
  \citenamefont {Quintanilla}}]{Tomasello:2015}%
  \BibitemOpen
  \bibfield  {author} {\bibinfo {author} {\bibfnamefont {B.}~\bibnamefont
  {Tomasello}}, \bibinfo {author} {\bibfnamefont {C.}~\bibnamefont
  {Castelnovo}}, \bibinfo {author} {\bibfnamefont {R.}~\bibnamefont
  {Moessner}}, \ and\ \bibinfo {author} {\bibfnamefont {J.}~\bibnamefont
  {Quintanilla}},\ }\href {http://link.aps.org/doi/10.1103/PhysRevB.92.155120}
  {\bibfield  {journal} {\bibinfo  {journal} {Physical Review B}\ }\textbf
  {\bibinfo {volume} {92}},\ \bibinfo {pages} {155120} (\bibinfo {year}
  {2015})}\BibitemShut {NoStop}%
\bibitem [{\citenamefont {Ruminy}\ \emph {et~al.}(2016)\citenamefont {Ruminy},
  \citenamefont {Pomjakushina}, \citenamefont {Iida}, \citenamefont {Kamazawa},
  \citenamefont {Adroja}, \citenamefont {Stuhr},\ and\ \citenamefont
  {Fennell}}]{Ruminy:2016}%
  \BibitemOpen
  \bibfield  {author} {\bibinfo {author} {\bibfnamefont {M.}~\bibnamefont
  {Ruminy}}, \bibinfo {author} {\bibfnamefont {E.}~\bibnamefont
  {Pomjakushina}}, \bibinfo {author} {\bibfnamefont {K.}~\bibnamefont {Iida}},
  \bibinfo {author} {\bibfnamefont {K.}~\bibnamefont {Kamazawa}}, \bibinfo
  {author} {\bibfnamefont {D.~T.}\ \bibnamefont {Adroja}}, \bibinfo {author}
  {\bibfnamefont {U.}~\bibnamefont {Stuhr}}, \ and\ \bibinfo {author}
  {\bibfnamefont {T.}~\bibnamefont {Fennell}},\ }\href
  {http://link.aps.org/doi/10.1103/PhysRevB.94.024430} {\bibfield  {journal}
  {\bibinfo  {journal} {Physical Review B}\ }\textbf {\bibinfo {volume} {94}},\
  \bibinfo {pages} {024430} (\bibinfo {year} {2016})}\BibitemShut {NoStop}%
\bibitem [{\citenamefont {Rosenkranz}\ \emph {et~al.}(2000)\citenamefont
  {Rosenkranz}, \citenamefont {Ramirez}, \citenamefont {Hayashi}, \citenamefont
  {Cava}, \citenamefont {Siddharthan},\ and\ \citenamefont
  {Shastry}}]{Rosenkranz:2000}%
  \BibitemOpen
  \bibfield  {author} {\bibinfo {author} {\bibfnamefont {S.}~\bibnamefont
  {Rosenkranz}}, \bibinfo {author} {\bibfnamefont {A.~P.}\ \bibnamefont
  {Ramirez}}, \bibinfo {author} {\bibfnamefont {A.}~\bibnamefont {Hayashi}},
  \bibinfo {author} {\bibfnamefont {R.~J.}\ \bibnamefont {Cava}}, \bibinfo
  {author} {\bibfnamefont {R.}~\bibnamefont {Siddharthan}}, \ and\ \bibinfo
  {author} {\bibfnamefont {B.~S.}\ \bibnamefont {Shastry}},\ }\href
  {http://link.aip.org/link/?JAP/87/5914/1} {\bibfield  {journal} {\bibinfo
  {journal} {Journal of Applied Physics}\ }\textbf {\bibinfo {volume} {87}},\
  \bibinfo {pages} {5914} (\bibinfo {year} {2000})}\BibitemShut {NoStop}%
\bibitem [{\citenamefont {Ehlers}\ \emph {et~al.}(2003)\citenamefont {Ehlers},
  \citenamefont {Cornelius}, \citenamefont {Orend{\'a}c}, \citenamefont
  {Kajnakov{\'a}}, \citenamefont {Fennell}, \citenamefont {Bramwell},\ and\
  \citenamefont {Gardner}}]{Ehlers:2003}%
  \BibitemOpen
  \bibfield  {author} {\bibinfo {author} {\bibfnamefont {G.}~\bibnamefont
  {Ehlers}}, \bibinfo {author} {\bibfnamefont {A.~L.}\ \bibnamefont
  {Cornelius}}, \bibinfo {author} {\bibfnamefont {M.}~\bibnamefont
  {Orend{\'a}c}}, \bibinfo {author} {\bibfnamefont {M.}~\bibnamefont
  {Kajnakov{\'a}}}, \bibinfo {author} {\bibfnamefont {T.}~\bibnamefont
  {Fennell}}, \bibinfo {author} {\bibfnamefont {S.~T.}\ \bibnamefont
  {Bramwell}}, \ and\ \bibinfo {author} {\bibfnamefont {J.~S.}\ \bibnamefont
  {Gardner}},\ }\href {http://stacks.iop.org/0953-8984/15/i=2/a=102} {\bibfield
   {journal} {\bibinfo  {journal} {Journal of Physics: Condensed Matter}\
  }\textbf {\bibinfo {volume} {15}},\ \bibinfo {pages} {L9} (\bibinfo {year}
  {2003})}\BibitemShut {NoStop}%
\bibitem [{Note1()}]{Note1}%
  \BibitemOpen
  \bibinfo {note} {The fourth spin {$S_{6}$} remains static because its
  reversal would produce higher energy defects (double monopoles)}\BibitemShut
  {NoStop}%
\bibitem [{pie()}]{pie}%
  \BibitemOpen
  \href@noop {} {}\bibinfo {note} {The configuration of 6 n.n. spins in
  Fig.~\ref{fig:TwoTetrahedra(c)} has inversion symmetry with respect to the
  central site and therefore cannot lead to any net effective field at that
  site except through spontaneous breaking of that symmetry. The latter would
  require coupling to other degrees of freedom which are not present in our
  model, e.g., lattice distortions or spins tilting away from the easy
  axes.}\BibitemShut {Stop}%
\bibitem [{\citenamefont {Isakov}\ \emph {et~al.}(2005)\citenamefont {Isakov},
  \citenamefont {Moessner},\ and\ \citenamefont {Sondhi}}]{Isakov:2005}%
  \BibitemOpen
  \bibfield  {author} {\bibinfo {author} {\bibfnamefont {S.~V.}\ \bibnamefont
  {Isakov}}, \bibinfo {author} {\bibfnamefont {R.}~\bibnamefont {Moessner}}, \
  and\ \bibinfo {author} {\bibfnamefont {S.~L.}\ \bibnamefont {Sondhi}},\
  }\href {http://link.aps.org/doi/10.1103/PhysRevLett.95.217201} {\bibfield
  {journal} {\bibinfo  {journal} {Physical Review Letters}\ }\textbf {\bibinfo
  {volume} {95}},\ \bibinfo {pages} {217201} (\bibinfo {year}
  {2005})}\BibitemShut {NoStop}%
\bibitem [{\citenamefont {Sala}()}]{Sala:MonteCarlo}%
  \BibitemOpen
  \bibfield  {author} {\bibinfo {author} {\bibfnamefont {G.}~\bibnamefont
  {Sala}},\ }\href@noop {} {}\bibinfo {note} {(private
  communication)}\BibitemShut {NoStop}%
\bibitem [{\citenamefont {Onoda}\ and\ \citenamefont
  {Tanaka}(2011)}]{Onoda:2011}%
  \BibitemOpen
  \bibfield  {author} {\bibinfo {author} {\bibfnamefont {S.}~\bibnamefont
  {Onoda}}\ and\ \bibinfo {author} {\bibfnamefont {Y.}~\bibnamefont {Tanaka}},\
  }\href {http://link.aps.org/doi/10.1103/PhysRevB.83.094411} {\bibfield
  {journal} {\bibinfo  {journal} {Physical Review B}\ }\textbf {\bibinfo
  {volume} {83}},\ \bibinfo {pages} {094411} (\bibinfo {year}
  {2011})}\BibitemShut {NoStop}%
\bibitem [{\citenamefont {Tomasello}(2014)}]{Tomasello:PhDThesis}%
  \BibitemOpen
  \bibfield  {author} {\bibinfo {author} {\bibfnamefont {B.}~\bibnamefont
  {Tomasello}},\ }\emph {\bibinfo {title} {A quantum-mechanical study of the
  dynamical properties of spin-ice materials}},\ \href
  {https://kar.kent.ac.uk/48015/} {Ph.D. thesis},\ \bibinfo  {school} {School
  of Physical Sciences, University of Kent} (\bibinfo {year}
  {2014})\BibitemShut {NoStop}%
\bibitem [{Note2()}]{Note2}%
  \BibitemOpen
  \bibinfo {note} {See Supplemental Material (SM) at [URL will be inserted by
  publisher] for more details.}\BibitemShut {Stop}%
\bibitem [{\citenamefont {Rau}\ and\ \citenamefont {Gingras}(2015)}]{Rau:2015}%
  \BibitemOpen
  \bibfield  {author} {\bibinfo {author} {\bibfnamefont {J.~G.}\ \bibnamefont
  {Rau}}\ and\ \bibinfo {author} {\bibfnamefont {M.~J.~P.}\ \bibnamefont
  {Gingras}},\ }\href {http://link.aps.org/doi/10.1103/PhysRevB.92.144417}
  {\bibfield  {journal} {\bibinfo  {journal} {Physical Review B}\ }\textbf
  {\bibinfo {volume} {92}},\ \bibinfo {pages} {144417} (\bibinfo {year}
  {2015})}\BibitemShut {NoStop}%
\bibitem [{\citenamefont {Snyder}\ \emph {et~al.}(2004)\citenamefont {Snyder},
  \citenamefont {Ueland}, \citenamefont {Slusky}, \citenamefont {Karunadasa},
  \citenamefont {Cava},\ and\ \citenamefont {Schiffer}}]{Snyder:2004}%
  \BibitemOpen
  \bibfield  {author} {\bibinfo {author} {\bibfnamefont {J.}~\bibnamefont
  {Snyder}}, \bibinfo {author} {\bibfnamefont {B.~G.}\ \bibnamefont {Ueland}},
  \bibinfo {author} {\bibfnamefont {J.~S.}\ \bibnamefont {Slusky}}, \bibinfo
  {author} {\bibfnamefont {H.}~\bibnamefont {Karunadasa}}, \bibinfo {author}
  {\bibfnamefont {R.~J.}\ \bibnamefont {Cava}}, \ and\ \bibinfo {author}
  {\bibfnamefont {P.}~\bibnamefont {Schiffer}},\ }\href
  {http://link.aps.org/doi/10.1103/PhysRevB.69.064414} {\bibfield  {journal}
  {\bibinfo  {journal} {Physical Review B}\ }\textbf {\bibinfo {volume} {69}},\
  \bibinfo {pages} {064414} (\bibinfo {year} {2004})}\BibitemShut {NoStop}%
\bibitem [{\citenamefont {Zhou}\ \emph {et~al.}(2008)\citenamefont {Zhou},
  \citenamefont {Wiebe}, \citenamefont {Janik}, \citenamefont {Balicas},
  \citenamefont {Yo}, \citenamefont {Qiu}, \citenamefont {Copley},\ and\
  \citenamefont {Gardner}}]{Zhou:2008}%
  \BibitemOpen
  \bibfield  {author} {\bibinfo {author} {\bibfnamefont {H.~D.}\ \bibnamefont
  {Zhou}}, \bibinfo {author} {\bibfnamefont {C.~R.}\ \bibnamefont {Wiebe}},
  \bibinfo {author} {\bibfnamefont {J.~A.}\ \bibnamefont {Janik}}, \bibinfo
  {author} {\bibfnamefont {L.}~\bibnamefont {Balicas}}, \bibinfo {author}
  {\bibfnamefont {Y.~J.}\ \bibnamefont {Yo}}, \bibinfo {author} {\bibfnamefont
  {Y.}~\bibnamefont {Qiu}}, \bibinfo {author} {\bibfnamefont {J.~R.~D.}\
  \bibnamefont {Copley}}, \ and\ \bibinfo {author} {\bibfnamefont {J.~S.}\
  \bibnamefont {Gardner}},\ }\href {\doibase 10.1103/PhysRevLett.101.227204}
  {\bibfield  {journal} {\bibinfo  {journal} {Physical Review Letters}\
  }\textbf {\bibinfo {volume} {101}},\ \bibinfo {pages} {227204} (\bibinfo
  {year} {2008})}\BibitemShut {NoStop}%
\bibitem [{\citenamefont {Kimura}\ \emph {et~al.}(2013)\citenamefont {Kimura},
  \citenamefont {Nakatsuji}, \citenamefont {Wen}, \citenamefont {Broholm},
  \citenamefont {Stone}, \citenamefont {Nishibori},\ and\ \citenamefont
  {Sawa}}]{Kimura:2013}%
  \BibitemOpen
  \bibfield  {author} {\bibinfo {author} {\bibfnamefont {K.}~\bibnamefont
  {Kimura}}, \bibinfo {author} {\bibfnamefont {S.}~\bibnamefont {Nakatsuji}},
  \bibinfo {author} {\bibfnamefont {J.-J.}\ \bibnamefont {Wen}}, \bibinfo
  {author} {\bibfnamefont {C.}~\bibnamefont {Broholm}}, \bibinfo {author}
  {\bibfnamefont {M.~B.}\ \bibnamefont {Stone}}, \bibinfo {author}
  {\bibfnamefont {E.}~\bibnamefont {Nishibori}}, \ and\ \bibinfo {author}
  {\bibfnamefont {H.}~\bibnamefont {Sawa}},\ }\href
  {http://dx.doi.org/10.1038/ncomms2914} {\bibfield  {journal} {\bibinfo
  {journal} {Nature Communications}\ }\textbf {\bibinfo {volume} {4}},\
  \bibinfo {pages} {1934 EP } (\bibinfo {year} {2013})}\BibitemShut {NoStop}%
\bibitem [{\citenamefont {Princep}\ \emph {et~al.}(2013)\citenamefont
  {Princep}, \citenamefont {Prabhakaran}, \citenamefont {Boothroyd},\ and\
  \citenamefont {Adroja}}]{Princep:2013}%
  \BibitemOpen
  \bibfield  {author} {\bibinfo {author} {\bibfnamefont {A.~J.}\ \bibnamefont
  {Princep}}, \bibinfo {author} {\bibfnamefont {D.}~\bibnamefont
  {Prabhakaran}}, \bibinfo {author} {\bibfnamefont {A.~T.}\ \bibnamefont
  {Boothroyd}}, \ and\ \bibinfo {author} {\bibfnamefont {D.~T.}\ \bibnamefont
  {Adroja}},\ }\href {http://link.aps.org/doi/10.1103/PhysRevB.88.104421}
  {\bibfield  {journal} {\bibinfo  {journal} {Physical Review B}\ }\textbf
  {\bibinfo {volume} {88}},\ \bibinfo {pages} {104421} (\bibinfo {year}
  {2013})}\BibitemShut {NoStop}%
\bibitem [{\citenamefont {Lago}\ \emph {et~al.}(2007)\citenamefont {Lago},
  \citenamefont {Blundell},\ and\ \citenamefont {Baines}}]{Lago:2007}%
  \BibitemOpen
  \bibfield  {author} {\bibinfo {author} {\bibfnamefont {J.}~\bibnamefont
  {Lago}}, \bibinfo {author} {\bibfnamefont {S.~J.}\ \bibnamefont {Blundell}},
  \ and\ \bibinfo {author} {\bibfnamefont {C.}~\bibnamefont {Baines}},\ }\href
  {http://stacks.iop.org/0953-8984/19/i=32/a=326210} {\bibfield  {journal}
  {\bibinfo  {journal} {Journal of Physics: Condensed Matter}\ }\textbf
  {\bibinfo {volume} {19}},\ \bibinfo {pages} {326210} (\bibinfo {year}
  {2007})}\BibitemShut {NoStop}%
\bibitem [{\citenamefont {Dunsiger}\ \emph {et~al.}(2011)\citenamefont
  {Dunsiger}, \citenamefont {Aczel}, \citenamefont {Arguello}, \citenamefont
  {Dabkowska}, \citenamefont {Dabkowski}, \citenamefont {Du}, \citenamefont
  {Goko}, \citenamefont {Javanparast}, \citenamefont {Lin}, \citenamefont
  {Ning}, \citenamefont {Noad}, \citenamefont {Singh}, \citenamefont
  {Williams}, \citenamefont {Uemura}, \citenamefont {Gingras},\ and\
  \citenamefont {Luke}}]{Dunsiger:2011}%
  \BibitemOpen
  \bibfield  {author} {\bibinfo {author} {\bibfnamefont {S.~R.}\ \bibnamefont
  {Dunsiger}}, \bibinfo {author} {\bibfnamefont {A.~A.}\ \bibnamefont {Aczel}},
  \bibinfo {author} {\bibfnamefont {C.}~\bibnamefont {Arguello}}, \bibinfo
  {author} {\bibfnamefont {H.}~\bibnamefont {Dabkowska}}, \bibinfo {author}
  {\bibfnamefont {A.}~\bibnamefont {Dabkowski}}, \bibinfo {author}
  {\bibfnamefont {M.~H.}\ \bibnamefont {Du}}, \bibinfo {author} {\bibfnamefont
  {T.}~\bibnamefont {Goko}}, \bibinfo {author} {\bibfnamefont {B.}~\bibnamefont
  {Javanparast}}, \bibinfo {author} {\bibfnamefont {T.}~\bibnamefont {Lin}},
  \bibinfo {author} {\bibfnamefont {F.~L.}\ \bibnamefont {Ning}}, \bibinfo
  {author} {\bibfnamefont {H.~M.~L.}\ \bibnamefont {Noad}}, \bibinfo {author}
  {\bibfnamefont {D.~J.}\ \bibnamefont {Singh}}, \bibinfo {author}
  {\bibfnamefont {T.~J.}\ \bibnamefont {Williams}}, \bibinfo {author}
  {\bibfnamefont {Y.~J.}\ \bibnamefont {Uemura}}, \bibinfo {author}
  {\bibfnamefont {M.~J.~P.}\ \bibnamefont {Gingras}}, \ and\ \bibinfo {author}
  {\bibfnamefont {G.~M.}\ \bibnamefont {Luke}},\ }\href
  {http://link.aps.org/doi/10.1103/PhysRevLett.107.207207} {\bibfield
  {journal} {\bibinfo  {journal} {Physical Review Letters}\ }\textbf {\bibinfo
  {volume} {107}},\ \bibinfo {pages} {207207} (\bibinfo {year}
  {2011})}\BibitemShut {NoStop}%
\bibitem [{\citenamefont {Blundell}(2012)}]{Blundell:2012}%
  \BibitemOpen
  \bibfield  {author} {\bibinfo {author} {\bibfnamefont {S.~J.}\ \bibnamefont
  {Blundell}},\ }\href {http://link.aps.org/doi/10.1103/PhysRevLett.108.147601}
  {\bibfield  {journal} {\bibinfo  {journal} {Physical Review Letters}\
  }\textbf {\bibinfo {volume} {108}},\ \bibinfo {pages} {147601} (\bibinfo
  {year} {2012})}\BibitemShut {NoStop}%
\bibitem [{\citenamefont {Shiddiq}\ \emph {et~al.}(2016)\citenamefont
  {Shiddiq}, \citenamefont {Komijani}, \citenamefont {Duan}, \citenamefont
  {Gaita-Ari{\~n}o}, \citenamefont {Coronado},\ and\ \citenamefont
  {Hill}}]{Shiddiq:2016}%
  \BibitemOpen
  \bibfield  {author} {\bibinfo {author} {\bibfnamefont {M.}~\bibnamefont
  {Shiddiq}}, \bibinfo {author} {\bibfnamefont {D.}~\bibnamefont {Komijani}},
  \bibinfo {author} {\bibfnamefont {Y.}~\bibnamefont {Duan}}, \bibinfo {author}
  {\bibfnamefont {A.}~\bibnamefont {Gaita-Ari{\~n}o}}, \bibinfo {author}
  {\bibfnamefont {E.}~\bibnamefont {Coronado}}, \ and\ \bibinfo {author}
  {\bibfnamefont {S.}~\bibnamefont {Hill}},\ }\href
  {http://dx.doi.org/10.1038/nature16984} {\bibfield  {journal} {\bibinfo
  {journal} {Nature}\ }\textbf {\bibinfo {volume} {531}},\ \bibinfo {pages}
  {348} (\bibinfo {year} {2016})}\BibitemShut {NoStop}%
\bibitem [{\citenamefont {Misra}\ and\ \citenamefont
  {Sudarshan}(1977)}]{Misra:1977}%
  \BibitemOpen
  \bibfield  {author} {\bibinfo {author} {\bibfnamefont {B.}~\bibnamefont
  {Misra}}\ and\ \bibinfo {author} {\bibfnamefont {E.~C.~G.}\ \bibnamefont
  {Sudarshan}},\ }\bibfield  {booktitle} {\emph {\bibinfo {booktitle} {Journal
  of Mathematical Physics}},\ }\href {\doibase 10.1063/1.523304} {\bibfield
  {journal} {\bibinfo  {journal} {Journal of Mathematical Physics}\ }\textbf
  {\bibinfo {volume} {18}},\ \bibinfo {pages} {756} (\bibinfo {year}
  {1977})}\BibitemShut {NoStop}%
\bibitem [{\citenamefont {Gherardini}\ \emph {et~al.}(2016)\citenamefont
  {Gherardini}, \citenamefont {Gupta}, \citenamefont {Cataliotti},
  \citenamefont {Smerzi}, \citenamefont {Caruso},\ and\ \citenamefont
  {Ruffo}}]{Gherardini:201601:NJP}%
  \BibitemOpen
  \bibfield  {author} {\bibinfo {author} {\bibfnamefont {S.}~\bibnamefont
  {Gherardini}}, \bibinfo {author} {\bibfnamefont {S.}~\bibnamefont {Gupta}},
  \bibinfo {author} {\bibfnamefont {F.~S.}\ \bibnamefont {Cataliotti}},
  \bibinfo {author} {\bibfnamefont {A.}~\bibnamefont {Smerzi}}, \bibinfo
  {author} {\bibfnamefont {F.}~\bibnamefont {Caruso}}, \ and\ \bibinfo {author}
  {\bibfnamefont {S.}~\bibnamefont {Ruffo}},\ }\href
  {http://stacks.iop.org/1367-2630/18/i=1/a=013048} {\bibfield  {journal}
  {\bibinfo  {journal} {New Journal of Physics}\ }\textbf {\bibinfo {volume}
  {18}},\ \bibinfo {pages} {013048} (\bibinfo {year} {2016})}\BibitemShut
  {NoStop}%
\bibitem [{\citenamefont {Jaubert}\ and\ \citenamefont
  {Holdsworth}(2009)}]{Jaubert:2009}%
  \BibitemOpen
  \bibfield  {author} {\bibinfo {author} {\bibfnamefont {L.~D.~C.}\
  \bibnamefont {Jaubert}}\ and\ \bibinfo {author} {\bibfnamefont {P.~C.~W.}\
  \bibnamefont {Holdsworth}},\ }\href {http://dx.doi.org/10.1038/nphys1227}
  {\bibfield  {journal} {\bibinfo  {journal} {Nat Phys}\ }\textbf {\bibinfo
  {volume} {5}},\ \bibinfo {pages} {258} (\bibinfo {year} {2009})}\BibitemShut
  {NoStop}%
\bibitem [{\citenamefont {Yaraskavitch}\ \emph {et~al.}(2012)\citenamefont
  {Yaraskavitch}, \citenamefont {Revell}, \citenamefont {Meng}, \citenamefont
  {Ross}, \citenamefont {Noad}, \citenamefont {Dabkowska}, \citenamefont
  {Gaulin},\ and\ \citenamefont {Kycia}}]{Yaraskavitch:2012}%
  \BibitemOpen
  \bibfield  {author} {\bibinfo {author} {\bibfnamefont {L.~R.}\ \bibnamefont
  {Yaraskavitch}}, \bibinfo {author} {\bibfnamefont {H.~M.}\ \bibnamefont
  {Revell}}, \bibinfo {author} {\bibfnamefont {S.}~\bibnamefont {Meng}},
  \bibinfo {author} {\bibfnamefont {K.~A.}\ \bibnamefont {Ross}}, \bibinfo
  {author} {\bibfnamefont {H.~M.~L.}\ \bibnamefont {Noad}}, \bibinfo {author}
  {\bibfnamefont {H.~A.}\ \bibnamefont {Dabkowska}}, \bibinfo {author}
  {\bibfnamefont {B.~D.}\ \bibnamefont {Gaulin}}, \ and\ \bibinfo {author}
  {\bibfnamefont {J.~B.}\ \bibnamefont {Kycia}},\ }\href
  {https://link.aps.org/doi/10.1103/PhysRevB.85.020410} {\bibfield  {journal}
  {\bibinfo  {journal} {Physical Review B}\ }\textbf {\bibinfo {volume} {85}},\
  \bibinfo {pages} {020410} (\bibinfo {year} {2012})}\BibitemShut {NoStop}%
\bibitem [{\citenamefont {Revell}\ \emph {et~al.}(2013)\citenamefont {Revell},
  \citenamefont {Yaraskavitch}, \citenamefont {Mason}, \citenamefont {Ross},
  \citenamefont {Noad}, \citenamefont {Dabkowska}, \citenamefont {Gaulin},
  \citenamefont {Henelius},\ and\ \citenamefont {Kycia}}]{Revell:2013}%
  \BibitemOpen
  \bibfield  {author} {\bibinfo {author} {\bibfnamefont {H.~M.}\ \bibnamefont
  {Revell}}, \bibinfo {author} {\bibfnamefont {L.~R.}\ \bibnamefont
  {Yaraskavitch}}, \bibinfo {author} {\bibfnamefont {J.~D.}\ \bibnamefont
  {Mason}}, \bibinfo {author} {\bibfnamefont {K.~A.}\ \bibnamefont {Ross}},
  \bibinfo {author} {\bibfnamefont {H.~M.~L.}\ \bibnamefont {Noad}}, \bibinfo
  {author} {\bibfnamefont {H.~A.}\ \bibnamefont {Dabkowska}}, \bibinfo {author}
  {\bibfnamefont {B.~D.}\ \bibnamefont {Gaulin}}, \bibinfo {author}
  {\bibfnamefont {P.}~\bibnamefont {Henelius}}, \ and\ \bibinfo {author}
  {\bibfnamefont {J.~B.}\ \bibnamefont {Kycia}},\ }\href
  {http://dx.doi.org/10.1038/nphys2466} {\bibfield  {journal} {\bibinfo
  {journal} {Nat Phys}\ }\textbf {\bibinfo {volume} {9}},\ \bibinfo {pages}
  {34} (\bibinfo {year} {2013})}\BibitemShut {NoStop}%
\bibitem [{Note3()}]{Note3}%
  \BibitemOpen
  \bibinfo {note} {Thus far, experimental AC susceptibility is only in coarse
  agreement with Monte Carlo simulations.}\BibitemShut {Stop}%
\bibitem [{\citenamefont {Breuer}\ and\ \citenamefont
  {Petruccione}(2002)}]{BreuerPetruccioneBook:2002}%
  \BibitemOpen
  \bibfield  {author} {\bibinfo {author} {\bibfnamefont {H.~P.}\ \bibnamefont
  {Breuer}}\ and\ \bibinfo {author} {\bibfnamefont {F.}~\bibnamefont
  {Petruccione}},\ }\href@noop {} {\emph {\bibinfo {title} {The theory of open
  quantum systems}}}\ (\bibinfo  {publisher} {Oxford University Press},\
  \bibinfo {address} {Great Clarendon Street},\ \bibinfo {year}
  {2002})\BibitemShut {NoStop}%
\bibitem [{Note4()}]{Note4}%
  \BibitemOpen
  \bibinfo {note} {On average, there are two fast spins per tetrahedron, and
  the corresponding quantum many body system is a network of coordination 2 (1D
  {on average}) with occasional 3-way branching and end points.}\BibitemShut
  {Stop}%
\bibitem [{\citenamefont {Khomskii}(2012)}]{Khomskii:2012}%
  \BibitemOpen
  \bibfield  {author} {\bibinfo {author} {\bibfnamefont {D.~I.}\ \bibnamefont
  {Khomskii}},\ }\href {https://doi.org/10.1038/ncomms1904} {\bibfield
  {journal} {\bibinfo  {journal} {Nature Communications}\ }\textbf {\bibinfo
  {volume} {3}},\ \bibinfo {pages} {904 EP } (\bibinfo {year}
  {2012})}\BibitemShut {NoStop}%
\end{thebibliography}%


\begin{thebibliography}{15}%
\makeatletter
\providecommand \@ifxundefined [1]{%
 \@ifx{#1\undefined}
}%
\providecommand \@ifnum [1]{%
 \ifnum #1\expandafter \@firstoftwo
 \else \expandafter \@secondoftwo
 \fi
}%
\providecommand \@ifx [1]{%
 \ifx #1\expandafter \@firstoftwo
 \else \expandafter \@secondoftwo
 \fi
}%
\providecommand \natexlab [1]{#1}%
\providecommand \enquote  [1]{``#1''}%
\providecommand \bibnamefont  [1]{#1}%
\providecommand \bibfnamefont [1]{#1}%
\providecommand \citenamefont [1]{#1}%
\providecommand \href@noop [0]{\@secondoftwo}%
\providecommand \href [0]{\begingroup \@sanitize@url \@href}%
\providecommand \@href[1]{\@@startlink{#1}\@@href}%
\providecommand \@@href[1]{\endgroup#1\@@endlink}%
\providecommand \@sanitize@url [0]{\catcode `\\12\catcode `\$12\catcode
  `\&12\catcode `\#12\catcode `\^12\catcode `\_12\catcode `\%12\relax}%
\providecommand \@@startlink[1]{}%
\providecommand \@@endlink[0]{}%
\providecommand \url  [0]{\begingroup\@sanitize@url \@url }%
\providecommand \@url [1]{\endgroup\@href {#1}{\urlprefix }}%
\providecommand \urlprefix  [0]{URL }%
\providecommand \Eprint [0]{\href }%
\providecommand \doibase [0]{http://dx.doi.org/}%
\providecommand \selectlanguage [0]{\@gobble}%
\providecommand \bibinfo  [0]{\@secondoftwo}%
\providecommand \bibfield  [0]{\@secondoftwo}%
\providecommand \translation [1]{[#1]}%
\providecommand \BibitemOpen [0]{}%
\providecommand \bibitemStop [0]{}%
\providecommand \bibitemNoStop [0]{.\EOS\space}%
\providecommand \EOS [0]{\spacefactor3000\relax}%
\providecommand \BibitemShut  [1]{\csname bibitem#1\endcsname}%
\let\auto@bib@innerbib\@empty
%</preamble>
\bibitem [{\citenamefont {Onoda}\ and\ \citenamefont
  {Tanaka}(2011)}]{Onoda:2011}%
  \BibitemOpen
  \bibfield  {author} {\bibinfo {author} {\bibfnamefont {S.}~\bibnamefont
  {Onoda}}\ and\ \bibinfo {author} {\bibfnamefont {Y.}~\bibnamefont {Tanaka}},\
  }\href {http://link.aps.org/doi/10.1103/PhysRevB.83.094411} {\bibfield
  {journal} {\bibinfo  {journal} {Physical Review B}\ }\textbf {\bibinfo
  {volume} {83}},\ \bibinfo {pages} {094411} (\bibinfo {year}
  {2011})}\BibitemShut {NoStop}%
\bibitem [{\citenamefont {Tomasello}\ \emph {et~al.}(2015)\citenamefont
  {Tomasello}, \citenamefont {Castelnovo}, \citenamefont {Moessner},\ and\
  \citenamefont {Quintanilla}}]{Tomasello:2015}%
  \BibitemOpen
  \bibfield  {author} {\bibinfo {author} {\bibfnamefont {B.}~\bibnamefont
  {Tomasello}}, \bibinfo {author} {\bibfnamefont {C.}~\bibnamefont
  {Castelnovo}}, \bibinfo {author} {\bibfnamefont {R.}~\bibnamefont
  {Moessner}}, \ and\ \bibinfo {author} {\bibfnamefont {J.}~\bibnamefont
  {Quintanilla}},\ }\href {http://link.aps.org/doi/10.1103/PhysRevB.92.155120}
  {\bibfield  {journal} {\bibinfo  {journal} {Physical Review B}\ }\textbf
  {\bibinfo {volume} {92}},\ \bibinfo {pages} {155120} (\bibinfo {year}
  {2015})}\BibitemShut {NoStop}%
\bibitem [{\citenamefont {Sala}()}]{Sala:MonteCarlo}%
  \BibitemOpen
  \bibfield  {author} {\bibinfo {author} {\bibfnamefont {G.}~\bibnamefont
  {Sala}},\ }\href@noop {} {}\bibinfo {note} {(private
  communication)}\BibitemShut {NoStop}%
\bibitem [{Note1()}]{Note1}%
  \BibitemOpen
  \bibinfo {note} {We choose without any loss of generality only the
  configurations where the central spin is a majority spin for the tetrahedron
  hosting the monopole}\BibitemShut {NoStop}%
\bibitem [{\citenamefont {Castelnovo}\ \emph {et~al.}(2010)\citenamefont
  {Castelnovo}, \citenamefont {Moessner},\ and\ \citenamefont
  {Sondhi}}]{Castelnovo:2010}%
  \BibitemOpen
  \bibfield  {author} {\bibinfo {author} {\bibfnamefont {C.}~\bibnamefont
  {Castelnovo}}, \bibinfo {author} {\bibfnamefont {R.}~\bibnamefont
  {Moessner}}, \ and\ \bibinfo {author} {\bibfnamefont {S.~L.}\ \bibnamefont
  {Sondhi}},\ }\href {http://link.aps.org/doi/10.1103/PhysRevLett.104.107201}
  {\bibfield  {journal} {\bibinfo  {journal} {Physical Review Letters}\
  }\textbf {\bibinfo {volume} {104}},\ \bibinfo {pages} {107201} (\bibinfo
  {year} {2010})}\BibitemShut {NoStop}%
\bibitem [{\citenamefont {Tomasello}(2014)}]{Tomasello:PhDThesis}%
  \BibitemOpen
  \bibfield  {author} {\bibinfo {author} {\bibfnamefont {B.}~\bibnamefont
  {Tomasello}},\ }\emph {\bibinfo {title} {A quantum-mechanical study of the
  dynamical properties of spin-ice materials}},\ \href
  {https://kar.kent.ac.uk/48015/} {Ph.D. thesis},\ \bibinfo  {school} {School
  of Physical Sciences, University of Kent} (\bibinfo {year}
  {2014})\BibitemShut {NoStop}%
\bibitem [{\citenamefont {Rau}\ and\ \citenamefont {Gingras}(2015)}]{Rau:2015}%
  \BibitemOpen
  \bibfield  {author} {\bibinfo {author} {\bibfnamefont {J.~G.}\ \bibnamefont
  {Rau}}\ and\ \bibinfo {author} {\bibfnamefont {M.~J.~P.}\ \bibnamefont
  {Gingras}},\ }\href {http://link.aps.org/doi/10.1103/PhysRevB.92.144417}
  {\bibfield  {journal} {\bibinfo  {journal} {Physical Review B}\ }\textbf
  {\bibinfo {volume} {92}},\ \bibinfo {pages} {144417} (\bibinfo {year}
  {2015})}\BibitemShut {NoStop}%
\bibitem [{\citenamefont {Abragam}\ and\ \citenamefont
  {Bleaney}(1987)}]{AbragamBleaneyBook:1987}%
  \BibitemOpen
  \bibfield  {author} {\bibinfo {author} {\bibfnamefont {A.}~\bibnamefont
  {Abragam}}\ and\ \bibinfo {author} {\bibfnamefont {B.}~\bibnamefont
  {Bleaney}},\ }\href@noop {} {\emph {\bibinfo {title} {Electron Paramagnetic
  Resonance of Transition Ions}}}\ (\bibinfo  {publisher} {Dover Publications
  Inc.},\ \bibinfo {year} {1987})\BibitemShut {NoStop}%
\bibitem [{\citenamefont {Ruminy}\ \emph {et~al.}(2016)\citenamefont {Ruminy},
  \citenamefont {Pomjakushina}, \citenamefont {Iida}, \citenamefont {Kamazawa},
  \citenamefont {Adroja}, \citenamefont {Stuhr},\ and\ \citenamefont
  {Fennell}}]{Ruminy:2016}%
  \BibitemOpen
  \bibfield  {author} {\bibinfo {author} {\bibfnamefont {M.}~\bibnamefont
  {Ruminy}}, \bibinfo {author} {\bibfnamefont {E.}~\bibnamefont
  {Pomjakushina}}, \bibinfo {author} {\bibfnamefont {K.}~\bibnamefont {Iida}},
  \bibinfo {author} {\bibfnamefont {K.}~\bibnamefont {Kamazawa}}, \bibinfo
  {author} {\bibfnamefont {D.~T.}\ \bibnamefont {Adroja}}, \bibinfo {author}
  {\bibfnamefont {U.}~\bibnamefont {Stuhr}}, \ and\ \bibinfo {author}
  {\bibfnamefont {T.}~\bibnamefont {Fennell}},\ }\href
  {http://link.aps.org/doi/10.1103/PhysRevB.94.024430} {\bibfield  {journal}
  {\bibinfo  {journal} {Physical Review B}\ }\textbf {\bibinfo {volume} {94}},\
  \bibinfo {pages} {024430} (\bibinfo {year} {2016})}\BibitemShut {NoStop}%
\bibitem [{\citenamefont {Princep}\ \emph {et~al.}(2013)\citenamefont
  {Princep}, \citenamefont {Prabhakaran}, \citenamefont {Boothroyd},\ and\
  \citenamefont {Adroja}}]{Princep:2013}%
  \BibitemOpen
  \bibfield  {author} {\bibinfo {author} {\bibfnamefont {A.~J.}\ \bibnamefont
  {Princep}}, \bibinfo {author} {\bibfnamefont {D.}~\bibnamefont
  {Prabhakaran}}, \bibinfo {author} {\bibfnamefont {A.~T.}\ \bibnamefont
  {Boothroyd}}, \ and\ \bibinfo {author} {\bibfnamefont {D.~T.}\ \bibnamefont
  {Adroja}},\ }\href {http://link.aps.org/doi/10.1103/PhysRevB.88.104421}
  {\bibfield  {journal} {\bibinfo  {journal} {Physical Review B}\ }\textbf
  {\bibinfo {volume} {88}},\ \bibinfo {pages} {104421} (\bibinfo {year}
  {2013})}\BibitemShut {NoStop}%
\bibitem [{\citenamefont {Kimura}\ \emph {et~al.}(2013)\citenamefont {Kimura},
  \citenamefont {Nakatsuji}, \citenamefont {Wen}, \citenamefont {Broholm},
  \citenamefont {Stone}, \citenamefont {Nishibori},\ and\ \citenamefont
  {Sawa}}]{Kimura:2013}%
  \BibitemOpen
  \bibfield  {author} {\bibinfo {author} {\bibfnamefont {K.}~\bibnamefont
  {Kimura}}, \bibinfo {author} {\bibfnamefont {S.}~\bibnamefont {Nakatsuji}},
  \bibinfo {author} {\bibfnamefont {J.-J.}\ \bibnamefont {Wen}}, \bibinfo
  {author} {\bibfnamefont {C.}~\bibnamefont {Broholm}}, \bibinfo {author}
  {\bibfnamefont {M.~B.}\ \bibnamefont {Stone}}, \bibinfo {author}
  {\bibfnamefont {E.}~\bibnamefont {Nishibori}}, \ and\ \bibinfo {author}
  {\bibfnamefont {H.}~\bibnamefont {Sawa}},\ }\href
  {http://dx.doi.org/10.1038/ncomms2914} {\bibfield  {journal} {\bibinfo
  {journal} {Nature Communications}\ }\textbf {\bibinfo {volume} {4}},\
  \bibinfo {pages} {1934 EP } (\bibinfo {year} {2013})}\BibitemShut {NoStop}%
\bibitem [{\citenamefont {Bramwell}\ and\ \citenamefont
  {Gingras}(2001)}]{Bramwell:2001}%
  \BibitemOpen
  \bibfield  {author} {\bibinfo {author} {\bibfnamefont {S.~T.}\ \bibnamefont
  {Bramwell}}\ and\ \bibinfo {author} {\bibfnamefont {M.~J.~P.}\ \bibnamefont
  {Gingras}},\ }\href
  {http://www.sciencemag.org/content/294/5546/1495.abstract} {\bibfield
  {journal} {\bibinfo  {journal} {Science}\ }\textbf {\bibinfo {volume}
  {294}},\ \bibinfo {pages} {1495} (\bibinfo {year} {2001})}\BibitemShut
  {NoStop}%
\bibitem [{\citenamefont {Misra}\ and\ \citenamefont
  {Sudarshan}(1977)}]{Misra:1977}%
  \BibitemOpen
  \bibfield  {author} {\bibinfo {author} {\bibfnamefont {B.}~\bibnamefont
  {Misra}}\ and\ \bibinfo {author} {\bibfnamefont {E.~C.~G.}\ \bibnamefont
  {Sudarshan}},\ }\bibfield  {booktitle} {\emph {\bibinfo {booktitle} {Journal
  of Mathematical Physics}},\ }\href {\doibase 10.1063/1.523304} {\bibfield
  {journal} {\bibinfo  {journal} {Journal of Mathematical Physics}\ }\textbf
  {\bibinfo {volume} {18}},\ \bibinfo {pages} {756} (\bibinfo {year}
  {1977})}\BibitemShut {NoStop}%
\bibitem [{\citenamefont {Gherardini}\ \emph {et~al.}(2016)\citenamefont
  {Gherardini}, \citenamefont {Gupta}, \citenamefont {Cataliotti},
  \citenamefont {Smerzi}, \citenamefont {Caruso},\ and\ \citenamefont
  {Ruffo}}]{Gherardini:201601:NJP}%
  \BibitemOpen
  \bibfield  {author} {\bibinfo {author} {\bibfnamefont {S.}~\bibnamefont
  {Gherardini}}, \bibinfo {author} {\bibfnamefont {S.}~\bibnamefont {Gupta}},
  \bibinfo {author} {\bibfnamefont {F.~S.}\ \bibnamefont {Cataliotti}},
  \bibinfo {author} {\bibfnamefont {A.}~\bibnamefont {Smerzi}}, \bibinfo
  {author} {\bibfnamefont {F.}~\bibnamefont {Caruso}}, \ and\ \bibinfo {author}
  {\bibfnamefont {S.}~\bibnamefont {Ruffo}},\ }\href
  {http://stacks.iop.org/1367-2630/18/i=1/a=013048} {\bibfield  {journal}
  {\bibinfo  {journal} {New Journal of Physics}\ }\textbf {\bibinfo {volume}
  {18}},\ \bibinfo {pages} {013048} (\bibinfo {year} {2016})}\BibitemShut
  {NoStop}%
\bibitem [{\citenamefont {Gherardini}\ \emph {et~al.}(2017)\citenamefont
  {Gherardini}, \citenamefont {Lovecchio}, \citenamefont {M{\"u}ller},
  \citenamefont {Lombardi}, \citenamefont {Caruso},\ and\ \citenamefont
  {Cataliotti}}]{Gherardini:2017}%
  \BibitemOpen
  \bibfield  {author} {\bibinfo {author} {\bibfnamefont {S.}~\bibnamefont
  {Gherardini}}, \bibinfo {author} {\bibfnamefont {C.}~\bibnamefont
  {Lovecchio}}, \bibinfo {author} {\bibfnamefont {M.~M.}\ \bibnamefont
  {M{\"u}ller}}, \bibinfo {author} {\bibfnamefont {P.}~\bibnamefont
  {Lombardi}}, \bibinfo {author} {\bibfnamefont {F.}~\bibnamefont {Caruso}}, \
  and\ \bibinfo {author} {\bibfnamefont {F.~S.}\ \bibnamefont {Cataliotti}},\
  }\href {http://stacks.iop.org/2058-9565/2/i=1/a=015007} {\bibfield  {journal}
  {\bibinfo  {journal} {Quantum Science and Technology}\ }\textbf {\bibinfo
  {volume} {2}},\ \bibinfo {pages} {015007} (\bibinfo {year}
  {2017})}\BibitemShut {NoStop}%
\end{thebibliography}%
%%\nocite{}

\end{document}

\title{Supplemental Material for `Correlated Quantum Tunnelling of Monopoles in Spin Ice'}

\author	{Bruno Tomasello}
\email	{brunotomasello83@gmail.com, tomasello@ill.fr}
\affiliation	{SEPnet and Hubbard Theory Consortium, University of Kent, Canterbury CT2 7NH, U.K.}
\affiliation	{ISIS facility, STFC Rutherford Appleton Laboratory, Harwell Campus, Didcot OX11 0QX,  U.K.}
\affiliation	{Institut Laue-Langevin, CS 20156, 71 avenue des Martyrs, 38042 Grenoble Cedex 9, France}

\author	{Claudio Castelnovo}
\affiliation	{TCM group, Cavendish Laboratory, University of Cambridge, Cambridge CB3 0HE, U.K.}

\author	{Roderich Moessner}
\affiliation	{Max-Planck-Institut f\"{u}r Physik komplexer Systeme, 01187 Dresden, Germany}

\author	{Jorge Quintanilla}
\email	{j.quintanilla@kent.ac.uk}
\affiliation	{SEPnet and Hubbard Theory Consortium, University of Kent, Canterbury CT2 7NH, U.K.}
\affiliation	{ISIS facility, STFC Rutherford Appleton Laboratory, Harwell Campus, Didcot OX11 0QX,  U.K.}

\begin{abstract} 
Details of the theoretical models used in the main text. More specifically, Sec. I defines the framework with notations and conventions; Sec. II derives the most convenient expression for the dipolar fields to adopt in our model; Sec. III derives, from many-body theory of electrons, the most realistic and general expression for (super)exchange interactions in pyrochlore rare earth magnets; Sec. IV describes in detail our model of decoherence via quantum Zeno effect.
\end{abstract}

\maketitle

%%%%%%%%%%%%%%%%%%%%%%%%%%%%%%%%%%%%%%%%%
\section{Notations and conventions}
\label{sec:NotationsConventions}

\begin{figure}[hb]
	\centering
	\subfloat	[\label{fig:FigRefFrames01}]	
			{\includegraphics[trim=	5mm		0mm		0mm		0mm, clip, width=.48\linewidth]{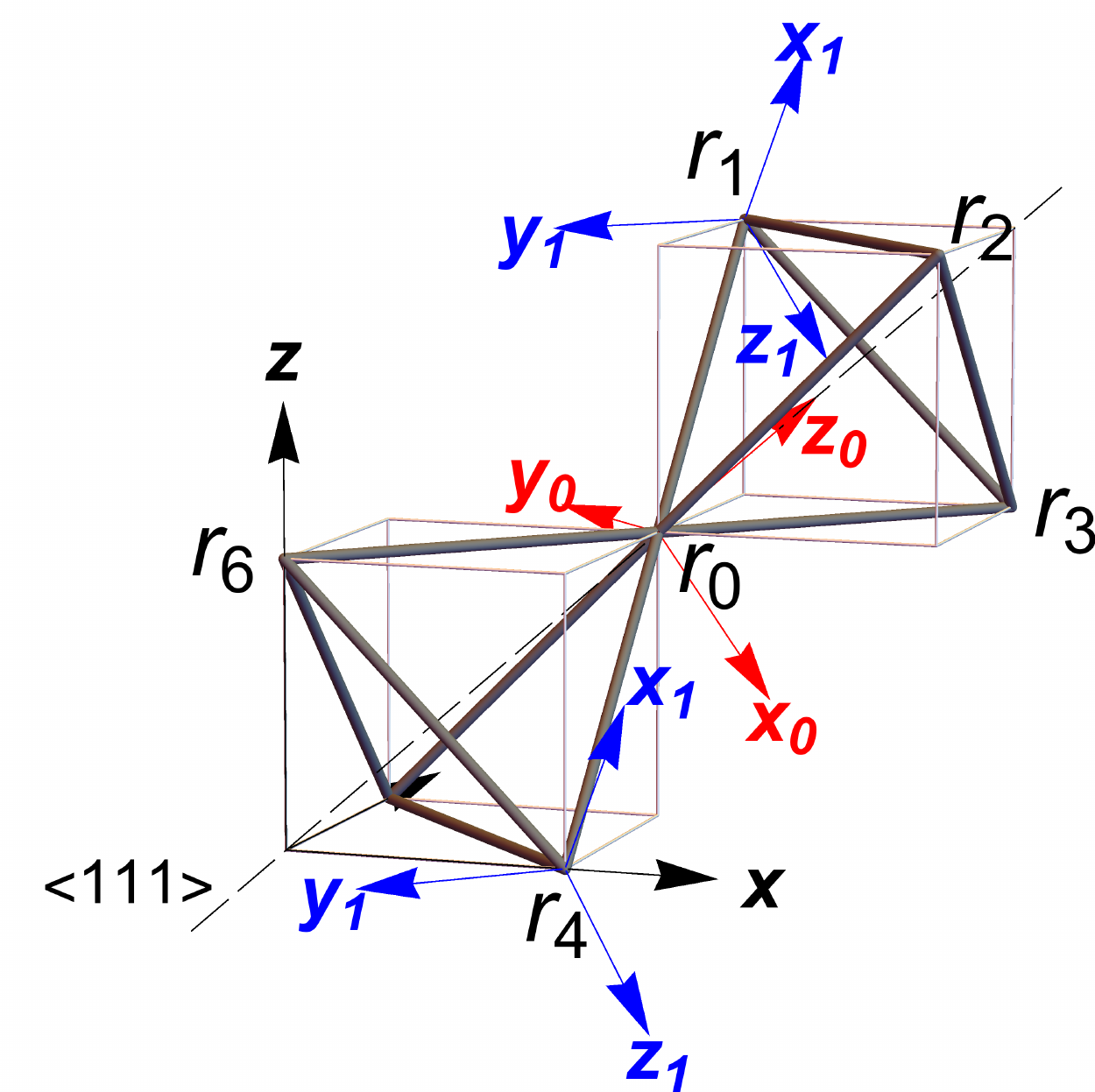}}
	\subfloat	[\label{fig:FigNLowNoFlipX}]
			{\includegraphics[trim=	0mm		0mm		0mm		0mm, clip, width=.48\linewidth]{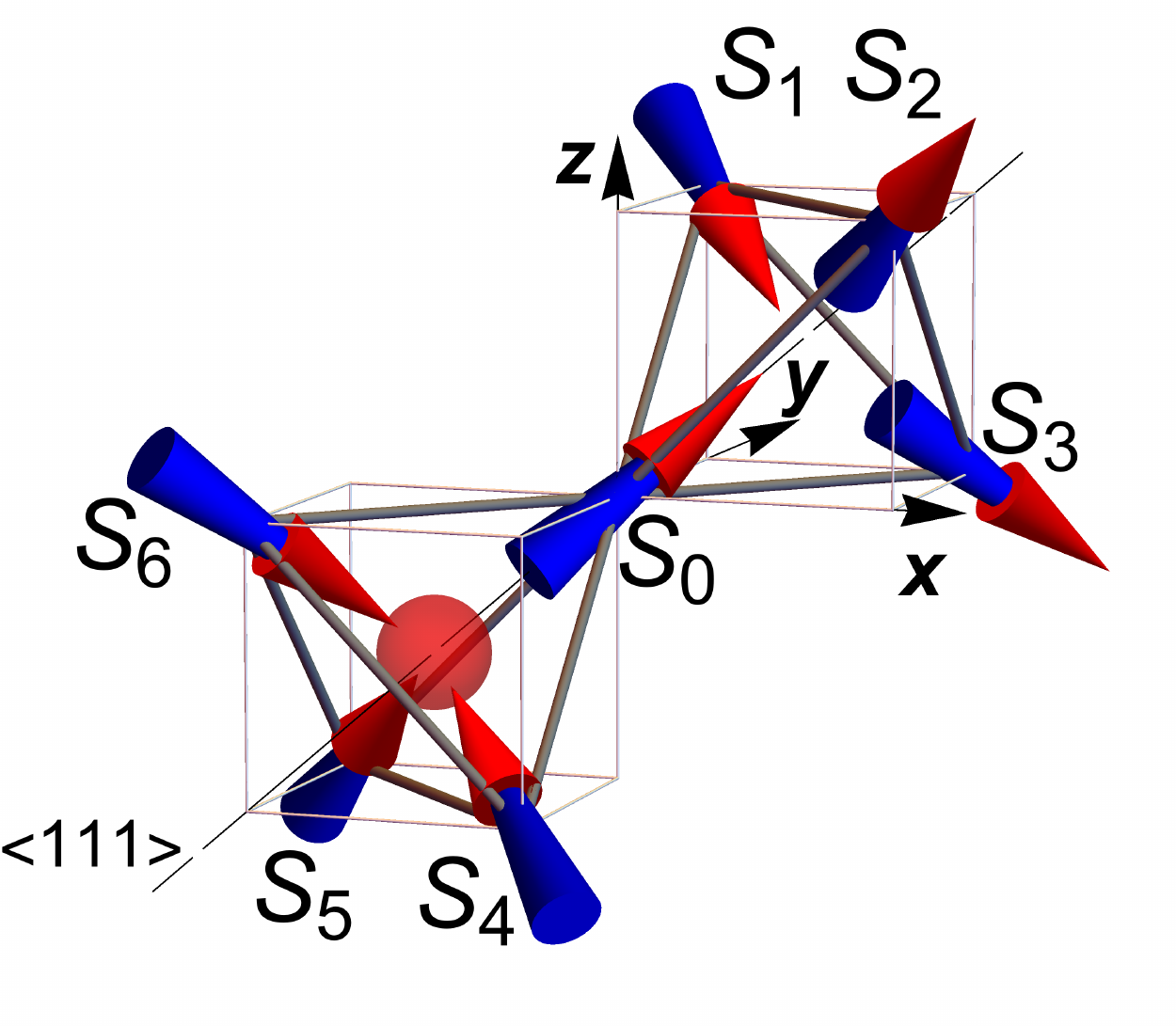}}
	\caption{
		(a) The crystal axes of the system, $\mathbf{x}, \mathbf{y}, \mathbf{z}$, are shown in black. The local axes $\mathbf{x}_{0}, \mathbf{y}_{0}, \mathbf{z}_{0}$ on the central site $0$, and $\mathbf{x}_{1}, \mathbf{y}_{1}, \mathbf{z}_{1}$ at both $\mathbf{r}_{01}$ and $\mathbf{r}_{04}=-\mathbf{r}_{01}$ are shown in blue (where $\mathbf{r}_{0j}$ is the displacement vector of site $j$ from $0$). See also Fig.~(4) in Ref.~\cite{Tomasello:2015}. (b) Example of a 2-tetrahedron configuration with a north monopole in the lower tetrahedron and the central spin as its minority one (the dipolar field in the centre -- not shown -- points along $\{4,3,3\}$).
	}
	\label{fig:FigSM2tetra}
\end{figure}

	A convenient set of local coordinate systems for a pyrochlore lattice is
\begin{subequations}
\begin{align}
	\mathbf{x}_{0} &= \frac{[1,1,\bar{2}]}{\sqrt{6}}, 	& \mathbf{y}_{0} &= \frac{[\bar{1},1,0]}{\sqrt{2}},		&  \mathbf{z}_{0} &= \frac{[1,1,1]}{\sqrt{3}},			\label{eq:LocCoo0}	\\
	\mathbf{x}_{1} &= \frac{[1,\bar{1},{2}]}{\sqrt{6}}, 	& \mathbf{y}_{1} &= \frac{[\bar{1},\bar{1},0]}{\sqrt{2}},	&  \mathbf{z}_{1} &= \frac{[1,\bar{1},\bar{1}]}{\sqrt{3}},	\label{eq:LocCoo1}	\\
	\mathbf{x}_{2} &= \frac{[\bar{1},1,{2}]}{\sqrt{6}}, 	& \mathbf{y}_{2} &= \frac{[1,1,0]}{\sqrt{2}},			&  \mathbf{z}_{2} &= \frac{[\bar{1},1,\bar{1}]}{\sqrt{3}},	\label{eq:LocCoo2}	\\
	\mathbf{x}_{3} &= \frac{[1,1,{2}]}{\sqrt{6}}, 		& \mathbf{y}_{3} &= \frac{[1,\bar{1},0]}{\sqrt{2}},		&  \mathbf{z}_{3} &= \frac{[\bar{1},\bar{1},1]}{\sqrt{3}}
\, ,	
\label{eq:LocCoo3}
\end{align}
	\label{eq:LocCoo}
\end{subequations}
illustrated in Fig.~\ref{fig:FigRefFrames01} (see also Refs.~\cite{Onoda:2011,Tomasello:2015}).

	Notice that spins at sites $j$ and $j+3$, with $j = 1, 2, 3$, have  $\mathbf{r}_{0j} \equiv \mathbf{r}_{j}-\mathbf{r}_{0} =-\mathbf{r}_{0(j+3)}$ and the same local axes. 

%%%%%%%%%%%%%%%%%%%%%%%%%%%%%%%%%%%%%%%%%
\section{Dipolar interactions}
\label{sec:Dipolar}

	We consider the classical dipolar Hamiltonian 
\begin{gather}
	H_{\rm dip}= 
	D r_{\rm nn}^3 \sum_{i,j} 
	\left[ 
		\frac{\mathbf{S}_{i} \cdot \mathbf{S}_{j}}{\vert \mathbf{r}_{ij} \vert^3}
		- 
		\frac{3\left(\mathbf{S}_{i} \cdot \mathbf{r}_{ij} \right)
		\left(\mathbf{S}_{j} \cdot \mathbf{r}_{ij} \right)}
		{\vert \mathbf{r}_{ij} \vert^5}
	\right],
\label{eq:HamiltonianDipolar}
\end{gather}
where $\mathbf{r}_{ij}$ is the  separation vector between sites $i$ and $j$, $D= \nicefrac{\mu_{0} |\mathbf{m}|^{2}}{4 \pi r_{\rm nn}^{3}}$ is the dipolar coupling constant between unit-length spins ($|\mathbf{S}_{i}|=|\mathbf{S}_{j}|=1$), and $r_{\rm nn}$ is their n.n. distance. 

	The magnetic dipole moment of a magnetic ion is $\mathbf{m}_{i}= |\mathbf{m}| \mathbf{S}_{i}$. More precisely, $\mathbf{m}_{i} = g_{J} \mu_{\mathrm{B}} \braket{\psi_{i}| \hat{\mathbf{J}} |\psi_{i}}$, where $\ket{\psi_{i}}$ is one of the two maximally polarised CEF ground states of the single-ion of interest and ${\hat{\mathbf{J}}} = (\hat{J}_{x}, \hat{J}_{y}, \hat{J}_{z})$ its total angular momentum operator. In HTO and DTO the strength of dipolar interactions originates from $|m| \approx 10 \mu_{\mathrm{B}}$ of Dy$^{3+}$ and Ho$^{3+}$ ions, respectively. Their Ising character derives from the axial anisotropy of the magnetic moment, i.e. $\braket{\hat{\mathbf{J}}}_{\psi_{i}} \approx \braket{\hat{J}_{z}}_{\psi_{i}}$, where $\hat{J}_{z}$ is defined with respect to the local axis $\mathbf{z}_{i}$. Therefore, it is convenient to write $\mathbf{m}_{i}= S_{i} |\mathbf{m}| \mathbf{z}_{i}$. 

	In a system of $N$ RE$^{3+}$ ions, the $N-1$ spins $S_{1}, S_{2}, \dots, S_{N-1}$ produce a dipolar field on site $i=0$, 
\begin{gather}
	\mathbf{B}_{\rm dip}^{\{N-1\}}(0) = 
								\frac{D r_{\rm nn}^3}{|\mathbf{m}|}
								\sum_{j=1}^{N-1}		
									\left[ 
										\frac	{\mathbf{z}_{j} - 3 \left( \hat{\mathbf{z}}_{j}\cdot \hat{\mathbf{r}}_{j} \right) \hat{\mathbf{r}}_{j}}
										{|\mathbf{r}_{j}|^{3}}
									 \right]
									 S_{j}
\, , 
\label{eq:DipFieldN}
\end{gather}
where we introduced for convenience the notation $\mathbf{r}_{0j} \equiv |\mathbf{r}_{0j}| \hat{\mathbf{r}}_{0j}$, with $|\hat{\mathbf{r}}_{0j}|=1$. 

	If we consider a 2-tetrahedron (7 spin) cluster (Fig.~\ref{fig:FigRefFrames01}), the 6 spins nearest-neighbours to the central one can be conveniently paired according to their easy axis: \{1,4\}, \{2,5\}, \{3,6\}. Each pair $j,j+3$ has $\mathbf{z}_{j} = \mathbf{z}_{j+3}$ and also opposite displacement vectors $\hat{\mathbf{r}}_{j} \equiv \hat{\mathbf{r}}_{0j}=-\hat{\mathbf{r}}_{0j+3}$. The resulting dipolar field at the central spin $0$ is then 
\begin{gather}
\mathbf{B}_{\rm dip}^{\{\rm 6\}}(0)	=	
				\frac{D}{|\mathbf{m}|}
				\sum_{j=1,2,3}
				\left[ \mathbf{z}_{j} + \sqrt{6} \, \hat{\mathbf{r}}_{j} \right]
				\left( S_{j} + S_{j+3}  \right),
\label{eq:DipFieldTwoTetra}
\end{gather}
where we used the fact that $3 \left(\mathbf{z}_{j}  \cdot \hat{\mathbf{r}}_{j} \right)= - \sqrt{6}$, for any $j$. % for $j=1,2,3$. 
If the two spins have opposite orientations ($S_{j} = -S_{j+3}$) their contribution to $\mathbf{B}_{\rm dip}^{\{\rm 6\}}(0)$ vanishes, while if they have the same orientation ($S_{j} = S_{j+3}$), it doubles. (For n.n. interactions, it is typical to define $D_{\mathrm{nn}}=5D/3$, from  $\mathbf{z}_{i} \cdot \left( \mathbf{z}_{j} + \sqrt{6} \, \hat{\mathbf{r}}_{j} \right)=5/3$, for any $i \neq j$.) 

	From Eq.~\eqref{eq:DipFieldTwoTetra} one sees readily that all 2in-2out states of a 2-tetrahedron system produce dipolar fields with strong longitudinal components along the local easy axis of the central spin. The situation is remarkably different in the case where one of the two tetrahedra hosts a monopole. Up to symmetries of the system, there are three such inequivalent configurations. The other configurations are obtained by global (clockwise and anticlockwise) rotations of 120 degrees around the $\langle111\rangle$ symmetry axis of the 2 tetrahedra, and by overall spin-reversal. 
Using the pairwise summation in Eq.~\eqref{eq:DipFieldTwoTetra}, one can immediately show that the field acting on the central spin vanishes in one such configuration, whereas it is finite (and purely transverse) in the remaining two. We find that the finite transverse fields point along $\phi_{n}=30 \,^{\circ} + n \, 60 \,^{\circ}$, $n=0,1, \dots,5$, namely the high-symmetry crystal-field directions in Ref.~\cite{Tomasello:2015}. This fact plays a crucial role in the extent of the spin precession. 

	Notice that there are other configurations featuring one monopole in a 2-tetrahedron system, not discussed in the main text, where the central spin is a minority spin in the tetrahedron hosting the monopole (minority with respect to the 3in-1out or 3out-1in configuration). One such configuration is shown for example in Fig.~\ref{fig:FigNLowNoFlipX}. In this case, the central spin experiences a large longitudinal field component that pins its direction and prevents any substantial precession. Indeed, the reversal of the central spin produces a tetrahedron with 4 spins pointing in (or out), which is a ``double-monopole'' state with higher energy in spin ice than 3in-1out or 3out-1in monopoles. 

	Although we have considered only a small cluster of spins surrounding the central one, we find that they provide a good indication of the behaviour of the internal fields even when farther neighbours are included. We verified this by comparing to a 25-spin cluster (discussed hereafter) and to large scale Monte Carlo simulations (not shown; we are grateful to G.~Sala for sharing with us Ewald-summed Monte Carlo data~\cite{Sala:MonteCarlo}). 

\begin{figure*}
	\centering
	\;	\subfloat	[\label{fig:ClusterNoMonop}]% Type sub-cation inside [ ] before \label
			{\includegraphics[trim = 	0mm		0mm		0mm 	0mm,	clip,	width=.3\linewidth]{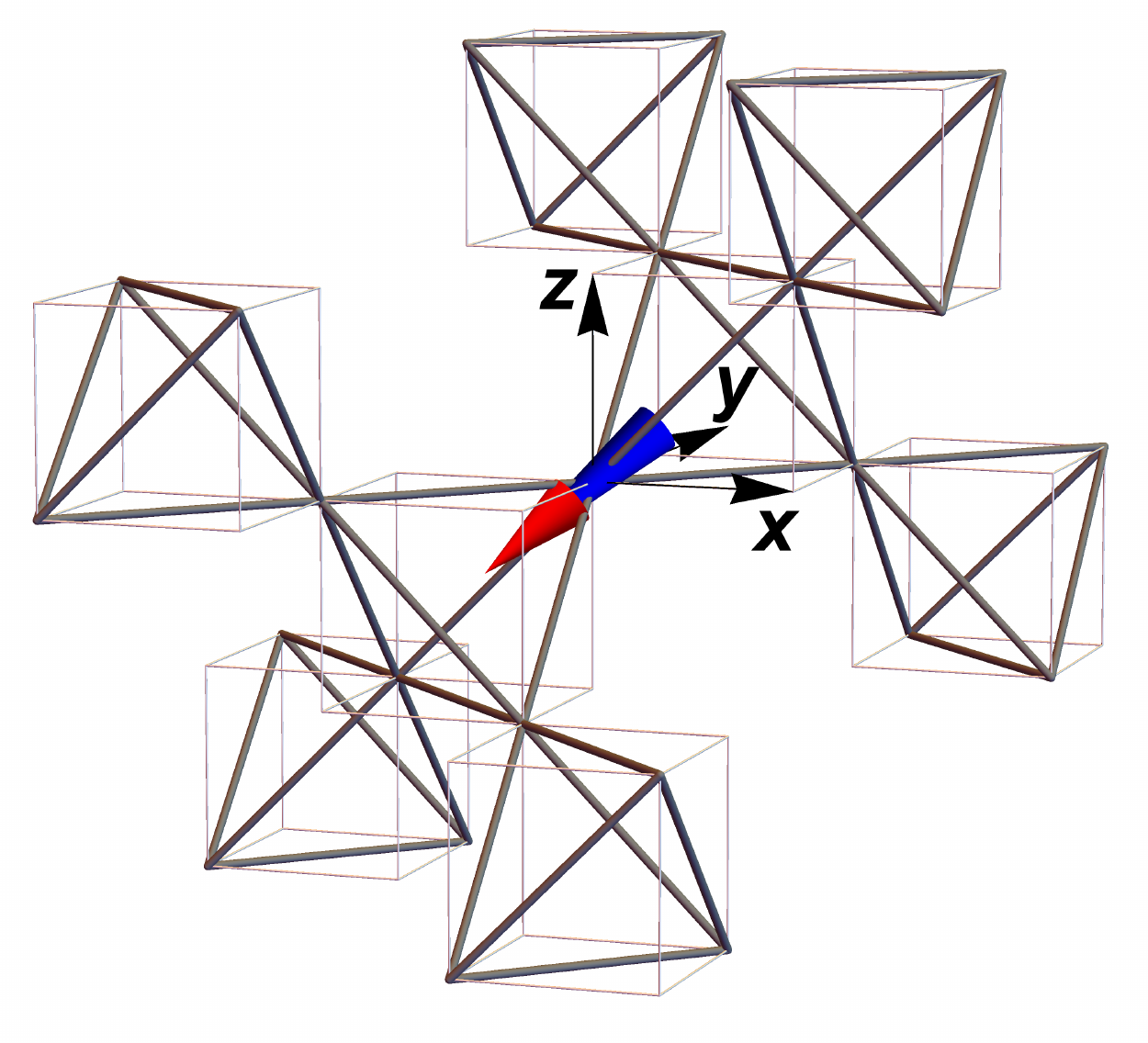}}
		\qquad
	\;	\subfloat	[\label{fig:ClusterMonopAway}]% Type sub-cation inside [ ] before \label
			{\includegraphics[trim = 	0mm		0mm		0mm 	0mm,	clip,	width=.3\linewidth]{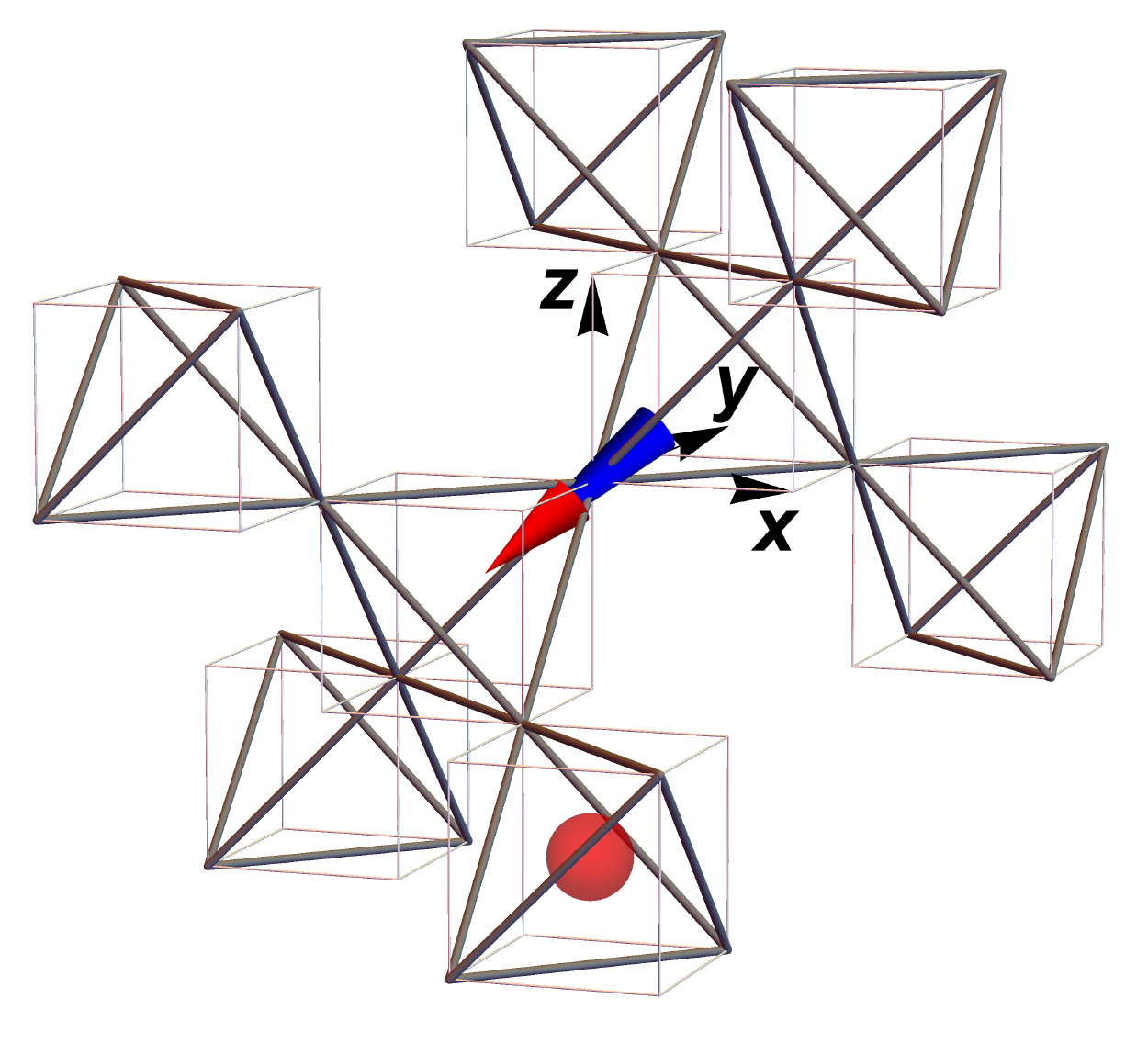}}
		\qquad
	\;	\subfloat	[\label{fig:ClusterMonopBelowCentre}]% Type sub-cation inside [ ] before \label
			{\includegraphics[trim = 	0mm		0mm		0mm 	0mm,	clip,	width=.3\linewidth]{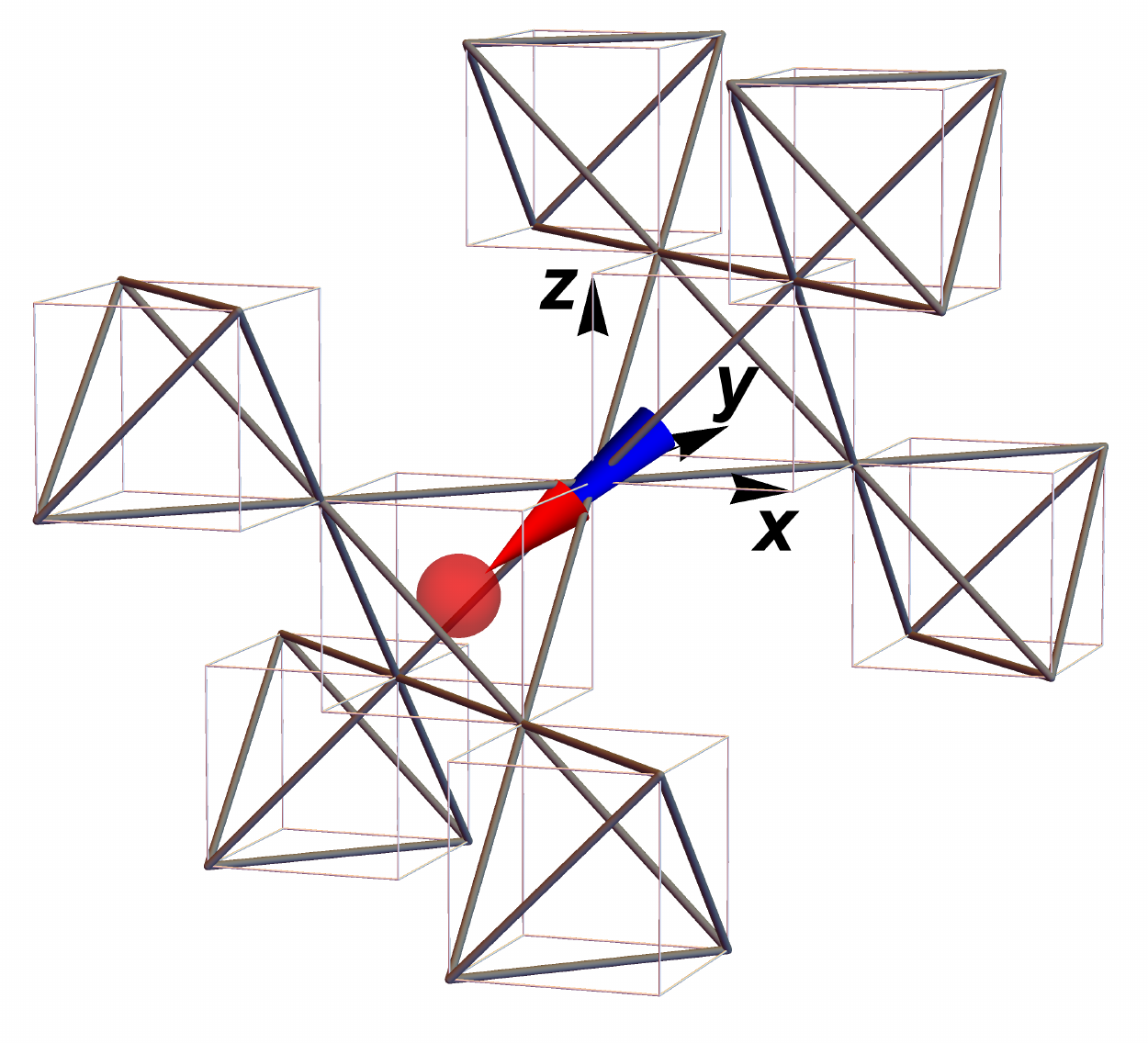}}

		\subfloat	[\label{fig:HistogramNoMonop}]% Type sub-cation inside [ ] before \label
			{\includegraphics[trim = 	0mm		0mm		0mm 	0mm,	clip,	width=.3\linewidth]{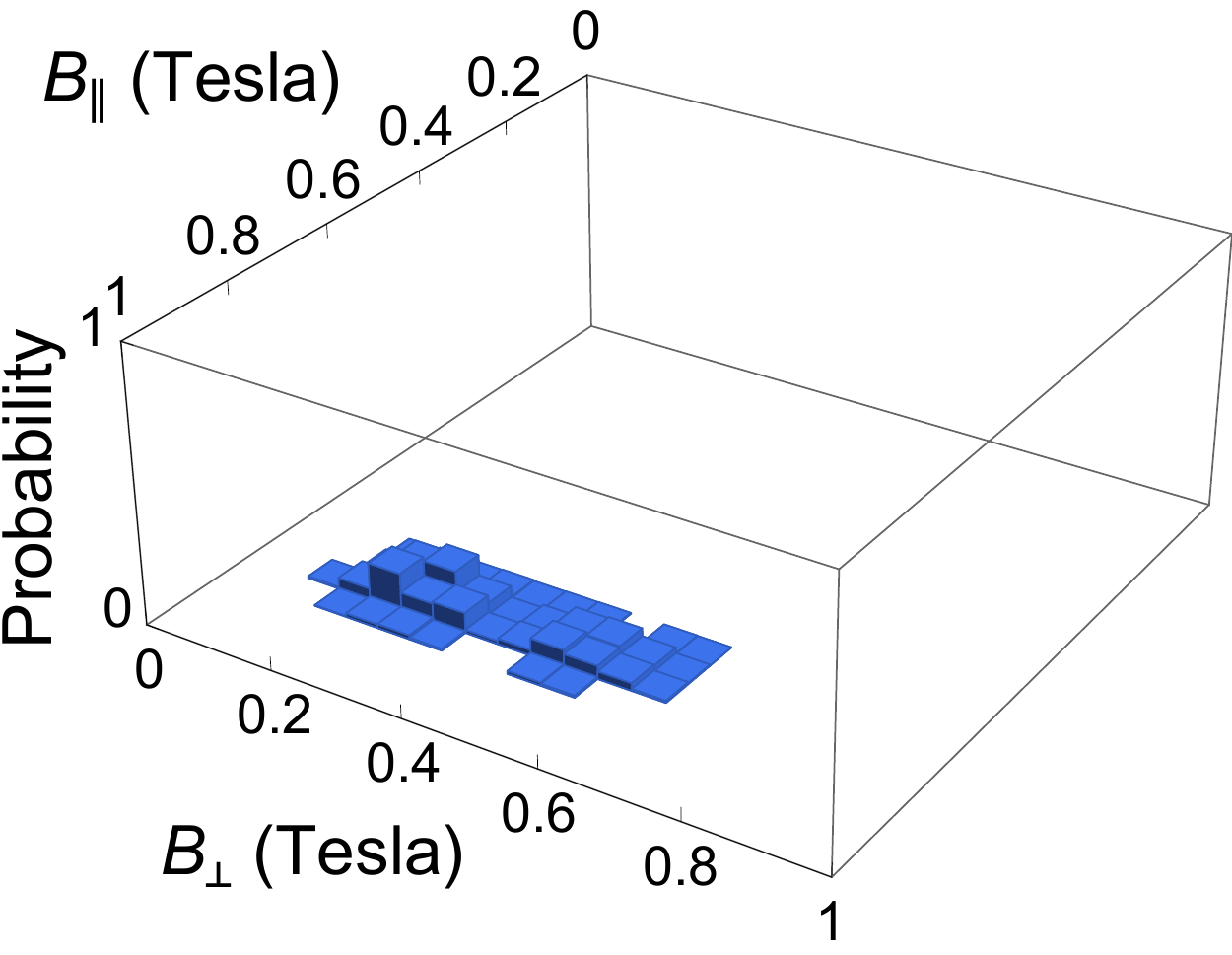}}	%trim option's order: left bottom right top
		\qquad
		\subfloat	[\label{fig:HistogramMonopAway}]% Type sub-cation inside [ ] before \label
			{\includegraphics[trim = 	0mm		0mm		0mm 	0mm,	clip,	width=.3\linewidth]{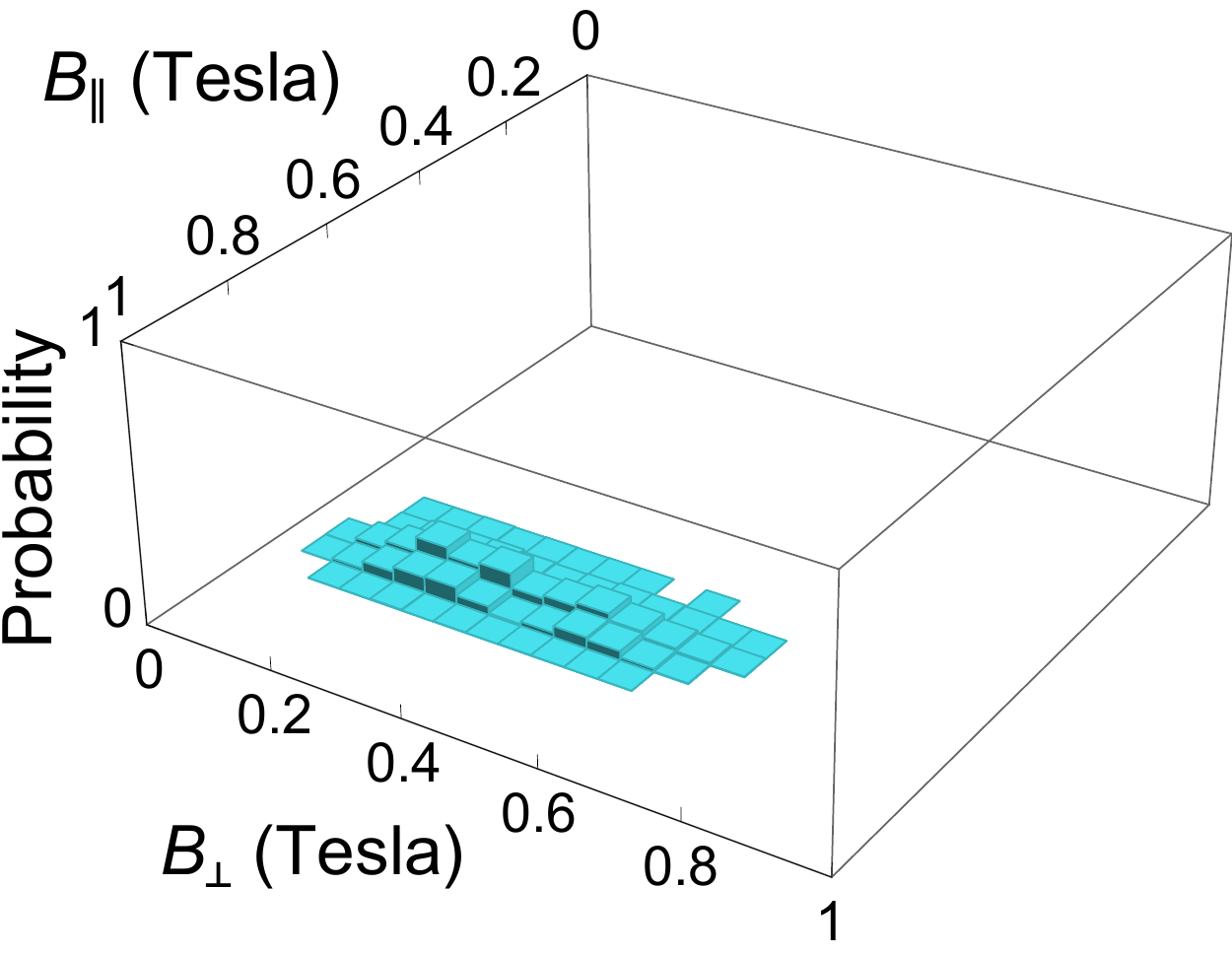}}
		\qquad
		\subfloat	[\label{fig:HistogramMonopBelowCentre}]% Type sub-cation inside [ ] before \label
			{\includegraphics[trim = 	0mm		11mm		0mm 	10mm,	clip,	width=.3\linewidth]{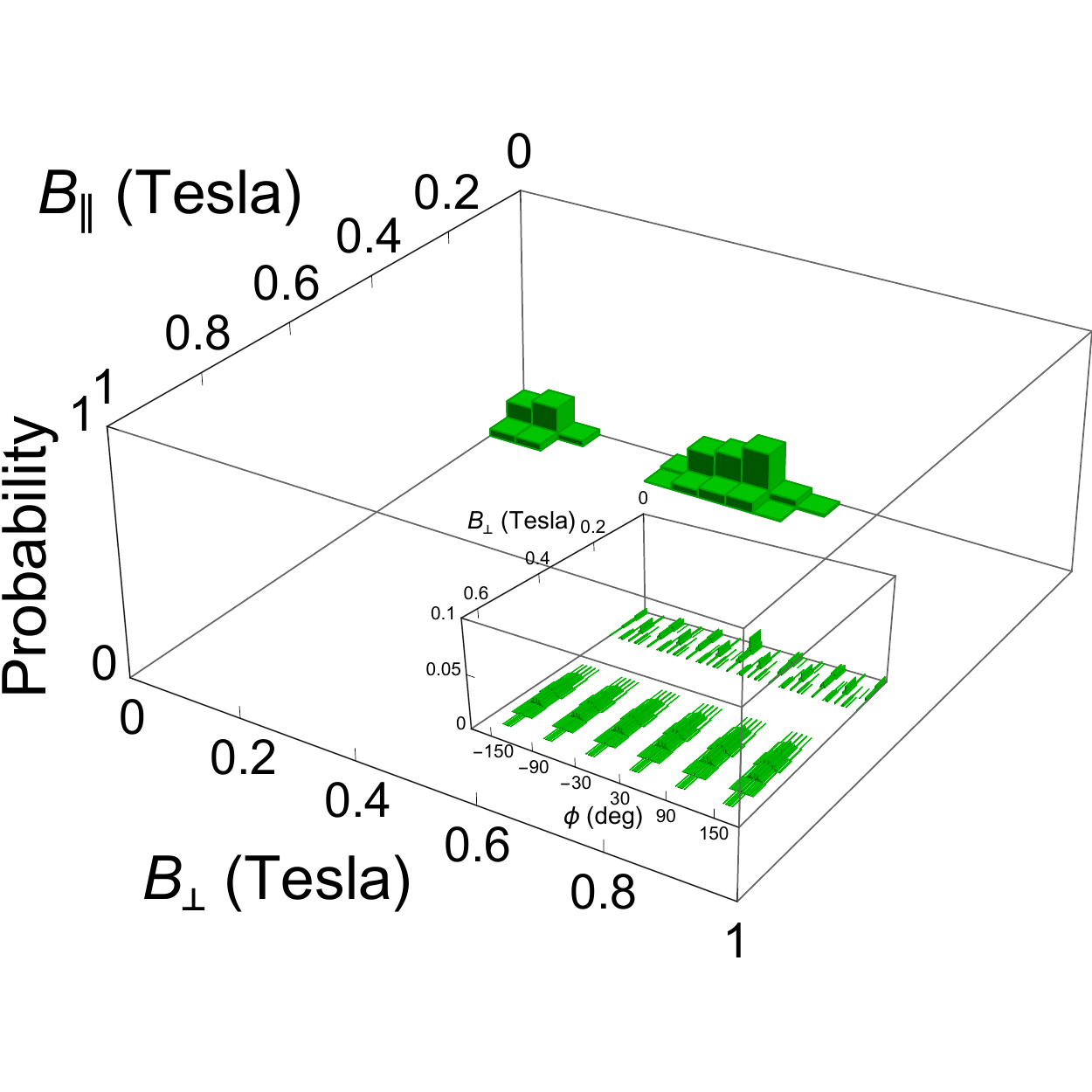}}	% all `green' values
	\caption{	
		Statistics of the dipolar field $\mathbf{B}$ at the site of the central spin in an 8-tetrahedron spin ice system. The top panels (a, b, c) illustrate the different cases corresponding to the respective histograms in the bottom panels (d, e, f). The central spin is kept fixed as we sample all the configurations of the surrounding 24 spins (not shown), consistently with the presence or absence of a monopole (red sphere). $B_{\|}$ and $B_{\bot}$ are, respectively, the field components parallel and perpendicular to the central easy axis. Note the dominant longitudinal component $B_{\|}$ (d, e), unless a single monopole is located next to the central spin (f); remarkably the fields are distributed along the high-symmetry directions $\phi_{n}$ on the transverse plane (see inset). 
	}
	\label{fig:HistogramsClusters}
\end{figure*}

	In the 25-spin cluster, we consider the system of 8 tetrahedra illustrated in Fig.~\ref{fig:HistogramsClusters}. Dipolar fields on the central site are sampled by considering exhaustively the configurations of the other 24 spins where all tetrahedra are in 2in-2out states, except for the one marked by a red sphere (panels~\ref{fig:ClusterMonopAway} and~\ref{fig:ClusterMonopBelowCentre}), which hosts a monopole~\footnote{We choose without any loss of generality only the configurations where the central spin is a majority spin for the tetrahedron hosting the monopole}. The resulting fields are then used to build the histograms in the bottom panels of Fig.~\ref{fig:HistogramsClusters}, illustrating the probabilities of the corresponding longitudinal ($B_{\|}$) and transverse ($B_{\bot}$) field components. 

	In the absence of monopoles, Fig.~\ref{fig:ClusterNoMonop}, the central spin is subject to a dominant longitudinal field $B_{\|}\approx0.8$~Tesla (Fig.~\ref{fig:HistogramNoMonop}). The small transverse field component is unlikely to induce appreciable quantum fluctuations. Crucially, the field distribution remains largely unchanged (Fig.~\ref{fig:HistogramMonopAway}) if a monopole is introduced in a tetrahedron that is not adjacent to the central spin (Fig~\ref{fig:ClusterMonopAway}). 

	In presence of a monopole adjacent to the central spin (Fig.~\ref{fig:ClusterMonopBelowCentre}), the situation is very different: the longitudinal component is suppressed ($B_{\|} \approx 0$) and the transverse field distribution becomes strikingly bimodal (peaked at $\approx 0.03$~Tesla and $0.45 $~Tesla), similarly to the 2-tetrahedron system. Once again, we find that, of the total number of configurations, $1/3$ have transverse field $0 \leq B \lesssim 0.15$~Tesla and $2/3$ have $0.3 \lesssim B \lesssim 0.6$~Tesla, exactly. This is also contrasted to the 2-tetrahedron case in Fig.~\ref{fig:PairedHistogram} in the main text.

	Finally, the inset to Fig.~\ref{fig:HistogramMonopBelowCentre} shows that the local fields are distributed on the transverse plane mainly along the high-symmetry crystal-field angles, $\phi_{n}$ as is the case (exactly) for n.n. interactions. Remarkably, these are the only transverse field directions that induce full-flip quantum dynamics~\cite{Tomasello:2015}. 

%\fi

%%%%%%%%%%%%%%%%%%%%%%%%%%%%%%%%%%%%%%%%%
\section{Exchange interactions}
\label{sec:Exchange}

	To calculate the exchange interaction, we follow the perturbative approach presented in Ref.~\cite{Onoda:2011} by Onoda and Tanaka. It considers the virtual superexchange of electrons between n.n. Pr$^{3+}$ ions, allowed by the hybridization of the $f$ orbitals of the RE with the $p$ orbitals of the O1 Oxygen sitting at the centre of the tetrahedra. 
	
%%%%%%%%%%%%%%%%%%%%%%%%%%%%
\subsection{$f$-electron superexchange}
\label{sec:Superexchange}

	We begin by generalising Eq.~(17) of Ref.~\cite{Onoda:2011} -- only for Pr$^{3+}$ pyrochlores -- to pyrochlore oxides with RE$^{3+}$ ions hosting any number $n$ of electrons in the $f$-shell (a detailed derivation can be found in Chap.~4 of Ref.~\cite{Tomasello:PhDThesis}):
\begin{widetext}
\begin{gather}	
	\begin{split}
		\hat{\mathcal{H}}_{f\!f}		
		=
		\,	
		\frac{ 2	}{	(nU-\Delta)^{2}	}		
		&
		\sum_{
					\langle
					\mathbf{r},	\mathbf{r'}
					\rangle
					}
		\sum_{\substack
						{
							m_{1},		m_{2},			m'_{1},			m'_{2} 			=	-1, 0,	1		\\
							\sigma_{1},	\sigma_{2},	\sigma'_{1},	\sigma'_{2}	=	-,+
						}
					}															
		V_{m_{1}}		\,	V_{m'_{1}}		\,	V_{m_{2}}		\,	V_{m'_{2}} 
		\;\;%\times
		\hat{f}^{\dagger	}_{\mathbf{r},	m_{1},\sigma_{1}}										\,
		\hat{f}^{				}_{\mathbf{r},	m_{2},\sigma_{2}}										\,
		\hat{f}^{\dagger	}_{\mathbf{r'},	m'_{1},\sigma'_{1}}								\,
		\hat{f}^{				}_{\mathbf{r'},	m'_{2},\sigma'_{2}}											\\[0.1cm]
		&		
		\quad
		\times
		\Bigg[
		-	\frac{1}{nU-\Delta}
			\delta_{  \substack{ m_{1},	m_{2}	 \\ \sigma_{1}, \sigma_{2} } }	\,
			\delta_{  \substack{ m'_{1},	m'_{2}	 \\ \sigma'_{1}, \sigma'_{2} } }						
			+
			\bigg(			
				\frac{1}{nU-\Delta}
			+	\frac{1}{U}
			\bigg)
			(	\mathcal{R}^{\dagger}_{\mathbf{r}}	\mathcal{R}_{\mathbf{r'}}	)_{	\substack{ m_{1},	m'_{2}	 \\ \sigma_{1}, \sigma'_{2} } }
			(	\mathcal{R}^{\dagger}_{\mathbf{r'}}	\mathcal{R}_{\mathbf{r}}	)_{	\substack{ m'_{1},	m_{2}	 \\ \sigma'_{1}, \sigma_{2} } }
		\Bigg]	
\, . 
	\end{split}
\label{eq:Gen17Onoda}
\end{gather}
\end{widetext}
where $\mathbf{r}$ and $\mathbf{r'}$ are the coordinates of the two neighbouring RE-sites. The energy scales regulating the virtual electron hopping are named according to Ref.~\cite{Onoda:2011}: $U$ is the Coulomb energy for the repulsion of two electrons on the same RE-site $\mathbf{r}$; $\Delta$ is the change in energy for the {RE-O1-RE$'$} system if an electron is removed from the O1 site; and $V_{m=\pm1}=V_{pf\pi}, V_{m=0}= V_{pf\sigma}$ are the Slater-Koster parameters for RE-O1 hybridisation~\cite{Onoda:2011,Tomasello:PhDThesis,Rau:2015}. The fermionic operator $\hat{f}^{\dagger}_{\mathbf{r}, m, \sigma}$ creates an $f$-electron with magnetic quantum numbers $m_{l}\equiv m$ and $m_{s} \equiv \nicefrac{\sigma}{2}$ for, respectively, the orbital and spin contribution, at site $\mathbf{r}$. Analogously, $\hat{f}^{}_{\mathbf{r}, m, \sigma}$ is the annihilation operator. The Wigner matrix elements $(\mathcal{R}_{\mathbf{r}})_{\substack{ m, m' \\ \sigma, \sigma'}} = \bra{m, \sigma} \hat{R}_{\mathbf{r}} \ket{m', \sigma'}$ rotate the representations of the electronic states between the local and global coordinate systems as defined in Eq.~(4) of Ref.~\cite{Tomasello:2015}. The matrices $\mathcal{R}^{\dagger}_{\mathbf{r}} \mathcal{R}_{\mathbf{r'}}$ therefore match the local representations between two $\mathbf{r},\mathbf{r'}$ RE-sites. (A list of convenient coordinate systems for pyrochlores is in Eq.~\eqref{eq:LocCoo} in Sec.~\ref{sec:NotationsConventions} -- also see Eqs.~(4.22) in Ref.~\cite{Tomasello:PhDThesis}.) 

	We only consider nearest-neighbour superexchange interactions involving the central spin of a 2-tetrahedron system. The summation in Eq.~\eqref{eq:Gen17Onoda} therefore has $\mathbf{r}=\mathbf{r}_{0}$ and $\mathbf{r'}=\mathbf{r}_{j}$,  $j=1,2, ..., 6$ being the 6 nearest neighbours. A complete quantum-mechanical treatment, in the $\ket{M} \equiv \ket{J,M}$ eigenbasis of $\hat{J}_{z}$ of the ground $J$-multiplet associated to a given RE$^{3+}$ ion, requires evaluating $\bra{\widetilde{M}'} \bra{\widetilde{M}_{}} \hat{\mathcal{H}}_{f\!f} \ket{M}\ket{M'}$ by means of the expansions 
\begin{gather}\begin{split}
	\ket{M_{}}	= 
	\sum_{\substack{m_{1},\dots ,m_{n} \\ \sigma_1,\dots ,\sigma_n}}
		\tilde{C}^{M_{}}_{\substack{m_{1},\dots ,m_{n} \\ \sigma_1,\dots ,\sigma_n}} 
		\,\prod_{i=1}^{n} \hat{f}^{\dagger}_{m_{i} \sigma_{i}} \ket{0}	,
\label{eq:MJ_fermions}
\end{split}\end{gather}
where $\ket{0}\equiv \ket{0}_{\text{RE$^{3+}$}}$ is the `vacuum' for the $f$-shell of a given RE$^{3+}$ ion, and $m_{i}, \sigma_{i}$ are the magnetic quantum numbers of the $i$-th $f$-electron~\cite{Tomasello:PhDThesis}. 

	A simple example is the case of two electrons in the $f$-shell in Pr$^{3+}$ ions. In Appendix B of Ref.~\cite{Onoda:2011} the $^{4}H_{3}$ ground state manifold of Pr$^{3+}$ is given in terms of $f$-electron  fermionic operators. Each of the Eqs.~(B1) therein gives in the first line  the $\ket{M_{}}$ eigenstates as functions of $\ket{L,M_L;S,M_S}$ (eigenstates of orbital $\hat{L}_{z}$ and spin $\hat{S}_{z}$ operators with $\mathbf{\hat{J}}=\mathbf{\hat{L}}+\mathbf{\hat{S}}$), and in the second line the same states as functions of the fermionic creation operators $\hat{f}^{\dagger}_{m_{},\sigma}$ acting on the vacuum $\ket{0}\equiv \ket{0}_{\text{Pr$^{3+}$}}$.
Eqs.~(B1) in Ref.~\cite{Onoda:2011} can be summarised as
\begin{gather}\begin{split}
	\ket{M_{}}_{\text{Pr$^{3+}$}}	&= \sum_{M_L,M_S} C_{M_{},M_L,M_S} \, \ket{L,M_L;S,M_S}_{\text{Pr$^{3+}$}} \\
			&=  \sum_{\substack{m_{},m'_{} \\ \sigma,\sigma'}}
				\tilde{C}^{M_{}}_{m_{},m'_{}, \sigma,\sigma'} \, \hat{f}^{\dagger}_{m_{},\sigma} 
					\hat{f}^{\dagger}_{m'_{},\sigma'} \ket{0}_{\text{Pr$^{3+}$}} ,
\label{eq:MJ_fermions_Pr}
\end{split}\end{gather}
where 
\begin{gather}\begin{split}
	\tilde{C}^{M_{}}_{m_{},m'_{}, \sigma,\sigma'} = C_{M_{},M_L,M_S} \,\, C_{M_L,m_{},m'_{}} \,\, C_{M_S,\sigma,\sigma'} \, ,
\label{eq:ClGo:product}
\end{split}\end{gather}
and the Clebsch-Gordan coefficients ($C_{M_{},M_L,M_S}=\langle M_{} \ket{L,M_L;S,M_S}$, $C_{M_L,m_{},m'_{}}=\langle M_L \ket{m_{},m'_{}}$, $C_{M_S,\sigma,\sigma'} \equiv C_{M_S,m_{s},m_{s}'}=\langle M_S \ket{m_{s},m_{s}'}$; $m_{s}=\nicefrac{\sigma}{2}$) 
dictate the combination of angular momenta in composite systems by ensuring
\begin{gather}\begin{split}
	| l - l' | \leq L\leq l + l'		,	\quad	\quad	&	M_L= m_{}+m_{l'} 	,	\\
	| s - s' | \leq S \leq s + s'	,	\quad	\quad	&	M_S= \frac{\sigma +\sigma'}{2}	,
\label{eq:LS_condition}
\end{split}\end{gather}
and, analogously,
\begin{gather}\begin{split}
| L - S | \leq J \leq L + S		,	\quad	\quad		M_{}= M_L + M_S 	.
\label{eq:J_condition}
\end{split}\end{gather}
These properties apply to any two angular momenta~\cite{AbragamBleaneyBook:1987}. 

	Despite its generality, this approach becomes cumbersome as soon as more than two electrons are present in the $f$-shell.
Indeed, Eq.~\eqref{eq:Gen17Onoda} relies on the decompositions of $\ket{M_{}}$ in terms of the many-body operators (similarly to Eq.~\eqref{eq:MJ_fermions_Pr} above and more explicitly to Eqs.~(B1) in Ref.~\cite{Onoda:2011}). In DTO and HTO, for example, Dy$^{3+}$ and Ho$^{3+}$ ions have, respectively, $9$ and $10$ electrons in the $f$-shell, and the coefficients $\tilde{C}^{M_{}}_{m_{},m'_{}, \sigma,\sigma'}$, are drastically more complex than Eq.~\eqref{eq:ClGo:product}. 

	Here we use an alternative approach that circumvents such difficulties. We obtain the many-body expansion for only one of the possible states $\ket{M}_{0}$, and then deduce the other $\ket{M_{}} \neq \ket{M}_{0}$ using ladder operators, 
\begin{gather}\begin{split}
	\ket{ M_{}} = \frac{ \hat{J}_{\pm} \ket{ M_{} \mp 1}}{\alpha_{\mp} \left( J, M \right)}
\, ,
\label{eq:ladder:ket}
\end{split}\end{gather}
where 
\begin{gather}
	\alpha_{\pm} \left( J, M \right)	=	\sqrt{J (J+1) - M_{} (M_{} \pm 1)}
\, . 
\label{eq:alpha:JM:function}
\end{gather}
Given $\ket{M}_{0}$ in terms of the fermionic operators acting on the vacuum, then the complete set of many-body states in Eq.~\eqref{eq:MJ_fermions} can be obtained thanks to Eq.~\eqref{eq:ladder:ket}, and 
\begin{gather}
	\begin{split}
		\hat{J}_{\pm} =	
		&
			\sum_{i=1}^{n}	\hat{L}_{\pm}^{i} + \hat{S}_{\pm}^{i}	,								\\
		\hat{L}_{\pm}^{i} =	
		&	
			\sum_{m_{i}=-l_{i}}^{l_{i}} 
				{\alpha_{\pm} \left( l_{i}, m_{i} \right)}
				\sum_{\sigma_{i}=-,+} 
					\hat{f}^{\dagger }_{m_{i} \pm 1, \sigma_{i}} \hat{f}^{ }_{m_{i}, \sigma_{i}}	,	 \\
		\hat{S}_{\pm}^{i} =	
		&	
			\sum_{\sigma_{i}=-,+} 
				{\alpha_{\pm} \left( s_{i}, \frac{\sigma_{i}}{2} \right)}
				\sum_{m_{i}=-l_{i}}^{l_{i}} 
					\hat{f}^{\dagger }_{m_{i}, \sigma_{i} \pm 1} \hat{f}^{ }_{m_{i}, \sigma_{i}}	\, ,
	\end{split}
\label{eq:Jpm:MB}
\end{gather}
with three constraints: i) the Pauli principle (any vector $\ket{M^{'}_{}} \neq \ket{M_{}}$ from Eqs.~(\ref{eq:ladder:ket}-\ref{eq:Jpm:MB}) cannot have two-fermions with the same quantum numbers); ii) Hund's rules (any $\ket{M^{'}_{}} \neq \ket{M_{}}$ must have the same total $J,L,S$ as the initial $\ket{M_{}}$); and iii) angular momenta of the $n$ electrons in the $f$-shell must satisfy the equivalent of Eqs.~(\ref{eq:LS_condition}-\ref{eq:J_condition}). 

\begin{figure} 
	\captionsetup[subfloat]{labelformat=empty}
	\centering
	\subfloat	[\label{fig:sub1}]
			{\includegraphics[trim=	0mm		0mm		0mm		0mm, clip, width=.7\linewidth]{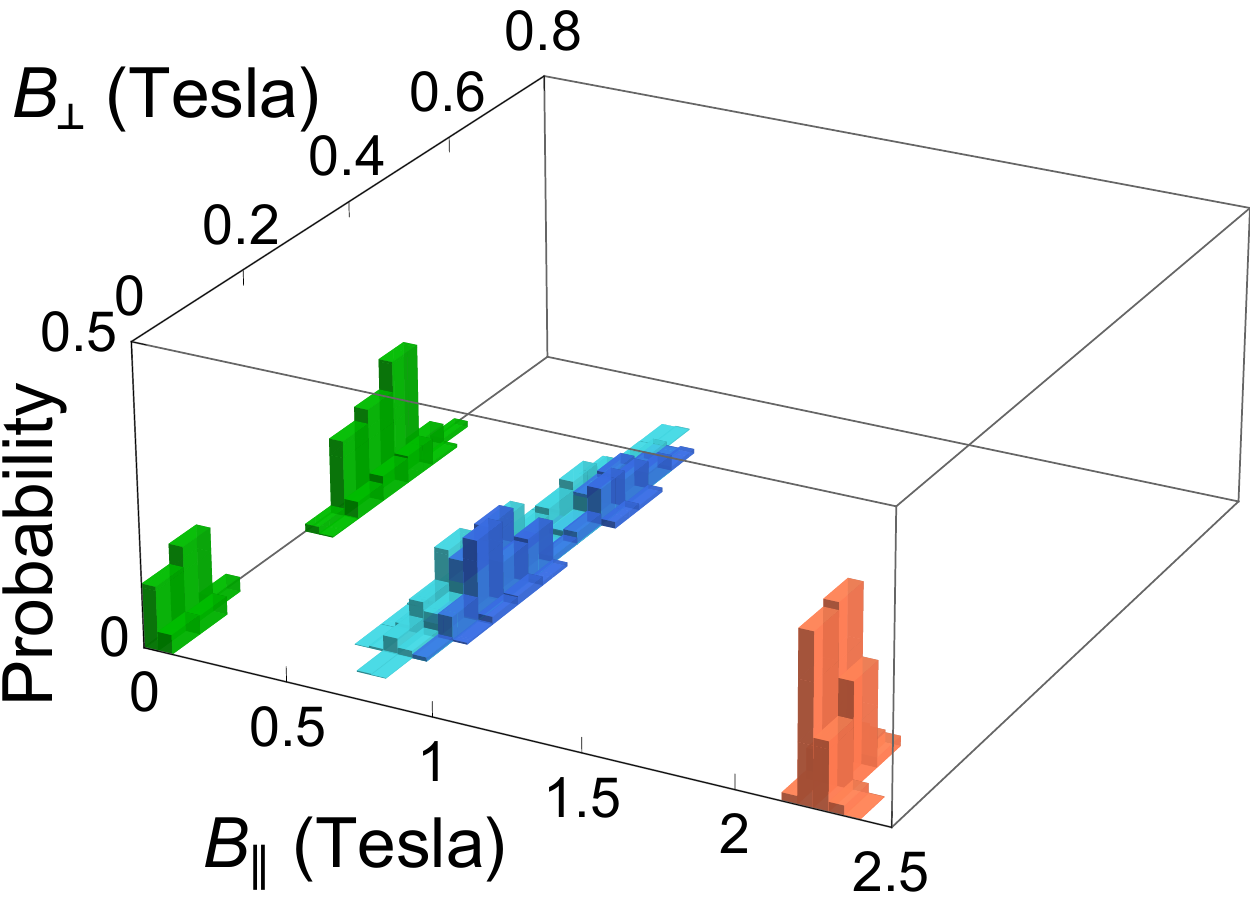}}
	\caption{
			Histograms of dipolar fields at the centre of a system of 25 spins.
			The distribution follows closely the values obtained for a 2-tetrahedron system. 
			% in Tab.~\ref{tab:TwoTetraConfigs}. 
			Legend: 
			\textcolor{NoMonopColor}{$\blacksquare$} no monopoles 
			(Figs.~\ref{fig:ClusterNoMonop},\ref{fig:HistogramNoMonop});
			\textcolor{MonopAwayColor}{$\blacksquare$} one monopole one-step away from the central site 
			(Figs.~\ref{fig:ClusterMonopAway},\ref{fig:HistogramMonopAway});
			\textcolor{MonopNearColor}{$\blacksquare$} one monopole next to the central site 
	(Figs.~\ref{fig:ClusterMonopBelowCentre},\ref{fig:HistogramMonopBelowCentre});
			\textcolor{NcPairColor}{$\blacksquare$} non contractible (n.c.) pair at the central site 
			(not shown in Fig.~\ref{fig:HistogramsClusters}, see Ref.~\cite{Castelnovo:2010}). 
			%Non contractible (n.c.) pair. \cite{Castelnovo:2010,Mostame:2014}.
	}
	\label{fig:HistogramsSummary}
\end{figure} 

	It is convenient to start from fully polarised states $\ket{M}_{0} = \ket{M= \pm J}$, where, for Ho$^{3+}$ and Dy$^{3+}$ ions, Hund's rules dictate a unique representation in terms of fermionic operators acting on the vacuum. For example, for Ho$^{3+}$ ions we have: 
\begin{widetext}
\begin{gather}
		\ket{ M_{}=8 }_{\text{Ho$^{3+}$}}	=
							\hat{f}^{\dagger }_{-3,\frac{1}{2}} 	\hat{f}^{\dagger }_{-2,\frac{1}{2}} 
							\hat{f}^{\dagger }_{-1,\frac{1}{2}} 	\hat{f}^{\dagger }_{0,\frac{1}{2}} 
							\hat{f}^{\dagger }_{1,-\frac{1}{2}} 	\hat{f}^{\dagger }_{1,\frac{1}{2}} 
							\hat{f}^{\dagger }_{2,-\frac{1}{2}} 	\hat{f}^{\dagger }_{2,\frac{1}{2}} 
							\hat{f}^{\dagger }_{3,-\frac{1}{2}} 	\hat{f}^{\dagger }_{3,\frac{1}{2}}		\ket{0}_{\text{Ho$^{3+}$}} .
\label{eq:ManyHo8}
\end{gather}
By applying the fermionic $\hat{J}_{-}$ operator, the state
\begin{gather}
	\begin{split}
		\ket{ M_{}=7 }_{\text{Ho$^{3+}$}}	=
		&	-\frac{1}{2} \sqrt{3} \hat{f}^{\dagger}_{-3,\frac{1}{2}} \hat{f}^{\dagger}_{-2,\frac{1}{2}} \hat{f}^{\dagger}_{-1,\frac{1}{2}} \hat{f}^{\dagger}_{0,\frac{1}{2}} \hat{f}^{\dagger}_{0,-\frac{1}{2}} \hat{f}^{\dagger}_{1,\frac{1}{2}} \hat{f}^{\dagger}_{2,\frac{1}{2}} \hat{f}^{\dagger}_{2,-\frac{1}{2}} \hat{f}^{\dagger}_{3,\frac{1}{2}} \hat{f}^{\dagger}_{3,-\frac{1}{2}}
\ket{0}_{\text{Ho$^{3+}$}}	\\
		&	\quad	+\frac{1}{4} \hat{f}^{\dagger}_{-3,-\frac{1}{2}} \hat{f}^{\dagger}_{-2,\frac{1}{2}} \hat{f}^{\dagger}_{-1,\frac{1}{2}} \hat{f}^{\dagger}_{0,\frac{1}{2}} \hat{f}^{\dagger}_{1,\frac{1}{2}} \hat{f}^{\dagger}_{1,-\frac{1}{2}} \hat{f}^{\dagger}_{2,\frac{1}{2}} \hat{f}^{\dagger}_{2,-\frac{1}{2}} \hat{f}^{\dagger}_{3,\frac{1}{2}} \hat{f}^{\dagger}_{3,-\frac{1}{2}}
\ket{0}_{\text{Ho$^{3+}$}}	\\
		&\quad	\quad	+\frac{1}{4} \hat{f}^{\dagger}_{-3,\frac{1}{2}} \hat{f}^{\dagger}_{-2,-\frac{1}{2}} \hat{f}^{\dagger}_{-1,\frac{1}{2}} \hat{f}^{\dagger}_{0,\frac{1}{2}} \hat{f}^{\dagger}_{1,\frac{1}{2}} \hat{f}^{\dagger}_{1,-\frac{1}{2}} \hat{f}^{\dagger}_{2,\frac{1}{2}} \hat{f}^{\dagger}_{2,-\frac{1}{2}} \hat{f}^{\dagger}_{3,\frac{1}{2}} \hat{f}^{\dagger}_{3,-\frac{1}{2}}
\ket{0}_{\text{Ho$^{3+}$}}	\\
		&\quad	\quad	\quad	+\frac{1}{4} \hat{f}^{\dagger}_{-3,\frac{1}{2}} \hat{f}^{\dagger}_{-2,\frac{1}{2}} \hat{f}^{\dagger}_{-1,-\frac{1}{2}} \hat{f}^{\dagger}_{0,\frac{1}{2}} \hat{f}^{\dagger}_{1,\frac{1}{2}} \hat{f}^{\dagger}_{1,-\frac{1}{2}} \hat{f}^{\dagger}_{2,\frac{1}{2}} \hat{f}^{\dagger}_{2,-\frac{1}{2}} \hat{f}^{\dagger}_{3,\frac{1}{2}} \hat{f}^{\dagger}_{3,-\frac{1}{2}}
\ket{0}_{\text{Ho$^{3+}$}}	\\
		&\quad	\quad	\quad	\quad	+\frac{1}{4} \hat{f}^{\dagger}_{-3,\frac{1}{2}} \hat{f}^{\dagger}_{-2,\frac{1}{2}} \hat{f}^{\dagger}_{-1,\frac{1}{2}} \hat{f}^{\dagger}_{0,-\frac{1}{2}} \hat{f}^{\dagger}_{1,\frac{1}{2}} \hat{f}^{\dagger}_{1,-\frac{1}{2}} \hat{f}^{\dagger}_{2,\frac{1}{2}} \hat{f}^{\dagger}_{2,-\frac{1}{2}} \hat{f}^{\dagger}_{3,\frac{1}{2}} \hat{f}^{\dagger}_{3,-\frac{1}{2}}
\ket{0}_{\text{Ho$^{3+}$}}
	\end{split}
\label{eq:ManyHo7}		
\end{gather}
\end{widetext}
can be obtained, and so on and so forth. (States $\ket{M<J}$ are very copious expansions; e.g., $\ket{M=0}_{\text{Ho$^{3+}$}}$ is a superposition of about fifty terms). 

%%%%%%%%%%%%%%%%%%%%%%%%%%%%
\subsection{Exchange parameters}
\label{sec:ExchParameters}

The first step in determining the exchange parameters consists of projecting the two-body exchange Hamiltonian onto the CEF ground state doublets of both spins, thus reducing it to a pseudospin-1/2 system. The diagonal part of the resulting Hamiltonian takes the generic form (for either Kramers or non-Kramers ions) 
\begin{equation}
	\mathcal{H}^{\rm diag}_{\mathrm{eff}} (\mathbf{r}, \mathbf{r'}) = 
		J_{\mathrm{nn}}
		\sigma_{\mathbf{r}}^{z} \otimes \sigma_{\mathbf{r'}}^{z}
\, .	
\label{eq:Eq18OnodaMatrixDiagonal}
\end{equation}
We shall thus require that the parameters in $\hat{\mathcal{H}}_{f\!f}$ 
also allow it to reduce to this projected form, namely that in the 
$\sigma^z$ product basis states, 
$
E_{\pm\pm} \equiv 
\langle \pm_{\mathbf{r}} \pm_{\mathbf{r'}} \vert 
\hat{\mathcal{H}}_{f\!f} 
\vert \pm_{\mathbf{r}} \pm_{\mathbf{r'}} \rangle 
= 
- \langle \pm_{\mathbf{r}} \mp_{\mathbf{r'}} \vert 
\hat{\mathcal{H}}_{f\!f} 
\vert \pm_{\mathbf{r}} \mp_{\mathbf{r'}} \rangle
\equiv 
-E_{\pm\mp}
$. 

Before explicitly projecting $\hat{\mathcal{H}}_{f\!f}(\mathbf{r}, \mathbf{r'})$ onto the GS-doublet, it is convenient to write it in a more compact notation as 
\begin{gather}	
	\begin{split}
		\hat{\mathcal{H}}_{f\!f} (\mathbf{r}, \mathbf{r'})
		&
		=
		\,	
		\mathcal{E}_{\rm exc}
		\sum_{ \mathbf{q}}
		\hat{f}^{\dagger	}_{\mathbf{r},	m_{1},\sigma_{1}}								
		\hat{f}^{				}_{\mathbf{r},	m_{2},\sigma_{2}}		
		\hat{f}^{\dagger	}_{\mathbf{r'},	m'_{1},\sigma'_{1}}				
		\hat{f}^{				}_{\mathbf{r'},	m'_{2},\sigma'_{2}}					
		\\
		\times
		&
		x^{\vert m_{1}\vert+\vert m'_{1}\vert+\vert m_{2}\vert+\vert m'_{2}\vert} 
		\Bigg[
		a	\;	
			\delta \left(  \mathbf{q} \right)
			+
			\rho \left(  \mathbf{q} \right)
		\Bigg]	
		,
	\end{split}
\label{eq:Gen17OnodaCompact}
\end{gather}
where we introduced 
\begin{subequations}	
	\begin{alignat}{1}% insert inside {} the number of `columns' per line (of equation)
		\sum_{ \mathbf{q}}
		&	\equiv	
			\sum_{\substack
						{
							m_{1},		m_{2},			m'_{1},			m'_{2} 			=	0,	\pm1		\\
							\sigma_{1},	\sigma_{2},	\sigma'_{1},	\sigma'_{2}	=	\pm
						}
			},	\label{eq:CompactNomenclature-sum_q}
		\\[0.1cm]					
		\delta \left(  \mathbf{q} \right) 	
		&	\equiv 
			\delta_{  \substack{ m_{1},	m_{2} \\ \sigma_{1}, \sigma_{2} } } \delta_{  \substack{ m'_{1},	m'_{2} \\ \sigma'_{1}, \sigma'_{2} } },
			\label{eq:CompactNomenclature-delta_q}
		\\[0.1cm]					
		\rho \left(  \mathbf{q} \right)	
		&	\equiv  
			(\mathcal{R}^{\dagger}_{\mathbf{r}}	\mathcal{R}_{\mathbf{r'}}	)_{\substack{ m_{1}, m'_{2} \\ \sigma_{1}, \sigma'_{2} } }
			(\mathcal{R}^{\dagger}_{\mathbf{r'}}	\mathcal{R}_{\mathbf{r}}	)_{\substack{ m'_{1},	m_{2} \\ \sigma'_{1}, \sigma_{2} } },
			\label{eq:CompactNomenclature-rho_q}
	\end{alignat}
	\label{eq:CompactNomenclature}
\end{subequations}
and the parameters
\begin{eqnarray}	
\mathcal{E}_{\rm exc}
&=& 2 \frac{V_{pf\sigma}^4}{(n U - \Delta)^2} 
  \left( \frac{1}{n U - \Delta}	+ \frac{1}{U} 	\right) 
\, , 
\label{eq:new_exch_constants-Eexc}
\\[0.1cm]
a	&=& 	\frac{U}{\Delta - U(n+1)}
\, ,
\label{eq:new_exch_constants-a}	
\\[0.1cm]
x &=& \frac{ V_{pf\pi}}{ V_{pf\sigma}} 
\, . 
\label{eq:new_exch_constants-x}
\end{eqnarray}
The ratio between the two Slater-Koster parameters $x$ allows to write more compactly 
\begin{equation}
\frac{V_{m_{1}}	\, V_{m'_{1}} \, V_{m_{2}}	\, V_{m'_{2}}}{ V_{pf\sigma}^{4}}
= 
x^{p},%x^{\vert m_{1}\vert+\vert m'_{1}\vert+\vert m_{2}\vert+\vert m'_{2}\vert} 
\end{equation}
where $p=\vert m_{1}\vert+\vert m'_{1}\vert+\vert m_{2}\vert+\vert m'_{2}\vert$. Upon projecting onto the CEF ground state doublet states, we find that $E_{++}=E_{--}$ and $E_{+-}=E_{-+}$ by symmetry. Therefore we are left with only one condition to impose: $E_{++}=-E_{+-}$. We note that, in all the equations, $\mathcal{E}_{\rm exc}$ cancels out and we are left with a relation between $a$ and $x$: 
\begin{gather}
	\begin{split}
		a(x) = \frac {\sum_{p=0}^{4} a_{p} \, x^{p}}{\sum_{p=0}^{4} d_{p} \, x^{p}} 
\, ,
	\end{split}
\label{eq:a_x}
\end{gather}
where the coefficients $(a_{p}, d_{p})$ are material-specific via the CEF parameters. As one may expect, the results are independent of the choice of nearest neighbour pair $\mathbf{r}$ and $\mathbf{r'}$. In Tab.~\ref{tab:apdp_coeffs} we list the values $(a_{p}, d_{p})$ of interest in the present work, obtained using the CEF parameters in Ref.~\cite{Ruminy:2016} (for HTO and DTO), and in Refs.~\cite{Princep:2013,Kimura:2013} (respectively, for PSO and PZO). 

\begin{table}[h!]
	\centering
	\setlength{\tabcolsep}{0.005pt}
	\def\arraystretch{1.2}%
	\begin{tabular}{  c | c | c | c | c | c | }
		&
			\begin{tabular}{ c }
				$a_{0}$
				\\%[-.02\columnwidth]
				$d_{0}$
			\end{tabular}					
		&
			\begin{tabular}{ c }
				$a_{1}$
				\\%[-.02\columnwidth]
				$d_{1}$
			\end{tabular}					
		&
			\begin{tabular}{ c }
				$a_{2}$
				\\%[-.02\columnwidth]
				$d_{2}$
			\end{tabular}					
		&
			\begin{tabular}{ c }
				$a_{3}$
				\\%[-.02\columnwidth]
				$d_{3}$
			\end{tabular}
		&
			\begin{tabular}{ c }
				$a_{4}$
				\\%[-.02\columnwidth]
				$d_{4}$
			\end{tabular}					\\
		\hline							
		\hline							
%		\begin{turn}{90}
			\textbf{HTO}
%		\end{turn}
		&
			\begin{tabular}{ c }
				$-1.14\times 10^{-1}$
				\\%[-.02\columnwidth]
				$2.06$
			\end{tabular}					
		&
			\begin{tabular}{ c }
				$0$
				\\%[-.02\columnwidth]
				$0$
			\end{tabular}					
		&
			\begin{tabular}{ c }
				$-2.71$
				\\%[-.02\columnwidth]
				$1.22\times 10^1$
			\end{tabular}					
		&
			\begin{tabular}{ c }
				$-1.20\times 10^{-2}$
				\\%[-.02\columnwidth]
				$0$
			\end{tabular}
		&
			\begin{tabular}{ c }
				$-2.55$
				\\%[-.02\columnwidth]
				$1.80\times 10^1$
			\end{tabular}					\\
		\hline							
%		\begin{turn}{90}
			\textbf{DTO}
%		\end{turn}
		&
			\begin{tabular}{ c }
				$-1.15\times 10^{-1}$
				\\%[-.02\columnwidth]
				$2.06$
			\end{tabular}					
		&
			\begin{tabular}{ c }
				$0$
				\\%[-.02\columnwidth]
				$0$
			\end{tabular}					
		&
			\begin{tabular}{ c }
				$-1.82$
				\\%[-.02\columnwidth]
				$8.20$
			\end{tabular}					
		&
			\begin{tabular}{ c }
				$-1.98\times 10^{-5}$
				\\%[-.02\columnwidth]
				$0$
			\end{tabular}
		&
			\begin{tabular}{ c }
				$-1.13$
				\\%[-.02\columnwidth]
				$8.14$
			\end{tabular}					\\
		\hline							
%		\begin{turn}{90}
			\textbf{PSO}
%		\end{turn}
		&
			\begin{tabular}{ c }
				$-1.18\times 10^{-3}$
				\\%[-.02\columnwidth]
				$2.13\times 10^{-2}$
			\end{tabular}					
		&
			\begin{tabular}{ c }
				$0$
				\\%[-.02\columnwidth]
				$0$
			\end{tabular}					
		&
			\begin{tabular}{ c }
				$-3.65\times 10^{-2}$
				\\%[-.02\columnwidth]
				$1.23 \times 10^{-1}$
			\end{tabular}					
		&
			\begin{tabular}{ c }
				$3.06\times 10^{-3}$
				\\%[-.02\columnwidth]
				$0$
			\end{tabular}
		&
			\begin{tabular}{ c }
				$-2.45\times 10^{-2}$
				\\%[-.02\columnwidth]
				$1.76\times 10^{-1}$
			\end{tabular}					\\
		\hline							
%		\begin{turn}{90}
			\textbf{PZO}
%		\end{turn}
		&
			\begin{tabular}{ c }
				$-1.74\times 10^{-4}$
				\\%[-.02\columnwidth]
				$3.14\times 10^{-3}$
			\end{tabular}					
		&
			\begin{tabular}{ c }
				$0$
				\\%[-.02\columnwidth]
				$0$
			\end{tabular}					
		&
			\begin{tabular}{ c }
				$1.00\times 10^{-2}$
				\\%[-.02\columnwidth]
				$3.30\times 10^{-2}$
			\end{tabular}					
		&
			\begin{tabular}{ c }
				$7.40\times 10^{-4}$
				\\%[-.02\columnwidth]
				$0$
			\end{tabular}
		&
			\begin{tabular}{ c }
				$-1.20\times 10^{-2}$
				\\%[-.02\columnwidth]
				$8.60\times 10^{-2}$
			\end{tabular}					\\
		\hline							
	\end{tabular}
	\caption
		{
		The $(a_{p}, d_{p})$ coefficients parametrising the $a=a(x)$ relationships for different pyrochlore systems.
		}
	\label{tab:apdp_coeffs}
\end{table}	

The second step consists of comparing the magnitude of the lowest energy gap of the projected $\hat{\mathcal{H}}_{f\!f} (\mathbf{r}, \mathbf{r'})$ with the magnitude of the corresponding gap in $\mathcal{H}^{\rm diag}_{\mathrm{eff}} (\mathbf{r}, \mathbf{r'})$. This allows us to find the dependence $\mathcal{E}_{\rm exc} = \mathcal{E}_{\rm exc} (x, J_{\mathrm{nn}})$. Note that the value of $J_{\rm nn}$ can be related to experimental measurements available in the literature, e.g. the Curie-Weiss temperature and the Schottky anomaly~\cite{Bramwell:2001,Kimura:2013,Princep:2013}. 

\begin{figure}
	\centering
	\subfloat	[\label{fig:ConfigAfPsi0}]
					{\includegraphics[trim=	0mm		0mm		0mm		0mm, clip, width=.48\linewidth]{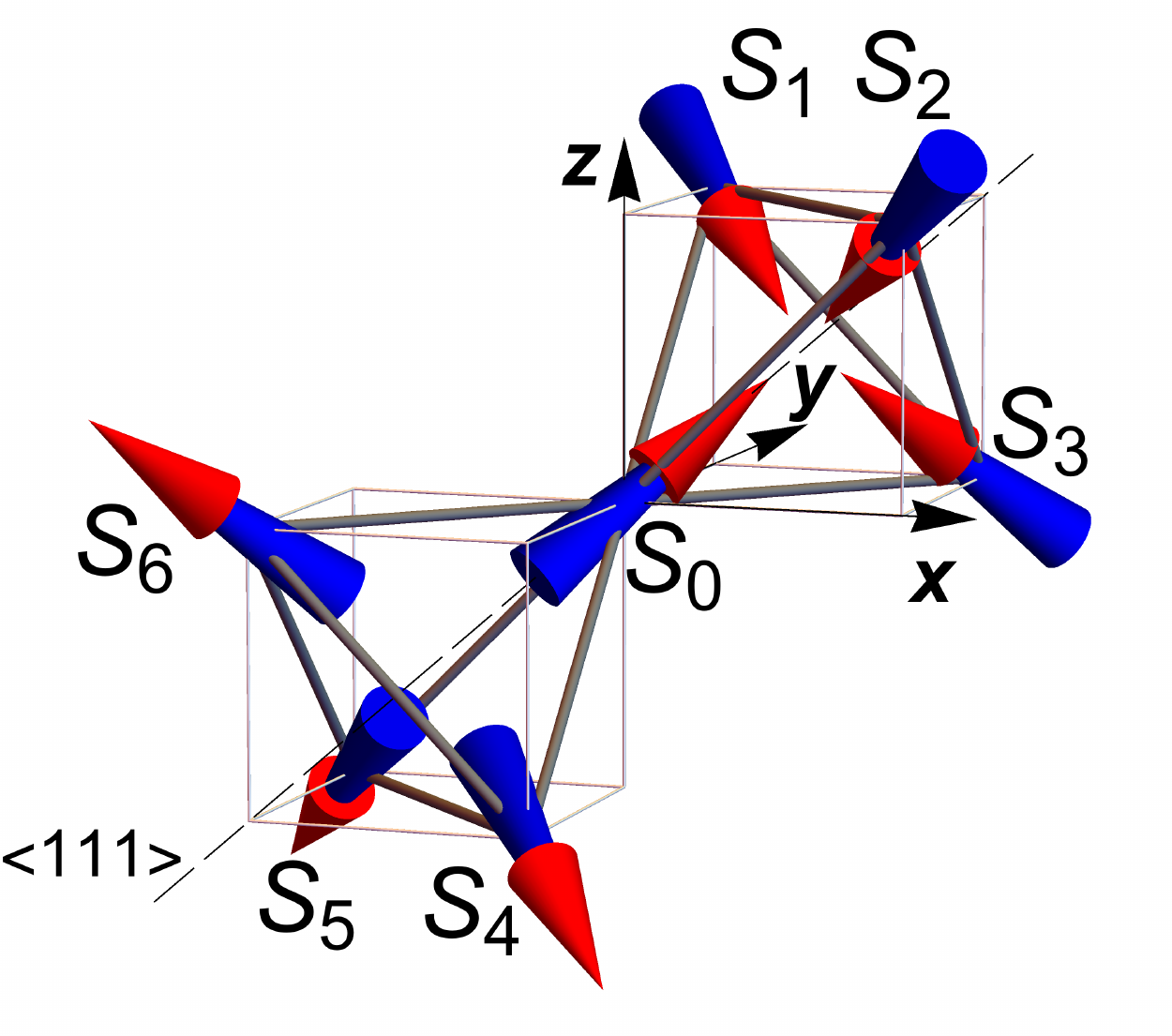}}
	%trim option's parameter order: 	left 		bottom 	right 		top
	\quad
	\subfloat	[\label{fig:ConfigAfPsi1}]
					{\includegraphics[trim=	0mm		0mm		0mm		0mm, clip, width=.48\linewidth]{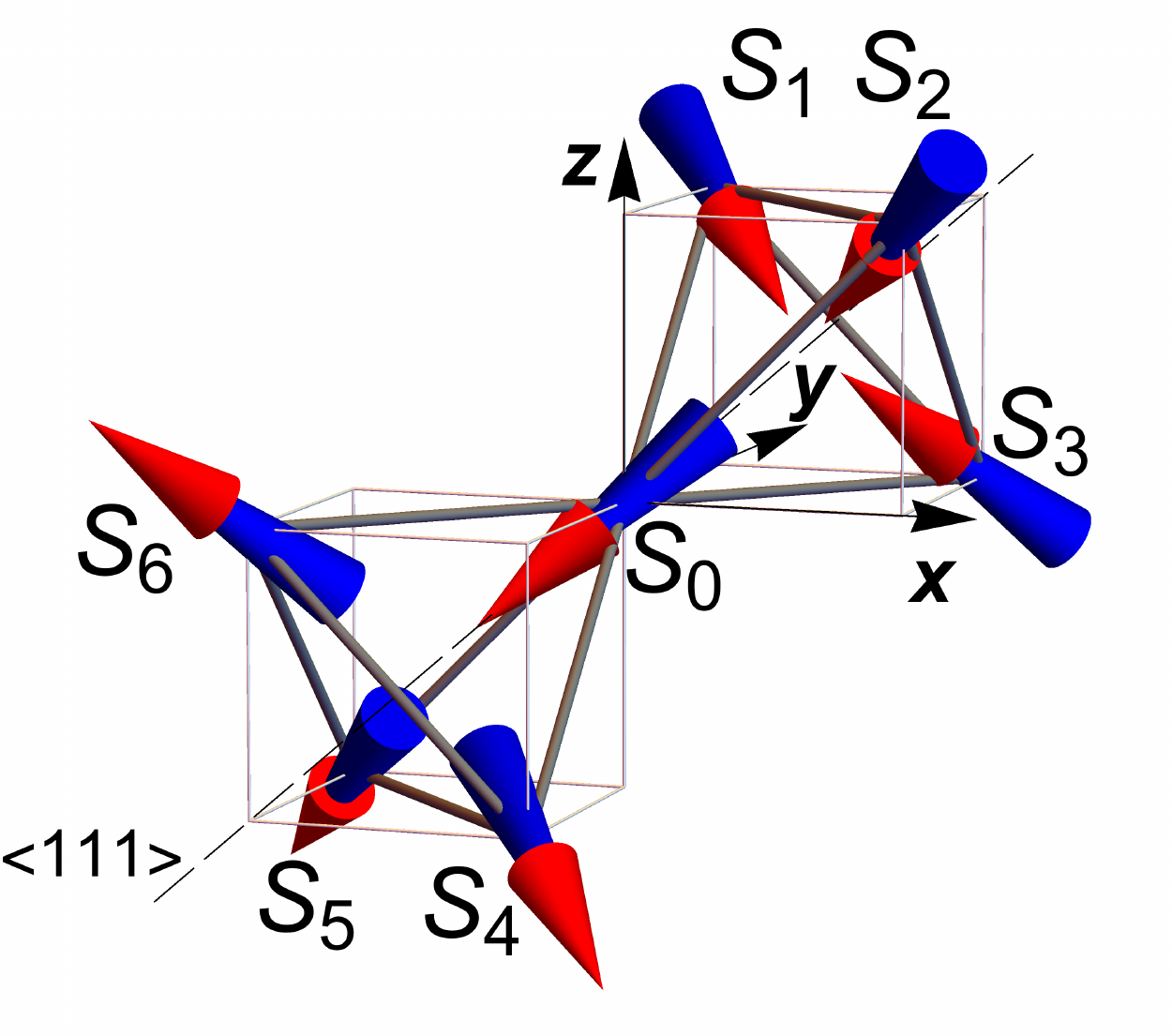}}
	%trim option's parameter order: 	left 		bottom 	right 		top
	\caption{	
		Examples of an all-in--all-out configuration  (a), and of a 3in-1out--3out-1in configuration (b) for a 2-tetrahedron system. The two panels have the same configuration of outer spins, and the direction of the central dipole moment in the lowest energy state tells us the ferro vs. antiferromagnetic nature of the exchange interaction. The direction in panel (a) is expected for HTO and DTO ($J_{\mathrm{nn}}<0$). Whereas the direction in panel (b) is expected for PSO and PZO ($J_{\mathrm{nn}}>0$). 
	}
	\label{fig:AFconfigs}
\end{figure}

	The third and final step is to fix the parameter $x$. Ref.~\cite{Rau:2015} argues that reasonable values for $x$ are in the range $x \in (-1,0)$, and Ref.~\cite{Onoda:2011} sets $x=-0.3$ for Pr$^{3+}$ pyrochlores (PSO and PZO included). Having obtained the relationships $a(x)$ and $\mathcal{E}_{\rm exc} (x)$ as above, we find that varying $x\in(-1,0)$ has essentially no effect for HTO and DTO parameters, and is consistent throughout with the known sign of the exchange coupling, $J_{\rm nn} < 0$ (i.e., favouring all-in and all-out states). On the contrary, varying $x$ in the same range does affect the behaviour for PSO and PZO parameters. This is most conveniently seen if we assemble the Hamiltonian for a 2-tetrahedron cluster, projected, as illustrated in Fig.~\ref{fig:AFconfigs}, onto a given choice of CEF ground states for the 6 outer spins, 
\begin{gather}
		\hat{\mathcal{H}}_{\rm exc}(0)=
		\sum_{j=1}^{6}
		\bra{\pm}_{j} \hat{\mathcal{H}}_{f\!f} (\mathbf{r}_{0},\mathbf{r}_{j}) \ket{\pm}_{j},
\, ,
\label{eq:HamExcQuasiQuantAppendix}
\end{gather}
which operates in the  $2 J+1$ dimensional Hilbert space of the central ion. Eq.~\eqref{eq:HamExcQuasiQuantAppendix} is then a single ion Hamiltonian that can be added to the CEF Hamiltonian for the central ion, and diagonalised to obtain, say, the behaviour of its ground state dipole moment $\mathbf{m}= g_{J} \mu_{\mathrm{B}} \braket{\hat{\mathbf{J}}}_0$. We find that the value of the moment depends on $x$, and more importantly it can invert its direction, thus changing the ferro/antiferromagnetic nature of the exchange interaction. From experiments we know that the magnetic dipole moment in PSO and PZO is, respectively, $\mathbf{m} \approx 2.5~\mu_{\mathrm{B}}$ and $\mathbf{m} \approx 3~\mu_{\mathrm{B}}$, and so we expect the exchange interactions to be dominant over the dipolar ones. Therefore, existing evidence of spin ice behaviour (2in-2out low energy states) implies that the nearest neighbour exchange coupling is frustrated (i.e., it has the opposite sign as in HTO and DTO). These conditions are generally verified in our approach for $x \in (-1,-0.3)$ (PSO) and $x \in (-1,-0.6)$ (PZO), although we noticed a remarkable sensitivity of the GS dipole moment $\mathbf{m}$ on the value of $x$ in PZO that may be worth investigating further in the future. In summary, we find that $x \approx -1$ is a good working value for all the systems considered in this study, and we therefore use it throughout the manuscript.

%%%%%%%%%%%%%%%%%%%%%%%%%%%%%%%%%%%%%%%%%
\section{Quantum Zeno effect}
\label{sec:Zeno}

	It is interesting to illustrate how the Quantum Zeno effect~\cite{Misra:1977} can lead to an increase in spin flip timescales using the following toy model. We consider a spin-1/2 degree of freedom initially prepared, say, in the `up' state, in presence of a transverse field that makes it precess on a timescale $\tau$. The system is coupled to a `model environment' that observes the state of the spin at random times with respect to the chosen initial basis, thus projecting it either onto the `up' or `down' state. The times between consecutive observations $\mu$ are drawn from an exponential 
%(\textcolor{red}{change\#A02}: exponential corrects Poissonian) 
distribution with characteristic time $\tau_{\rm o}$, 
\begin{gather}
	\begin{split}
		p \left( \mu \right)	=	\frac	{\exp \left( -\mu / \tau_{\rm o} \right)}
						{\tau_{\rm o}}
	\end{split}
\, . 
\label{eq:p_mu}
\end{gather}

Let us define $\ket{\psi_{0}}$ to be the initial state of the system, $\hat{U}_{j}\equiv \exp \left( -i \hat{H} \mu_{j} / \hbar \right)$ to be the unitary time evolution operator due the chosen transverse field Hamiltonian $\hat{H}$ over a time $\mu_{j}$, and for convenience we introduce the notation 
\begin{eqnarray}
	q_{j} = \left| \braket{ \psi_{0} | \hat{U}_{j} | \psi_{0} }	\right|^{2} 
	&=& \cos^{2}	\left( \pi \mu_{j}/ \tau\right)		
%	\qquad	&	\text{no-flip    (survival)},	 
\nonumber \\
	1- q_{j} &=& 	\sin^{2}	\left( \pi \mu_{j}/ \tau\right)	
%	\qquad	&	\text{flip} . 
\, . 
\label{eq:qjCos2Sin2}
\end{eqnarray}
The survival probability associated with a sequence of observations at times $\left\{ \mu_{j} \right\} = \left\{\mu_{1}, \mu_{2}, \dots, \mu_{m} \right\}$, stochastically drawn from $p(\mu_{j})$, can be expressed as~\cite{Gherardini:201601:NJP, Gherardini:2017}: 
\begin{gather}
	\begin{split}
		\mathcal{P}(\left\{ \mu_{j} \right\})	=	\prod_{j=1}^{m}		q(\mu_{j})
	\end{split}
\, . 
\label{eq:SurvivalProb}
\end{gather}
The average value of the survival probability after $m$ observations is therefore~\cite{Gherardini:201601:NJP}
\begin{gather}
	\begin{split}
		\braket{ \mathcal{P}(m) }	
	&	=	
			\left[ 
				\int_{\mu}	d\mu		\,	p \left( \mu \right)	q(\mu)
			\right]^{m}			
	\\
	&	=
			\exp
			\left( 
			m	\ln	\int_{\mu}	d\mu		\,	p \left( \mu \right)	q(\mu)
			\right) \, . 
	\end{split}
\label{eq:AverageSurvProb}
\end{gather}
Substituting $p \left( \mu \right)$ and $q(\mu)$ in Eq.~\eqref{eq:AverageSurvProb}, we obtain the (no-flip) survival probability
\begin{gather}
	\begin{split}
		\braket{ \mathcal{P}(m) }
	&		=	
			\exp
				\left[ 
					m	\ln	\int_{0}^{\infty}	d\mu		\,	
			    \frac
			    	{
				e^{-\mu / \tau_{\rm o}}
				}
				{\tau_{\rm o}}					
				\cos^{2} \left(\pi \mu/\tau\right)
				\right]
	\\
	&		=
			\exp
				\left[
					m	\ln	
						\left( 
						  \frac
						  	{\tau^{2} +2 \pi^{2} \tau_{\rm o}^{2}}
						  	{\tau^{2} +4 \pi^{2} \tau_{\rm o}^{2}}
 						\right)
 				\right] 
				\, . 
	\end{split}
\label{eq:AverageSurvProbTau0Tau1}
\end{gather}

This result can be readily modified to obtain the probability that the spin survives $m-1$ observations and flips on the following ($m$-th) one: 
\begin{eqnarray}
\braket{ \mathcal{P}_{\rm flip}(m) } \! 
			&=& \!\! 
			 \left[
			  \int_{\mu}
			   d\mu	\,
			   p	\left( \mu \right)	
			   \left[ 1- q \left( \mu \right) \right]
			 \right] \!\!\! 
			 \left[
			  \int_{\mu}
			   d\mu	\,			  
			   p	\left( \mu \right)	
			   q \left( \mu \right)
			 \right]^{m-1}
\nonumber \\ 
			&=&
			 \exp
			  \left[
			   \ln	
				 \frac{2 \pi^2 \tau_{\rm o}^2}
				      {\tau^2 + 4 \pi^2 \tau_{\rm o}^2}
			  \right]	
\nonumber \\
	&&
			\exp
				\left[
					(m-1)	\ln	
						\left( 
						  \frac
						  	{\tau^{2} +2 \pi^{2} \tau_{\rm o}^{2}}
						  	{\tau^{2} +4 \pi^{2} \tau_{\rm o}^{2}}
 						\right)
 				\right] 
\, . 
\label{eq:AverageSurvProbTau0Tau1}
\end{eqnarray}

	The average time for $m$ observations is $\Delta t = m \tau_{\rm o}$, and its distribution becomes progressively more peaked the larger the number of observations $m$, by the central limit theorem. Therefore, for sufficiently large values of $m$, it is reasonable to carry out the approximate change of variable $m = \Delta t / \tau_{\rm o}$, 
\begin{eqnarray}
		\braket{ \mathcal{P}_{\rm flip}(m) }	dm
		&=& 
		\braket{ \mathcal{P}_{\rm flip}(\Delta t / \tau_{\rm o}) }	d \left( \Delta t / \tau_{\rm o} \right)
\nonumber \\ 	
	&\equiv&
		P	\left( \Delta t \right)	d \left( \Delta t \right) 
\, , 
\label{eq:AverageSurvProbDeltat}
\end{eqnarray}
and obtain the probability distribution of flipping in a time interval $\Delta t $: 
\begin{eqnarray}
&&
		P	\left( \Delta t \right)
		=
		\frac
		{1}
		{\tau_{\rm o}}
		{\braket{ \mathcal{P}_{\rm flip}(\Delta t / \tau_{\rm o})}}
\label{eq:ProbDensDeltat} \\ 
		&& \qquad =
		\frac
		{2 \pi^2 \tau_{\rm o}}
		{\tau^{2}+ 2 \pi^2 \tau_{\rm o}^{2}}
		 \exp
		  \left[
		   \frac{\Delta t}{\tau_{\rm o}} 
				\ln	\displaystyle{
					 \frac
					 {\tau^{2}+ 2 \pi^2 \tau_{\rm o}^{2}}
					 {\tau^{2}+ 4 \pi^2 \tau_{\rm o}^{2}} 
					}
			  \right]
\, 		.
\nonumber
\end{eqnarray}
From it, we finally obtain the average time to flip a spin 
\begin{gather}
	\begin{split}
		\braket{\Delta t}
		=
		\frac
		{2 \pi^2 \tau_{\rm o}^{3}}
		{(\tau^{2}+ 2 \pi^2 \tau_{\rm o}^{2})
			\left(
				\ln \displaystyle{\frac{\tau^{2}+ 2 \pi^2 \tau_{\rm o}^{2}}{\tau^{2}+ 4 \pi^2 \tau_{\rm o}^{2}} }
			\right)^{2}
		} 
	\, , 
	\end{split}
\label{eq:ExpectDeltat_mod}
\end{gather}
which is more conveniently expressed in units of $\tau_{\rm o}$ and as a function of $x=\tau_{\rm o}/\tau$: 
\begin{gather}
	\begin{split}
\frac{\braket{ \Delta  t }}{\tau_{\rm o}}	=
	\frac{2 \pi ^2}{\left(1/x^2+2 \pi ^2\right) \left( \ln \displaystyle{\frac{1/x^2+2 \pi ^2}{1/x^2+4 \pi ^2}}\right)^2}
	\, .
	\end{split}
\label{eq:ExpectDeltatScaled_mod}
\end{gather}
By looking at the asymptotic behaviour, 
\begin{subequations}	
	\begin{alignat}{2}% insert inside {} the number of ``columns'' per line (of equation)
		\frac{\braket{ \Delta  t }}{\tau_{\rm o}} 
		 \approx  
		 \frac{1}{2 \pi^{2} x^{2}} 		
				\qquad	&	\text{for}		\quad	x \ll	1	,	\quad	\tau_{\rm o} \ll	\tau \, ,	\label{eq:AsymptoticsLowx}	\\[0.2cm]
		\frac{\braket{ \Delta  t }}{\tau_{\rm o}} 
		 \approx
		 \frac{1}{(\ln2)^{2}}
				\qquad	&	\text{for}		\quad	x \gg1	,	\quad	\tau_{\rm o} \gg		\tau \, ,	\label{eq:AsymptoticsHighx}
	\end{alignat}
	\label{eq:Asymptotics}%	This ``%'' is to avoid wrong indentation of the text following
\end{subequations}
we immediately recognise the Quantum Zeno effect in the divergence of the spin flip timescale in the limit $x \to 0$. 

\begin{figure} 
	\centering
	{\includegraphics[trim=	0mm		0mm		0mm		0mm, clip, width=1\linewidth]{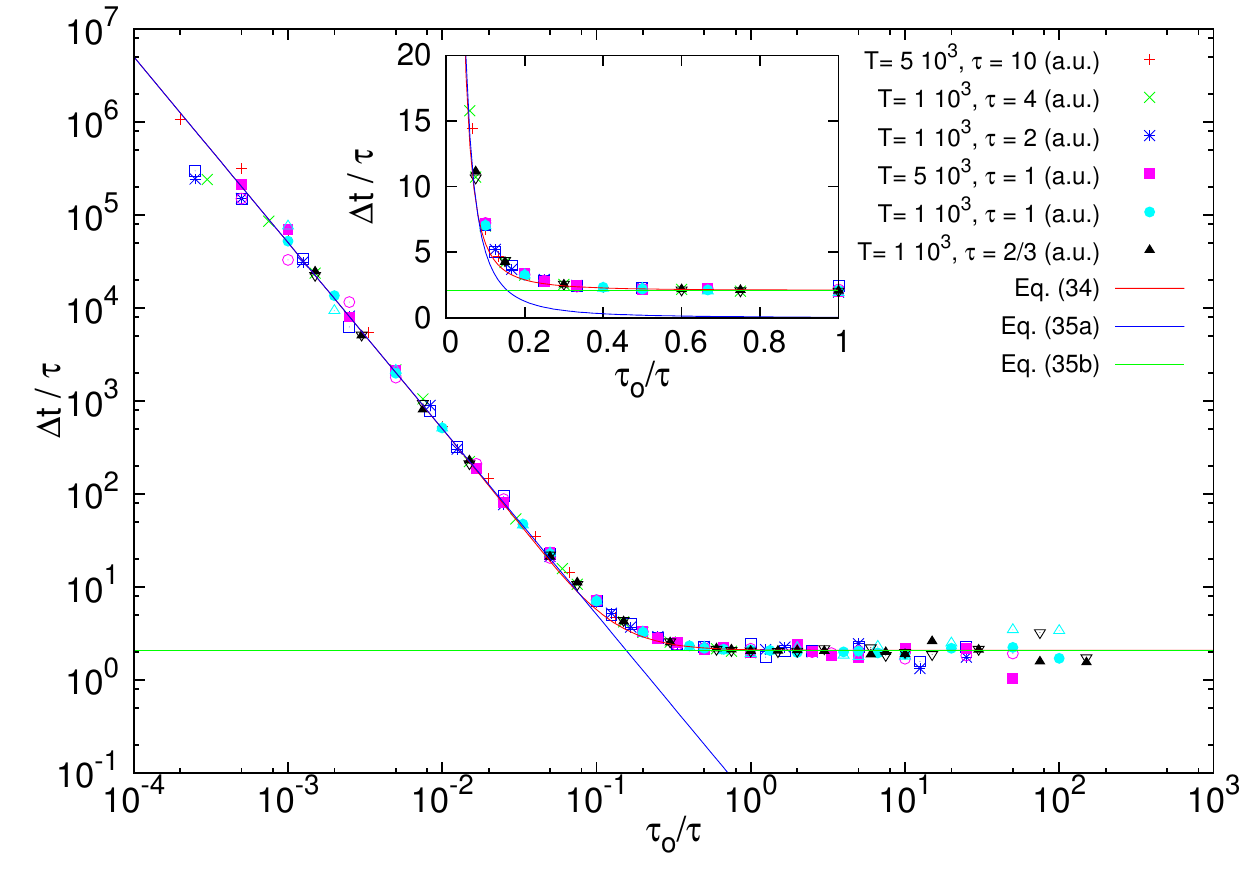}}
	\caption{	
			Comparison between the analytical expression Eq.~\eqref{eq:ExpectDeltatScaled_mod} for the average time of a spin flip (red solid line) and numerical simulations of the stochastic process. For each data point, we simulated observations at $T/\tau_{\rm o}$ randomly-selected times within a finite interval of duration $T$ (see legend).  Note the good agreement not only asymptotically, Eq.~\eqref{eq:Asymptotics} (straight green and blue solid lines), but also for finite values of $x=\tau_{\rm o}/\tau$. The devitations at small and large values of $x$ are statistical fluctuations due to the finite value of $T$. The inset shows the same data on a linear scale.  
	}
	\label{fig:ZenoAnalyticsVsNumerics}
\end{figure} 

	We note for completeness that the interpretation of $\Delta t$ as a spin flip timescale in the opposite limit of $x \to \infty$ is arguably questionable, as it corresponds to the case of a spin completing a large number of precessions between consecutive observations by the environment. 

	In Fig.~\ref{fig:ZenoAnalyticsVsNumerics} we compare the analytical result in Eq.~\eqref{eq:ExpectDeltatScaled_mod} with a numerical simulation of the quantum stochastic system. Notice the very good agreement already for a relatively small number of observations, $m \sim 25$. 

%\bibliography{BiblioPaper02v17}{}
%%\nocite{}